\documentclass[iop,referee]{emulateapj}
\usepackage{natbib}
\usepackage{graphicx,color,rotating}
\usepackage{footnote,lineno}
\usepackage{ulem} 
\usepackage{graphicx,amssymb,amsmath,amsfonts,times,hyperref}
\usepackage{subfigure}

\def \aap  {A\&A}

\def \apj  {ApJ}
\def \apjs  {ApJS}
\def \apjl  {ApJL}

\newcommand{\Fermi}{{\it Fermi}}

\shorttitle{Diffuse $\gamma$-ray emission from galactic pulsars}
\shortauthors{F. Calore et al.}

\begin{document}
\title{\boldmath Diffuse $\gamma$-ray emission from galactic pulsars}

   \author{F. Calore\altaffilmark{1}}
   \affil{GRAPPA Institute, University of Amsterdam, Science Park 904, 1090 GL Amsterdam, The Netherlands}
\author{M. Di Mauro\altaffilmark{2}}
\affil{Dipartimento di Fisica, Torino University and INFN, Sezione di Torino, Via P. Giuria 1, 10125 Torino, Italy,\\ Laboratoire d'Annecy-le-Vieux de Physique Th\'eorique (LAPTh), Univ. de Savoie, CNRS, B.P.110, Annecy-le-Vieux F-74941, France}
   \author{ F. Donato\altaffilmark{3}}
   \affil{Dipartimento di Fisica, Torino University and INFN, Sezione di Torino, Via P. Giuria 1, 10125 Torino, Italy}

\altaffiltext{1}{f.calore@uva.nl}
\altaffiltext{2}{mattia.dimauro@to.infn.it}
\altaffiltext{3}{donato@to.infn.it}

\text{LAPTH-037/14}

  \begin{abstract}
Millisecond Pulsars (MSPs) are old fast-spinning neutron stars and they represent the second most abundant source population discovered by the Large Area Telescope (LAT) on board of the {\it Gamma-Ray Space Telescope} ({\it Fermi}). As guaranteed $\gamma$-ray emitters, they might contribute non-negligibly to the diffuse emission measured at high latitudes by {\it Fermi}-LAT, i.e. the Isotropic Diffuse $\gamma$-Ray Background (IDGRB), which is believed to arise from the superposition of several components of galactic and extragalactic origin. Additionally, $\gamma$-ray sources also contribute to the anisotropy of the IDGRB measured on small scales by {\it Fermi}-LAT.
In this manuscript, we aim to assess the contribution of the unresolved counterpart of the detected MSPs population to the IDGRB and the maximal fraction of the measured anisotropy produced by this source class.
To this end, we model the MSPs spatial distribution in the Galaxy and the $\gamma$-ray emission parameters by considering observational constraints coming from the Australia Telescope National Facility pulsar catalog and the Second {\it Fermi}-LAT Catalog of $\gamma$-ray pulsars. 
By simulating a large number of MSPs populations through a Monte Carlo simulation, we compute the average diffuse emission and the anisotropy 1$\sigma$ upper limit.
We find that the emission from unresolved MSPs at 2 GeV, where the peak of the spectrum is located, is at most 0.9\% of the measured IDGRB above $10^{\circ}$ in latitude. 
The 1$\sigma$ upper limit on the angular power for unresolved MSP sources turns out to be about a factor of 60 smaller than {\it Fermi}-LAT measurements above $30^{\circ}$.
Our results indicate that this galactic source class represents a negligible contributor to the high-latitude $\gamma$-ray sky and confirm that most of the intensity and geometrical properties of the measured diffuse emission are imputable to other extragalactic source classes, e.g. blazars, mis-aligned active galactic nuclei or star-forming galaxies.
Nevertheless, because of the fact that MSPs are more concentrated towards the galactic center, we expect them to contribute significantly to the $\gamma$-ray diffuse emission at low latitudes. Since, along the galactic disk, the population of young pulsars overcomes in number the one of MSPs, we compute the $\gamma$-ray emission from the whole population of unresolved pulsars, both young and millisecond, in two low-latitude regions: the inner Galaxy and the galactic center.
  \end{abstract}


   \maketitle


\section{Introduction}
\label{sec:intro}
A guaranteed component of the $\gamma$-ray sky is represented by a faint and (almost) isotropic emission at latitudes $|b| \geq 10^{\circ}$. Such an Isotropic Diffuse $\gamma$-Ray Background (IDGRB) has been first suggested by the OSO-3 satellite \citep{1972ApJ...177..341K} and then measured by SAS-2
\citep{1975ApJ...198..163F} and EGRET \citep{1998ApJ...494..523S}.
The Large Area Telescope (LAT) on board of the {\it Gamma-Ray Space Telescope} ({\it Fermi}) has published a precise measurement of the IDGRB \citep{IDGRB} in the 200 MeV-100 GeV energy range, describing it with a single power-law with an index of $-2.41\pm0.15$. 
Recently, the {\it Fermi}-LAT $\gamma$-ray data have unveiled in the IDGRB an anisotropy signal at small scales \citep{Ackermann:2012uf}. Thus, while being still isotropic on large scales, the IDGRB presents fluctuations at $\theta \leq 2^{\circ}$, that are consistent with a population of point-like sources. One of the main puzzles for current $\gamma$-ray astrophysics is to understand the origin of this emission and its anisotropy, giving a coherent picture for those two measurements.
\newline
Typically, the IDGRB is thought to arise from the superposition of several contributions that can be ascribed to two main categories \citep{2012PhRvD..85b3004C}: the emission from the unresolved counterpart of known $\gamma$-ray point source emitters and the emission from diffuse processes involving interstellar gas and radiation fields.
As for the former, extragalactic and galactic source classes may participate in producing the measured IDGRB flux. In particular, active galactic nuclei (AGN), which represents the population with the largest detected counterpart, are believed to explain the most of the IDGRB, as it has been estimated for example in \cite{Collaboration:2010gqa,2012ApJ...751..108A,DiMauro:2013zfa,2011PhRvD..84j3007A,2014ApJ...780...73A} for blazars (BL Lacs objects and FSRQs) and in \cite{2011ApJ...733...66I,  RG} for mis-aligned AGN. 
Another extragalactic source of guaranteed diffuse emission is the unresolved population of star-forming galaxies, normal and starburst, that may even dominate the emission at few GeV because of the hadronic origin of the $\gamma$ rays \citep{2012ApJ...755..164A,Tamborra:2014xia}.
\\
The second most abundant population detected by the LAT is represented by galactic young pulsars and Millisecond Pulsars (MSPs). In particular, pulsars have been established as $\gamma$-ray emitters by the first observations of {\it Fermi}-LAT. Since the starting of the mission, the number of pulsars detected by LAT has increased significantly and the most up-to-date catalog of such objects is the Second {\it Fermi}-LAT Catalog of $\gamma$-ray pulsars (2FPC) \citep{2013ApJS..208...17A}. 

We mention here that, in addition, truly diffuse processes may contribute to the IDGRB. Among others we quote the $\gamma$-ray production from the interaction of ultra high energy cosmic rays with the CMB \citep{Kalashev:2007sn, 2011PhLB..695...13B}, the emission originating from photons up scattered through an Inverse Compton process by a population of highly relativistic electrons created during clusters mergers \citep{2007IJMPA..22..681B}, and $\gamma$ rays produced by the annihilation of Dark Matter (DM) particles in the Milky Way or in external galaxies, e.g.~\cite{1998APh.....9..137B,2010JCAP...11..041A,2013MNRAS.429.1529F,Calore:2014hna,Calore:2013yia}.

We do not aim in this context to give an extensive discussion of the different contributions to the high-latitude diffuse emission and we refer to \cite{2012PhRvD..85b3004C,Calore:2013yia,DiMauro:2013zfa,Cholis:2013ena} for more detailed explanations.
Nevertheless, we stress that in this papers it has been shown that current predictions of the unresolved emission from blazars, mis-aligned AGN, star-forming galaxies and MSPs could fully explain the IDGRB data in the {\it Fermi}-LAT whole energy range.

This work will focus on the galactic pulsars population and aims to assess the contribution to the IDGRB arising from the unresolved counterpart of this source class.
In particular, we are mostly interested to the high latitude $\gamma$-ray flux in the analysis of MSPs instead of young pulsars since such a population is expected to dominate the $\gamma$-ray emission in this region. MSPs are old, rapidly spinning neutron stars (with rotation period $P \leq 15$ ms) that are usually found (about 80 \% of MSPs) in binary systems and accrete matter from a companion \citep{1982Natur.300..728A}. 
Pulsars are believed to emit $\gamma$ rays from the conversion of their rotational kinetic energy. The initial rotational period (when the pulsar is born) slows down as a consequence of the magnetic-dipole braking \citep{Ng:2014qja,2000RSPTA.358..831L}. This decline is measured by the time derivative of the period, $\dot{P}$, which is related to the spin period, $P$, and the surface magnetic field, $B$ \citep{2013AA...554A..62G,2010JCAP...01..005F,2013ApJS..208...17A}:
\begin{eqnarray}
\label{eq:pdot}
\dot{P} = 9.8\cdot 10^{-26} \; \left(\frac{B}{\rm G} \right)^2  \left( \frac{P}{\rm{s}} \right)^{-1}  \,. 
\end{eqnarray}
As a consequence the loss energy rate, $\dot{E}$, i.e. spin-down luminosity, is \citep{2013AA...554A..62G}:
\begin{eqnarray}
\label{eq:edot}
  \dot{E} = 4 \pi^2 M \frac{\dot{P}}{P^3},
\end{eqnarray}
where $M$ is the moment of inertia of the star assumed to be $10^{45}$ g cm$^2$ \citep{2013AA...554A..62G,2010JCAP...01..005F,2013ApJS..208...17A}.
The spin-down luminosity is then converted with some efficiency into radiation.

For young pulsars that typically have periods of hundreds of ms, the slowing down of the period is rapid and they lose their energy very fast such that their $\gamma$-ray emission is substantially smaller than their older and faster spinning companions.
Indeed, assuming that the $\gamma$-ray luminosity follows the same relation $L_{\gamma} \propto {\dot{P}}^{1/2}P^{-3/2}$ \citep{2010JCAP...01..005F} for all pulsars we can write:
\begin{eqnarray}
\label{eq:confr}
  \frac{L^{\rm{MSP}}_{\gamma}}{L^{\rm{young}}_{\gamma}} = \left(\frac{\dot{P}_{\rm{MSP}}}{\dot{P}_{\rm{young}}}\right)^{1/2} \left(\frac{P_{\rm{MSP}}}{P_{\rm{young}}}\right)^{-3/2},
\end{eqnarray}
where typical values for the rotation period $P$ and the time derivative of the period $\dot{P}$ are: $P_{\rm{MSP}}=3$ ms, $P_{\rm{young}}=0.5$ s, $\dot{P}_{\rm{MPS}}=10^{-19}$ and $\dot{P}_{\rm{young}}=10^{-15}$ \citep{2004hpa..book.....L}. Therefore, $L^{\rm{MSP}}_{\gamma}/L^{\rm{young}}_{\gamma}\approx20$ meaning that the average $\gamma$-ray luminosity of MSPs is much higher than the one of young pulsars.
Moreover, due to their age MSPs are expected to distribute at higher latitudes with respect to young pulsars, that are instead concentrated along the galactic disk, within $|b| = 15^{\circ}$ \citep{2010JCAP...01..005F,2013ApJS..208...17A}. 

In the present analysis we derive the main characteristics of the pulsars population, namely the spatial distribution and the $\gamma$-ray emission parameters, using radio and $\gamma$-rays catalogs. We build a model for the pulsar emission that we use to generate Monte Carlo (MC) simulations of the population in order to predict the diffuse $\gamma$-ray flux originated from the non-detected source counterpart and to estimate the relevant theoretical uncertainty affecting our results.
The paper is organised as follows. In Sec.~\ref{sec:distribution} we describe the properties of the galactic MSPs observed in radio and we model their luminosity and spatial distributions performing fits based on radio observations. In Sec.~\ref{sec:gammaMSP}, we move to $\gamma$-ray observations and we individuate the main spectral and luminosity characteristics of the MSPs as detected by {\it Fermi}-LAT. With those ingredients we are able in Sec.~\ref{sec:MC} to set up a MC simulation of the MSP population in the Galaxy and to generate mock $\gamma$-ray emission from the unresolved counterpart of MSPs. Besides computing the diffuse emission coming from this population, we also calculate the anisotropy signal ascribable to such sources. Sec.~\ref{sec:results} is dedicated to the presentation of the results and their discussion. The $\gamma$-ray emission from unresolved MSPs at latitudes above $10^{\circ}$ is derived in Sec.~\ref{subsec:IDGRB}, while we estimate the contribution to the emission in the 
innermost part of the Galaxy in Sec.~\ref{subsec:Inner}. Since at low latitudes the population of young pulsars is more abundant than the MSP one, we study the spatial and $\gamma$-ray emission properties of young sources and we take into account also the contribution from young pulsars when analysing low-latitude regions.
The conclusions are presented in Sec.~\ref{sec:conclusions}.

\section{MSP distribution in the Galaxy}
\label{sec:distribution}
\begin{figure*}
 \begin{centering}
\includegraphics[width=0.60\textwidth]{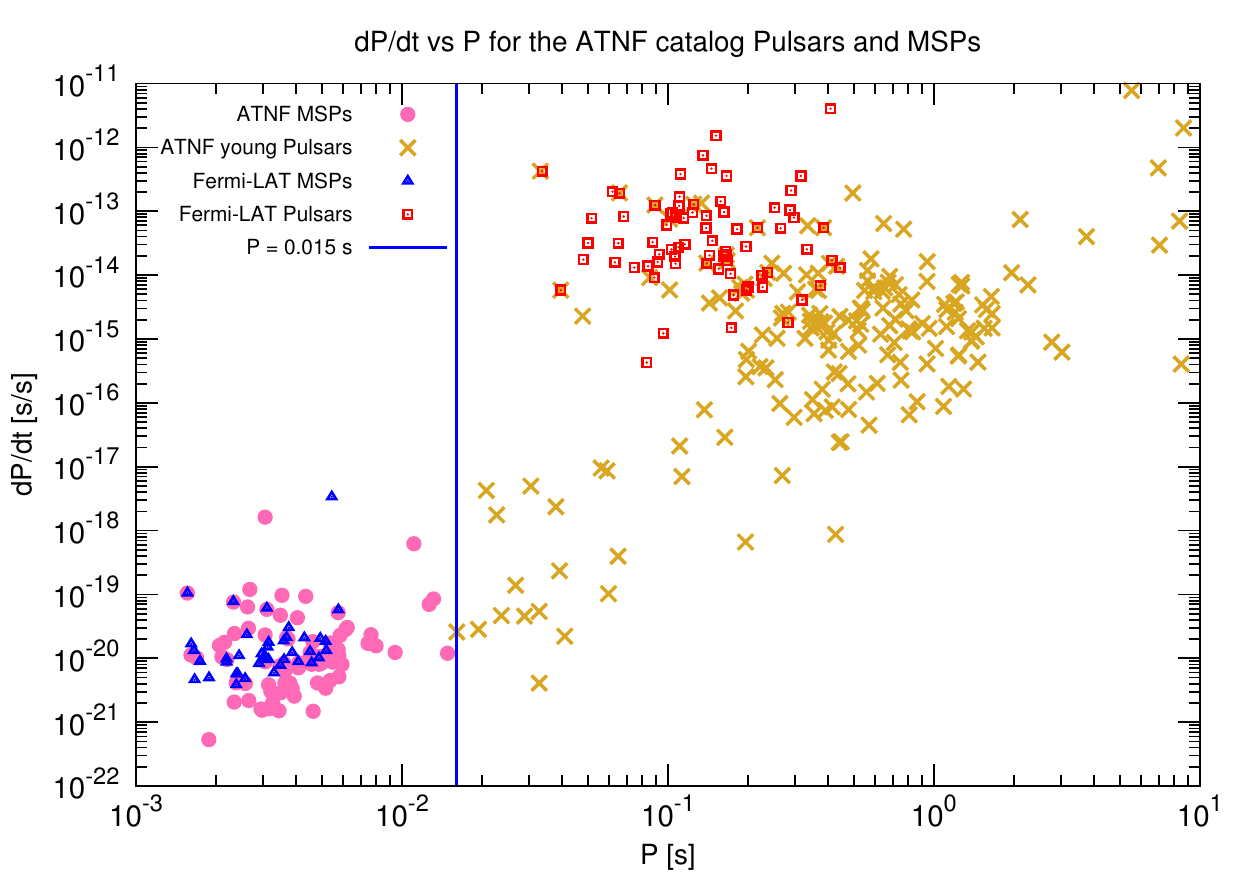}
\caption{Period $P$ and derivative of the period $\dot{P}$ for the MSPs (pink circles points) and young pulsars (gold crosses points) selected from the ATNF catalog  \citep{2005AJ....129.1993M}. We display also the {\it Fermi}-LAT MSPs (blue triangles points) and young pulsars (red squares points) selected from the 2FPC catalog  \citep{2013ApJS..208...17A}. The solid blue line sets the threshold value $P=15$ ms which separates the population into MSPs and young pulsars.}
\label{fig:popdot} 
\end{centering}
\end{figure*}

\begin{figure*}
 \begin{centering}
\includegraphics[width=0.60\textwidth]{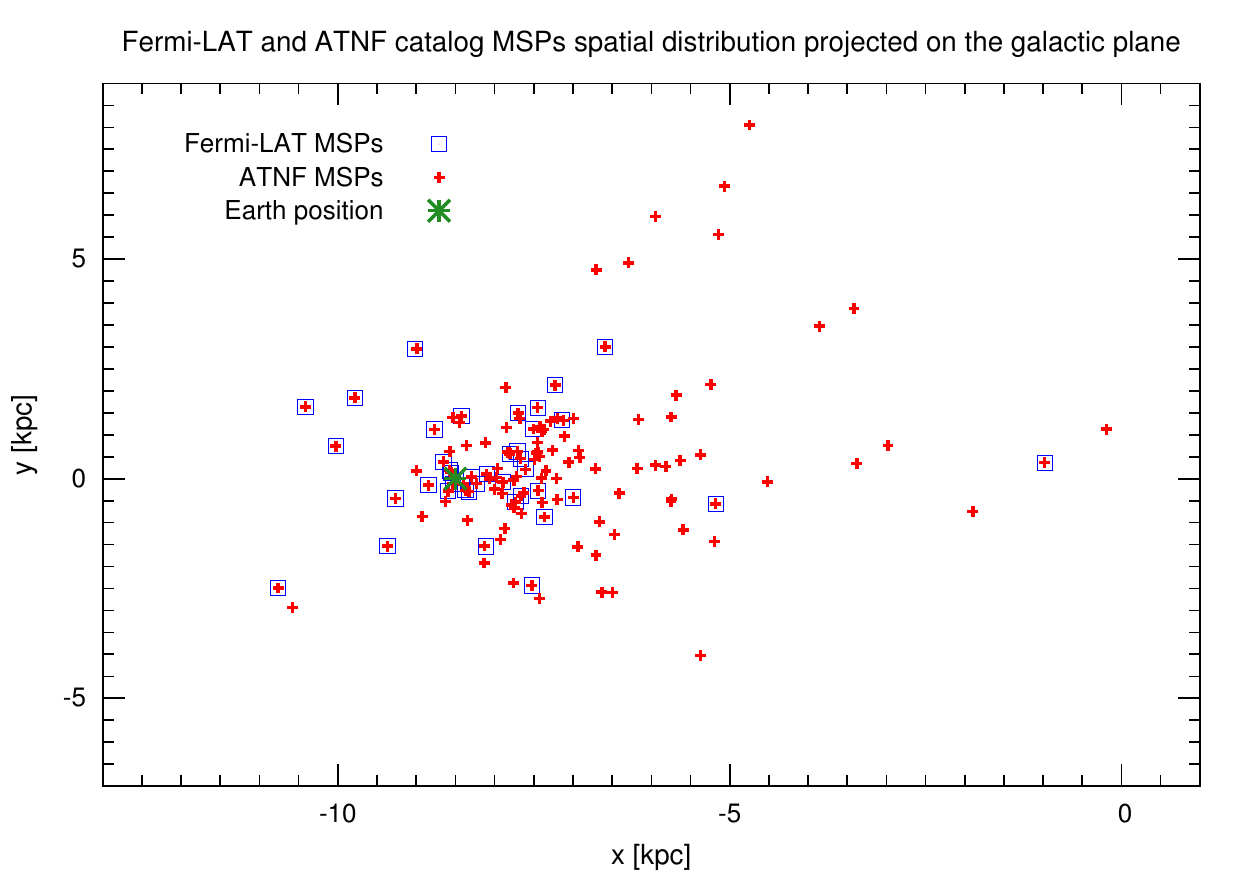}
\caption{MSP spatial distribution projected on the galactic plane for sources selected from the ATNF catalog \citep{2005AJ....129.1993M}. The sample has $P\leq15$ ms and is shown with red crosses, while the Earth position is displayed by the green point. For comparison, we also overlap the projected distribution of MSPs resolved by the {\it Fermi}-LAT and listed in the 2FPC catalog \citep{2013ApJS..208...17A} (blue squares).}
\label{fig:mapmsps} 
\end{centering}
\end{figure*}

In order to model the MSP population in our Galaxy we rely on the Australia Telescope National Facility (ATNF) pulsar catalog \citep{2005AJ....129.1993M}. It contains 1509 pulsars with published information, a huge improvement with respect to the previous avalilable catalog \citep{1993ApJ...411..674T} containing  558 radio sources.
We use the continuously updated on-line version of the catalog \footnote{\url{http://www.atnf.csiro.au/people/Pulsar/psrcat/}} to compile the list of MSPs.
In order to build our MSP sample we select, from the whole catalog, those objects with a period $P\leq15$ ms. This upper limit on the $P$ distribution is usually set to distinguish MSPs and young pulsar populations \citep{2013ApJS..208...17A,1997ApJ...482..971C,Lorimer:2012hy}. 
We display in Fig.~\ref{fig:popdot} the $\dot{P}-P$ plane with all the sources of the ANTF catalog divided into MSPs and young pulsars.
We show also the threshold $P=15$ ms and the {\it Fermi}-LAT detected MSPs and young pulsars.
The majority of MSPs have a period in the range $P\in[1,10]$ ms and there is a small number of sources with a period larger than 10 ms. Therefore, the MSP selected sample weakly depends on the $P$ upper bound.  
Considering this threshold, the number of MSPs selected from the ATNF catalog for our analysis is 132. In Fig.~\ref{fig:mapmsps} we show the position of the selected radio sources in the galactic plane, highlighting the Earth position, around which the sources are concentrated. For the sake of comparison, we also show the distribution in the galactic plane of MSPs resolved by the {\it Fermi}-LAT as 
reported in the 2FPC catalog \citep{2013ApJS..208...17A}. 
Since the distances claimed in the ATNF and 2PFC catalogs differ significantly for several sources, we have fixed 
the distance as reported in \cite{2013ApJS..208...17A} (Table 6) whenever dealing with
$\gamma$-ray MSPs, and fixed it to the value declared in the ATNF catalog for all the other sources. 

The physical observables which are directly measured are the period $P$, the derivative of the period $\dot{P}$, the distance $d$, the longitude $l$ and the latitude $b$.
However, the physical parameters which are generally used to derive the emission from pulsars are the surface magnetic field $B$, the rotation period $P$, the distance from the galactic plane $z$ and the distance from the galactic center projected on the galactic plane $r$, hereafter the radial distance.

\begin{figure*}
     \begin{centering}
        \subfigure[]{
            \label{fig:distr_b}
            \includegraphics[width=0.4\textwidth]{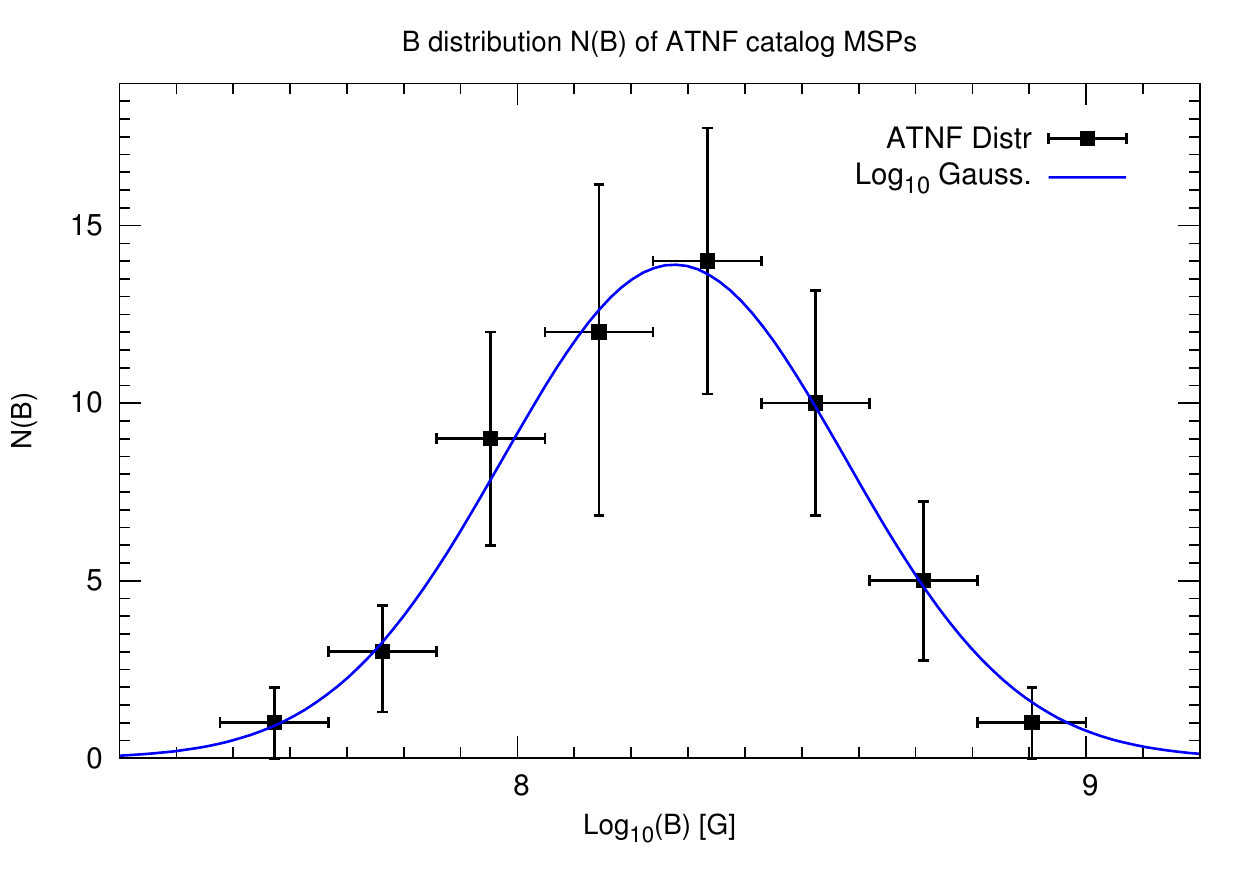}
        }%
        \subfigure[]{
           \label{fig:distr_P}
           \includegraphics[width=0.4\textwidth]{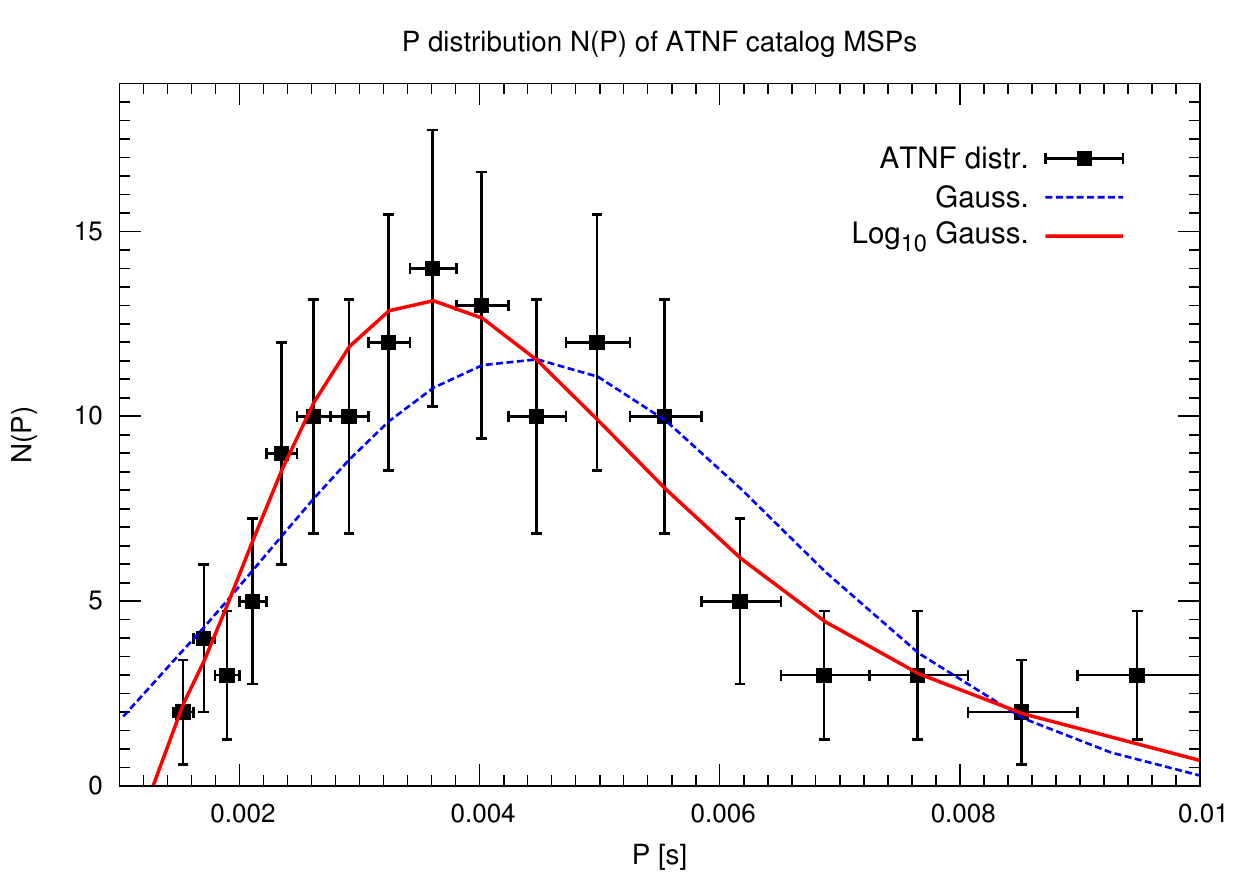}
        }\\
        \subfigure[]{
            \label{fig:distr_z}
            \includegraphics[width=0.4\textwidth]{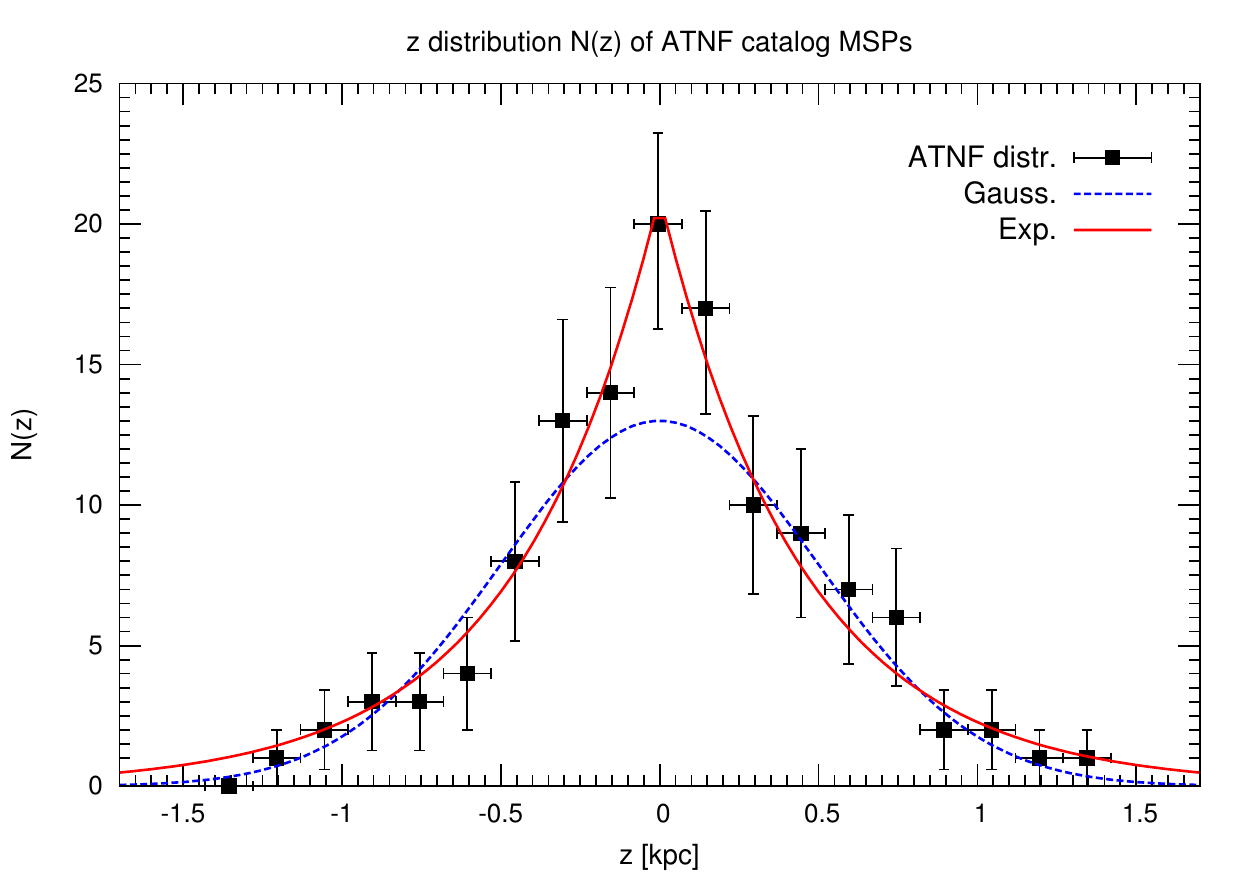}
        }%
        \subfigure[]{
            \label{fig:distr_r}
            \includegraphics[width=0.4\textwidth]{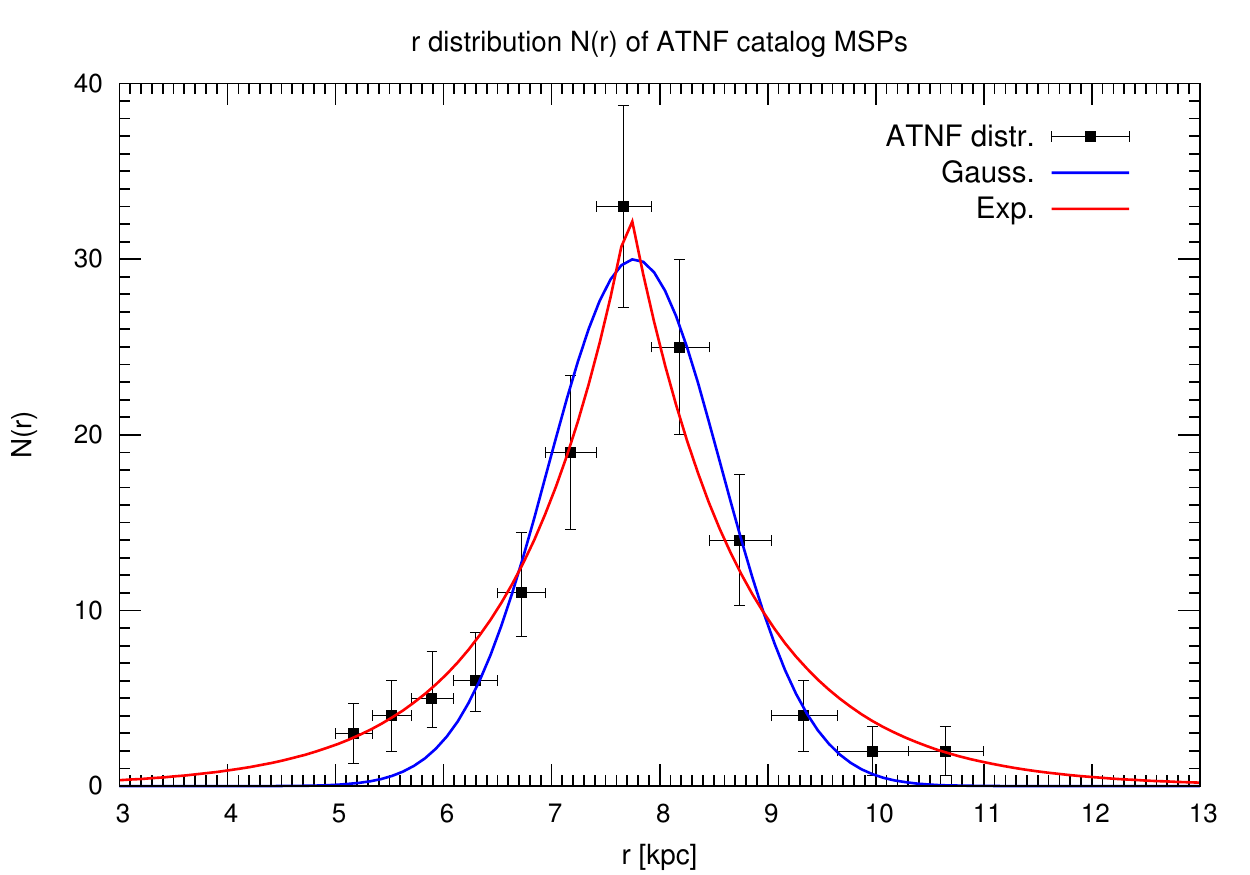}
        }%
    \caption{Cumulative distributions from top-left to bottom-right of the magnetic field $B$, the period $P$, the distance from the galactic plane $z$ and the radial distance $r$ for ATNF catalog MSPs.  
    Together with data we also plot the fitting functions in Eq.~\ref{eq:bdistr} - \ref{eq:rdistrtwo} for the best fit parameters.}
   \label{fig:distr}
       \end{centering}

\end{figure*}

The magnetic field can be derived from $P$ and $\dot{P}$ with Eq.~\ref{eq:pdot}, while $z$ and $r$ are connected to $d$, $l$, $b$ by:
\begin{eqnarray}
\label{eq:zrdef}
 z = d \sin{b} \;  &;&  \; r  = \sqrt{x^2+y^2} \nonumber \\
\; x = d \cos{b}\cos{l}-r_{\rm{sun}}\; &;& \; y = d \cos{b}\sin{l} ,
\end{eqnarray}
where $r_{\rm{sun}}$ is the distance of the Sun from the galactic center and it is fixed to be 8.5 kpc \citep{2010MNRAS.402..934M,2011AN....332..461B,2009ApJ...692.1075G}.
We use the sample of the 132 MSPs selected from the ATNF catalog in order to derive the distributions of $B$, $P$, $r$ and $z$ and we fit them with different functions in order to assess the distribution function of each observable.
For the magnetic field $B$ distribution we use a Log$_{10}$ Gaussian function, similarly to \cite{2011MNRAS.415.1074S,2013AA...554A..62G,2010JCAP...01..005F}:
\begin{equation}
\label{eq:bdistr}
  \displaystyle \frac{dN}{d\log_{10} B} \propto \exp{\left(-\frac{(\log_{10} B - \langle \log_{10} B \rangle)^2}{2\sigma^2_{\log_{10} B}}\right)},
\end{equation}
where $\langle \log_{10} B \rangle$ and $\sigma_{\log_{10} B}$ are the mean and the dispersion value of the Log$_{10}$ of surface magnetic field, respectively.
For the period $P$ distribution we consider a Log$_{10}$ and a linear Gaussian function:
\begin{eqnarray}
\label{eq:Pdistr}
  \frac{dN}{d\log_{10} P} &\propto& \exp{\left(-\frac{(\log_{10} P - \langle \log_{10} P\rangle)^2}{2\sigma^2_{\log_{10} P}}\right)}  \label{eq:Pdistrlog10}\\
  \frac{dN}{dP} &\propto& \exp{\left(-\frac{(P - \langle P \rangle)^2}{2\sigma^2_{P}}\right)},
\end{eqnarray}
where $\langle\log_{10} P\rangle$ ($\langle P\rangle$) and $\sigma_{\log_{10} P}$ ($\sigma_{P}$) are the mean and the dispersion value of the Log$_{10}$ (linear) value of the period.
On the other hand, we use an exponential and a Gaussian function for the distance from the galactic plane $z$:
\begin{eqnarray}
\label{eq:zdistrone}
  \frac{dN}{dz} &\propto& \exp{\left(-\frac{(z - \langle z \rangle)}{z_0}\right)}, \\
\label{eq:zdistrtwo}
  \frac{dN}{dz} &\propto& \exp{\left(-\frac{(z - \langle z \rangle)^2}{2\sigma^2_{z}}\right)}
\end{eqnarray}
where $\langle z\rangle$, $\sigma_{z}$ and $z_0$ are the mean, the dispersion and the width of the $z$ distribution.
Finally, we try to explain the radial distribution $r$ with an exponential and a linear Gaussian function:
\begin{eqnarray}
\label{eq:rdistrone}
  \frac{dN}{dr} &\propto& \exp{\left(-\frac{(r - \langle r \rangle)}{r_0}\right)} \\
\label{eq:rdistrtwo}
  \frac{dN}{dr} &\propto& \exp{\left(-\frac{(r - \langle r \rangle)^2}{2\sigma^2_{r}}\right)}\, ,
\end{eqnarray}
where $\langle r\rangle$, $\sigma_{r}$ and $r_0$ are the mean, the dispersion and the width of the $r$ distribution.
The best fit functions turn out to be a Log$_{10}$ Gaussian distribution for the period $P$, Eq.~\ref{eq:Pdistrlog10},  and an exponential for the distance from the galactic plane $z$, Eq.~\ref{eq:zdistrone},  and the radial distance $r$, Eq.~\ref{eq:rdistrone}.
The best fit parameters and the 1$\sigma$ errors are quoted in Tab.~\ref{tab:par}.
\begin{table*}
     \begin{centering}

\begin{tabular}  {||c|c|c||c|c|c||}
\hline
\hline
$B$ &     $\langle \log_{10} (B/\rm G) \rangle$   &   $\sigma_{\log_{10} B}$   &   $P$   &   $\langle \log_{10} (P/\rm s) \rangle$  &  $\sigma_{\log_{10} P}$  \\
\hline
       &     $8.27\pm0.09$    &    $0.30\pm0.12$                   &             &      $-2.54\pm0.05$  &  $0.19\pm0.03$                 \\ 
\hline
\hline
$r$  &     $ \langle r \rangle  [{\rm kpc}] $  & $r_0  [{\rm kpc}]$    &   $z$         &  $ \langle z \rangle [{\rm kpc}]$ &     $z_0  [{\rm kpc}]$                                                                \\
\hline
       &     $7.42\pm0.28$    &   $1.03\pm0.35$                   &             &      $0.00\pm0.14$    &  $0.67\pm0.11$		\\
\hline
\hline
\end{tabular}
\caption{Best fit parameters for the Log$_{10}$ Gaussian distribution 
for the surface magnetic field $B$ and the period $P$ and the exponential distribution for the distance from the galactic plane $z$ and the radial distance $r$ for the MSP population.}
\label{tab:par}
    \end{centering}
\end{table*} 
In Fig.~\ref{fig:distr}, we display the magnetic field $B$, period $P$, distance from the galactic plane $z$ and radial distance $r$ distributions for the MSPs of our ATNF catalog sample, together with the theoretical expectations of the same quantities using Eq.~\ref{eq:bdistr} - \ref{eq:rdistrtwo}.

We find that the mean value of the $\log_{10}( B/\rm G)$ Gaussian distribution is 8.3. This result is compatible with \cite{2013PhRvD..88h3009H}, although often a lower value for $\langle \log_{10}{(B/ \rm G)} \rangle$, i.e. 8, has been assumed, \citep{2011MNRAS.415.1074S,2013AA...554A..62G,2010JCAP...01..005F} .
The rotation period distribution turns out to be compatible with a $\log_{10}$ 
Gaussian\footnote{We obtain that a $\log_{10}$ Gaussian gives a reduced $\chi^2/d.~o.~f. = 0.21$, 
while $\chi^2/d.~o.~f. = 0.28$ for a linear Gaussian. 
Additionally, we notice that when increasing the MSP maximal period up to 30 ms, the sample is contaminated by the low-period young pulsars. 
This high-period tail in the MSPs' period distribution makes both parameterisations badly fitted.}, 
Eq.~\ref{eq:Pdistrlog10} with  $\langle \log_{10}{(P/ \rm s)} \rangle= -2.54$. The results for the $P$ distribution represent a novelty with respect to previous works. Indeed, in \cite{2011MNRAS.415.1074S,2013AA...554A..62G,2010JCAP...01..005F} a power-law has been used, $N(P)\propto P^{-\alpha}$. This assumption is based on the results of \cite{1997ApJ...482..971C} where the $P$ distribution of 22 MSPs is fitted with a power-law. It is clearly shown in Fig.~\ref{fig:distr} that 
the distribution at small values of $P$ ($1-4$ ms) is not compatible with a power-law; rather, it is explained fairly well by a Gaussian. At a few ms, indeed, there is a drop of the distribution because the number of sources with such a small rotation period decreases. This trend (less sources at smaller $P$) is not an experimental bias associated to the difficulty of detecting such small rotation periods. Actually, the uncertainties on the measurement of $P$ from radio telescopes are much smaller than a few ms (see e.g. \cite{Keith:2011km,2013MNRAS.429..579B,2001MNRAS.326..274L}).  Hence the decreasing of the shape at small rotation periods is physical and not due to an experimental bias.

As for the $z$ distribution, an exponential function with $z_0 = 0.67$, Eq.~\ref{eq:zdistrone}, fits the data well, as it has been derived in \cite{Levin:2013usa,2007ApJ...671..713S} for MSPs. 
Similarly the result for the $r$ distribution, i.e. an exponential with $\langle r \rangle \sim 8$ kpc, Eq.~\ref{eq:rdistrtwo}, is in agreement with the previous literature  \citep{Levin:2013usa,2007ApJ...671..713S}.
As one can see in Fig.~\ref{fig:distr} the radial distance distribution is not centered in zero but peaks at $\sim$ 8 kpc. This is due to a bias in the detection of pulsars. 
The displacement $\langle r \rangle$ corresponds to the distance of the Earth from the galactic center meaning that the most of the sources should be placed around the Earth. This is exactly what we show in the map of Fig.~\ref{fig:mapmsps} where the location of the ATNF MSPs in the galactic plane displays a clustering around the Earth position. This result indicates that the closer to us are the sources, the more easily they are detected. This thus represents an experimental bias we have to take into account when dealing with the radial distribution.

Models of birth and evolution of radio pulsars \citep{Lorimer:2006qs,FaucherGiguere:2005ny,Yusifov:2004fr,2010JCAP...01..005F} have found that evolved distributions of pulsars are peaked at about 3-4 kpc away from the galactic center in the direction of the Earth. Nevertheless, MSPs, with typical ages of 1 Gyr or more \citep{2005AJ....129.1993M}, are expected to have completed many orbits of the Galaxy. The position of such old sources is therefore believed to be uncorrelated with the birth position and the evolved distribution \citep{2010JCAP...01..005F}.
For this reason, the radial distribution is usually centered in the galactic center $ \langle r \rangle = 0$ kpc and modelled with an exponential distribution, e.g.~in \cite{2007ApJ...671..713S}:
\begin{equation}
\label{eq:rhoone}
  \displaystyle \frac{dN}{dr} \propto \exp{\left(-\frac{r}{r_0}\right)}\, ,
\end{equation}
where $r_0$ is the radial distance width, or with a Gaussian density profile \cite{2010JCAP...01..005F}:
\begin{equation}
\label{eq:rhogauss}
  \displaystyle \frac{dN}{dr} \propto \exp{\left(-\frac{r^2}{2\sigma_r^2}\right)} \, ,
\end{equation}
with $\sigma_r$ is the distance dispersion. 
We adopt in the rest of the article a radial distribution given by Eq.~\ref{eq:rhogauss} with $\sigma_r=10$ kpc and we quantify in Appendix \ref{app:uncertainties} the uncertainty on the diffuse $\gamma$-ray emission given by the choice of other radial distributions.

\section{The population of $\gamma$-ray MSPs}\label{sec:gammaMSP}
The first discovery of a radio MSP dates back to 1982 \citep{Backer82}. Since then, the studies on this new source class were focused on radio emission, although it was suddenly realised that MSPs could efficiently shine in $\gamma$ ray as well \citep{Usov83}. 

Before {\it Fermi}-LAT operation, only few young, radio-loud young pulsars were detected in $\gamma$ ray, while MSPs $\gamma$-ray emission was finally confirmed by  the LAT observations \citep{2009ApJ...699.1171A}. 
The 2FPC \citep{2013ApJS..208...17A} lists 117 $\gamma$-ray pulsars detected during the first three years of the mission. 
The 117 pulsars are classified into three groups: MSPs, young radio-loud pulsars, and young radio-quiet pulsars.
Out of the $\gamma$-ray pulsars in this catalog, roughly half (41 young pulsars and 20 MSPs) were already known in radio and/or X-rays. 
The remaining pulsars were discovered by or with the aid of the LAT, with 36 being young pulsars found in blind searches of LAT data 
and the remaining ones being MSPs found in radio searches of unassociated LAT sources.
The pulsars of the 2FPC are thus divided into 77 young pulsars and 40 MSPs.
All young pulsars except one have latitude $|b|\leq15^{\circ}$, while 31 MSPs, out of 40, have $|b|\geq10^{\circ}$ mainly because of the poor sensitivity to MSPs detection in the inner part of the Galaxy (see Fig.~17 of \cite{2013ApJS..208...17A}).

\begin{table*}
\begin{centering}
\begin{tabular}  {||c|c|c|c|c|c||}
\hline
\hline
PSR & $l[^\circ]$ & $b[^\circ]$  & $F_{\gamma}[10^{-8}\rm ph \, cm^{-2} \,s^{-1}]$ & $\Gamma$ & $E_{\rm cut} [\rm GeV]$ \\
 \hline
 \hline
  J0023+0923 &  111.15 & $-$53.22 & 	  $1.2 \pm 0.4 $ &   $1.4 \pm 0.4 $ & $1.4 \pm 0.6 $ \\ 
 J0030+0451 &  113.14 & $-$57.61 & 	  $6.6 \pm 0.3 $ &   $1.2 \pm 0.1 $ & $1.8 \pm 0.2 $ \\
 J0034$-$0534 & 111.49 & $-$68.07 &	  $2.2 \pm 0.3 $ &   $1.4 \pm 0.2 $ & $1.8 \pm 0.4 $ \\
 J0101$-$6422 &  301.19 & $-$52.72 &	  $0.75 \pm 0.14 $ &   $0.7 \pm 0.3 $ & $1.5 \pm 0.4 $ \\
 J0102+4839 &  124.93 & $-$14.83 &	  $1.3 \pm 0.3 $ &    $1.4 \pm 0.3 $ & $3.2 \pm 1.1 $ \\ 
 J0218+4232 & 139.51 & $-$17.53 & 	  $7.7 \pm 0.7 $ &    $2.0 \pm 0.1 $ & $4.6 \pm 1.2 $ \\ 
 J0340+4130 & 154.04 & $-$11.47 & 	  $1.5 \pm 0.2 $ &    $1.1 \pm 0.2 $ & $2.6 \pm 0.6 $ \\
 J0437$-$4715 & 253.39 & $-$41.96 &	  $2.7 \pm 0.3 $ &   $1.4 \pm 0.2 $ & $1.1 \pm 0.3 $ \\
 J0610$-$2100 & 227.75 & $-$18.18 &	  $0.78 \pm 0.25 $ &   $1.2 \pm 0.4 $ & $1.6 \pm 0.8 $ \\
 J0613$-$0200 & 210.41 & $-$9.30 & 	  $2.7 \pm 0.4 $ & $1.2 \pm 0.2 $ & $2.5 \pm 0.5 $ \\ 
 J0614$-$3329 & 240.50 & $-$21.83 &	  $8.5 \pm 0.3 $ & $1.3 \pm 0.1 $ & $3.9 \pm 0.3 $ \\
 J0751+1807 &  202.73 & 21.09 &  	  $1.1 \pm 0.2 $ & $1.1 \pm 0.2 $ & $2.6 \pm 0.7 $ \\ 
 J1024$-$0719 & 251.70 & 40.52 & 	  $0.2 \pm 0.2 $ &   \nodata &  \nodata \\ 
 J1124$-$3653 & 283.74 & 23.59 & 	  $0.94 \pm 0.23 $ &    $1.1 \pm 0.3 $ & $2.5 \pm 0.7 $ \\
 J1125$-$5825 & 291.89 & 2.60 &  	  $1.1 \pm 0.5 $ &    $1.7 \pm 0.2 $ & $4.8 \pm 2.4 $ \\
 J1231$-$1411 & 295.53 & 48.39 & 	  $9.2 \pm 0.4 $ &   $1.2 \pm 0.1 $ & $2.7 \pm 0.2 $ \\ 
 J1446$-$4701 & 322.50 & 11.43 & 	  $0.73 \pm 0.31 $ &   $1.4 \pm 0.4 $ & $3.0 \pm 1.7 $ \\
 J1514$-$4946 & 325.22 & 6.84 &  	  $4.1 \pm 0.6 $ &  $1.5 \pm 0.1 $ & $5.3 \pm 1.1 $ \\
 J1600$-$3053 & 344.09 & 16.45 & 	  $0.22 \pm 0.16 $ &   $0.40 \pm 0.47 $ & $2.0 \pm 0.7 $ \\
 J1614$-$2230 & 352.64 & 20.19 & 	 $2.0 \pm 0.4 $ &   $0.96 \pm 0.22 $ & $1.9 \pm 0.4 $ \\
 J1658$-$5324 & 334.87 & $-$6.63 &	  $5.7 \pm 0.7 $ &  $1.8 \pm 0.2 $ & $1.4 \pm 0.4 $ \\
 J1713+0747 &  28.75 & 25.22 &  	  $1.3 \pm 0.4 $ & $1.6 \pm 0.3 $ & $2.7 \pm 1.2 $ \\
 J1741+1351 &  37.90 & 21.62 &  	  $0.12 \pm 0.04 $  & \nodata &  \nodata  \\
 J1744$-$1134 & 14.79 & 9.18 & 	  $4.6 \pm 0.7 $ &   $1.3 \pm 0.2 $ & $1.2 \pm 0.3 $ \\
 J1747$-$4036 &  350.19 & $-$6.35 &	  $1.5 \pm 0.7 $ &  $1.9 \pm 0.3 $ & $5.4 \pm 3.3 $ \\
 J1810+1744 &  43.87 & 16.64 &  	  $4.2 \pm 0.5 $ &  $1.9 \pm 0.2 $ & $3.2 \pm 1.1 $ \\
 J1823$-$3021A & 2.79 & $-$7.91 & 	  $1.5 \pm 0.4 $ &   $1.6 \pm 0.2 $ & $2.5 \pm 0.6 $ \\
 J1858$-$2216 &  13.55 & $-$11.45 & 	  $0.55 \pm 0.28 $ &  $0.84 \pm 0.74 $ & $1.7 \pm 1.1 $ \\
 J1902$-$5105 &  345.59 & $-$22.40 & 	  $3.1 \pm 0.4 $ &   $1.7 \pm 0.2 $ & $3.4 \pm 1.1 $ \\
 J1939+2134 &  57.51 & $-$0.29 &  	  $1.5 \pm 0.8 $ &   \nodata &  \nodata \\
 J1959+2048 &  59.20 & $-$4.70 &  	  $2.4 \pm 0.5 $ &    $1.4 \pm 0.3 $ & $1.4 \pm 0.4 $ \\
 J2017+0603 &  48.62 & $-$16.03 & 	  $2.0 \pm 0.3 $ &   $1.0 \pm 0.2 $ & $3.4 \pm 0.6 $ \\
 J2043+1711 &  61.92 & $-$15.31 & 	  $2.7 \pm 0.3 $ &    $1.4 \pm 0.1 $ & $3.3 \pm 0.7 $ \\
 J2047+1053 &  57.06 & $-$19.67 & 	 $0.83 \pm 0.36 $ &    $1.5 \pm 0.5 $ & $2.0 \pm 1.1 $ \\
 J2051$-$0827 & 39.19 & $-$30.41 &	  $0.24 \pm 0.13 $ & $0.50 \pm 0.76$ & $1.3 \pm 0.7 $ \\
 J2124$-$3358 & 10.93 & $-$45.44 & 	  $2.7 \pm 0.3 $ &  $3.68 \pm 0.16 $ & $1.63 \pm 0.19 $ \\
 J2214+3000 &  86.86 & $-$21.67 & 	  $3.0 \pm 0.3 $ & $1.2 \pm 0.1 $ & $2.2 \pm 0.3 $ \\
 J2215+5135 &  99.46 & $-$4.60 &  	  $1.0 \pm 0.3 $ & $1.3 \pm 0.3 $ & $3.4 \pm 1.0 $ \\
 J2241$-$5236 & 337.46 & $-$54.93 &	  $3.0 \pm 0.3 $ &   $1.3 \pm 0.1 $ & $3.0 \pm 0.5 $ \\
 J2302+4442 &  103.40 & $-$14.00 & 	  $2.6 \pm 0.3 $ &  $0.94 \pm 0.12 $ & $2.1 \pm 0.3 $ \\
 \hline
 \hline
\end{tabular}
\caption{Relevant parameters of the {\it Fermi}-LAT detected MSPs in the 2FPC. 
Column 1: Pulsar name. 
Columns 2 and 3: galacto-centric longitude and latitude. 
Column 4: photon flux in the 0.1 - 100 GeV energy band, $F_{\gamma}$. 
Columns 5 and 6: spectral index $\Gamma$ and cutoff energy $E_{\rm cut}$. 
}
\label{tab:sample}
\end{centering}
\end{table*} 

Further studies of the 2FPC and multi-wavelengths analyses shed light on the nature of the $\gamma$-ray emission. 
The spectral cutoff shown by most sources at few GeV is consistent with curvature radiation as the dominant $\gamma$-ray production mechanism: electrons and positrons emit $\gamma$ rays as a consequence of their acceleration along magnetic field lines by the rotationally-induced electric field. Inverse Compton (IC) scattering could also participate as an alternative emission mechanism, mainly from synchrotron seed photons (self-synchrotron Compton - SSC). In this case, no strong cutoff at GeV energies is present making curvature radiation more likely \citep{2013IAUS..291..307K}.  Nevertheless, part of the very high energy emission of the Crab pulsar, whose flux has been measured at E $\geq$ 100 GeV from VERITAS \citep{Aliu11}, MAGIC \citep{Aleksic11, Aleksic12} and HESS \citep{Aharonian:2006pe}, is believed to arise from IC processes \citep{Lyutikov12}.
From $\gamma$-ray studies it is also possible to infer where the emission takes place; observations currently favour the outer magnetosphere location, but the full radiation model is still a matter of debate (see \cite{2014arXiv1404.2264J,2014arXiv1405.2148N} for an updated analysis).

We consider the sample of MSPs as reported in the 2FPC. The spectral energy distribution is fitted by a power-law with an exponential cutoff in the form:
\begin{eqnarray}
\label{eq:expcut}
   \frac{dN}{dE}\,=\,K\,\left(\frac{E}{E_{0}}\right)^{-\Gamma}\exp{\left(-\frac{E}{E_{\rm{cut}}}\right)},
\end{eqnarray}
where $K$ is a normalization factor, $\Gamma$ is the photon spectral index and $E_{\rm{cut}}$ is the energy cutoff.
For convenience, we quote in Tab.~\ref{tab:sample} the main parameters of those 40 objects: association name, galacto-centric longitude $l$ and latitude $b$, photon flux $F_{\gamma}$ integated between 100 MeV and 100 GeV, spectral index $\Gamma$, cutoff energy $E_{\rm cut}$. 

\begin{figure*}
     \begin{centering}
        \subfigure[]{%
            \label{fig:distr_g}
            \includegraphics[width=0.5\textwidth]{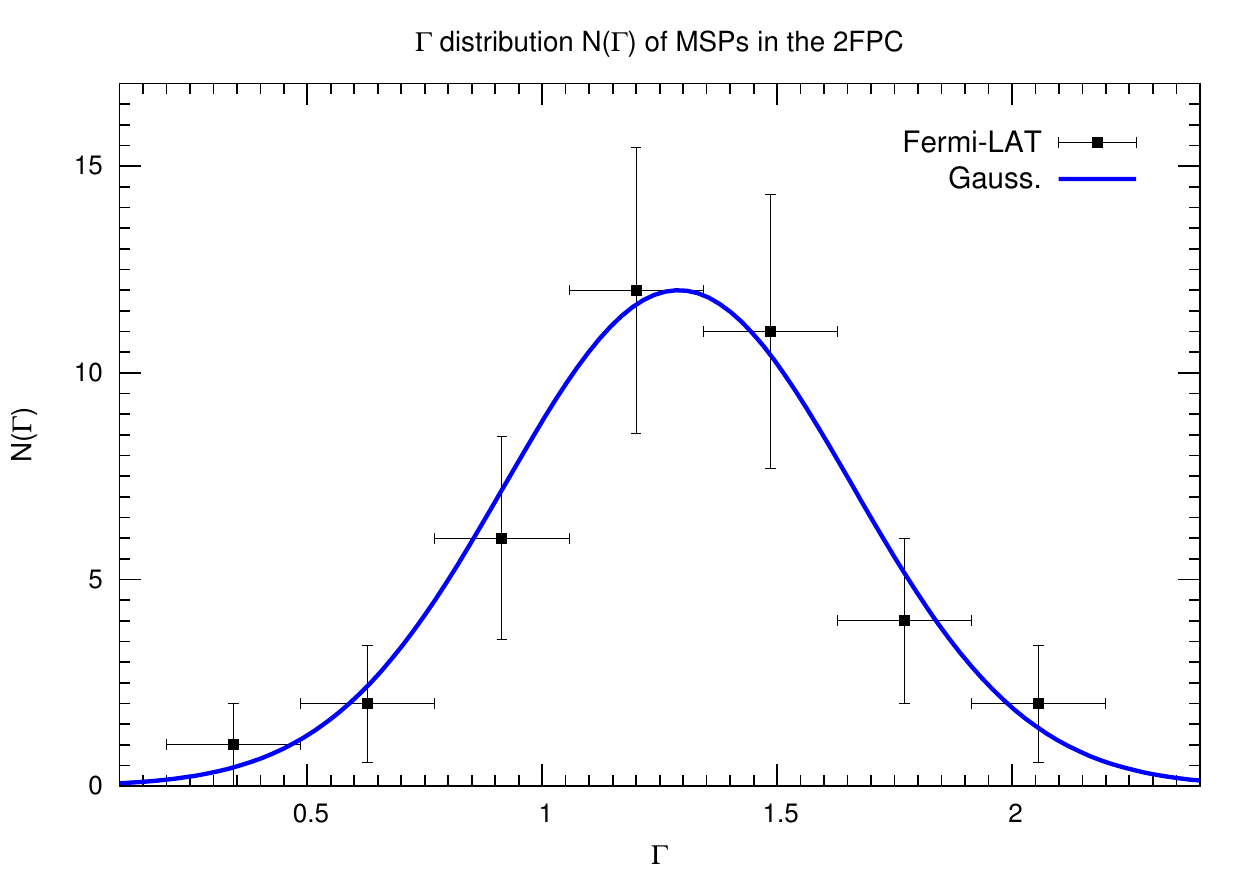}
        }%
        \subfigure[]{%
           \label{fig:distr_Ec}
           \includegraphics[width=0.5\textwidth]{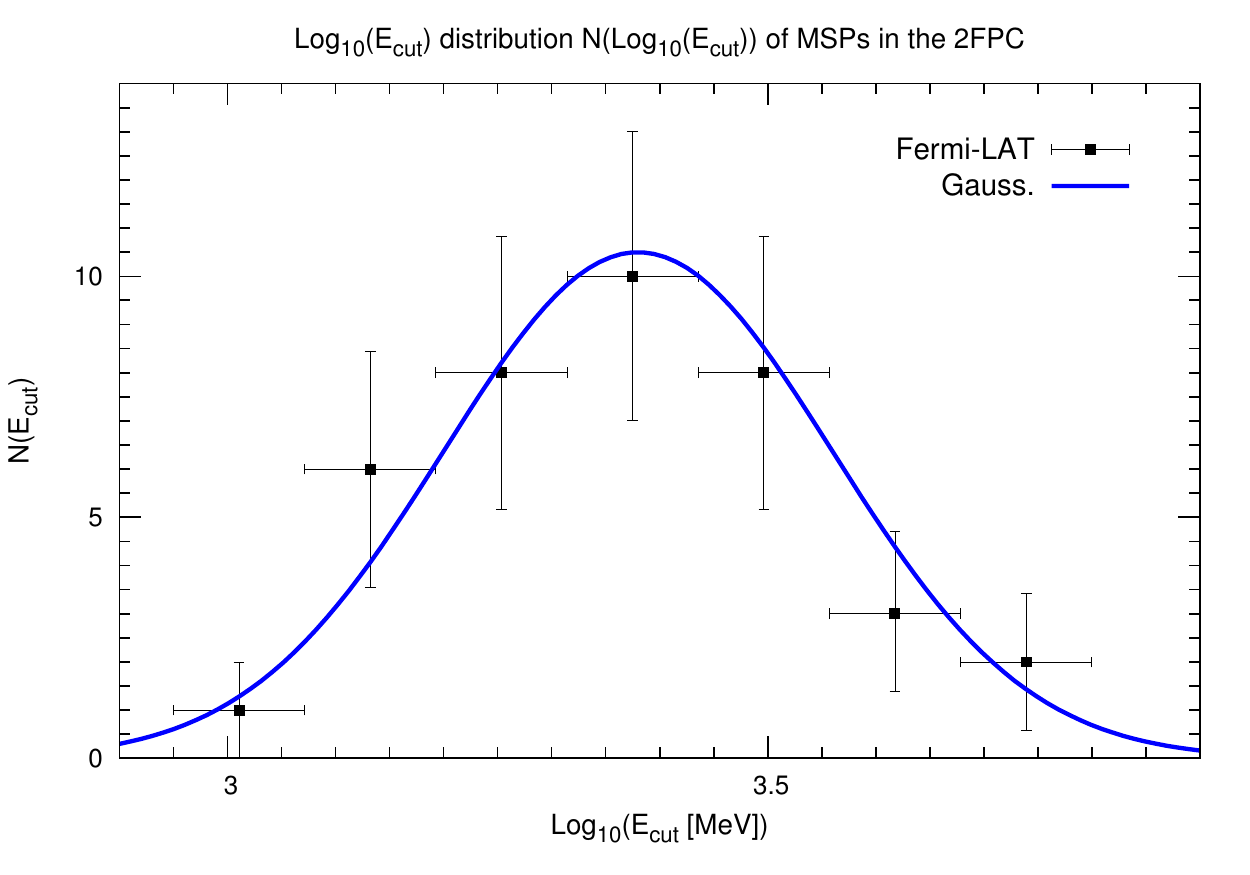}
        }
    \end{centering}
    \caption{Distribution of spectral index $\Gamma$ (left panel) and cutoff energy $E_{\rm cut}$ (right panel) of {\it Fermi}-LAT MSPs in Tab.~\ref{tab:sample}. The solid blue line refers to the fit with a Gaussian function with best fit parameters $[\Gamma,\sigma_{\Gamma}]=[1.29, 0.37]$ and $[ \tilde{E}_{\rm{cut}}, \sigma_{\tilde{E}_{\rm{cut}}} ] = [3.38, 0.18]$, respectively.}
   \label{fig:GEdistr}
\end{figure*}

Spectral index and cutoff energy distributions are consistent with a Gaussian function:
\begin{eqnarray}
\label{eq:gauss1}
   \frac{dN}{d\Gamma} &=& \exp{\left( \frac{(\Gamma-\langle \Gamma \rangle)^2}{2{\sigma_{\Gamma}}^2} \right)}  , \\
\label{eq:gauss2}
\frac{dN}{d\tilde{E}_{\rm{cut}}} &=& \exp{\left( \frac{(\tilde{E}_{\rm{cut}}-\langle \tilde{E}_{\rm{cut}} \rangle)^2}{2 \sigma_{\tilde{E}_{\rm{cut}}}^2} \right)},
\end{eqnarray}
where  $\tilde{E}_{\rm{cut}} \equiv \log_{10}(E_{\rm{cut}}/{\rm MeV})$. $\langle \Gamma \rangle$ and $\sigma_{\Gamma}$ ($\langle \tilde{E}_{\rm{cut}} \rangle$ and $\sigma_{\tilde{E}_{\rm{cut}}}$) are the mean and the dispersion values for the photon index (Log$_{10}$ of the energy cutoff) distribution.
Best fit parameters for the sample in Tab.~\ref{tab:sample} (excluding the three sources without spectral information) are: $[\Gamma,\sigma_{\Gamma}]=[1.29, 0.37]$ and $[ \tilde{E}_{\rm{cut}}, \sigma_{\tilde{E}_{\rm{cut}}} ] = [3.38, 0.18]$.
Fig.~\ref{fig:GEdistr} shows that the two distributions are well fitted by the functions in Eqs.~\ref{eq:gauss1} - \ref{eq:gauss2}.

\begin{figure*}
     \begin{centering}
        \subfigure[]{%
            \label{fig:distr_dNdr}
            \includegraphics[width=0.5\textwidth]{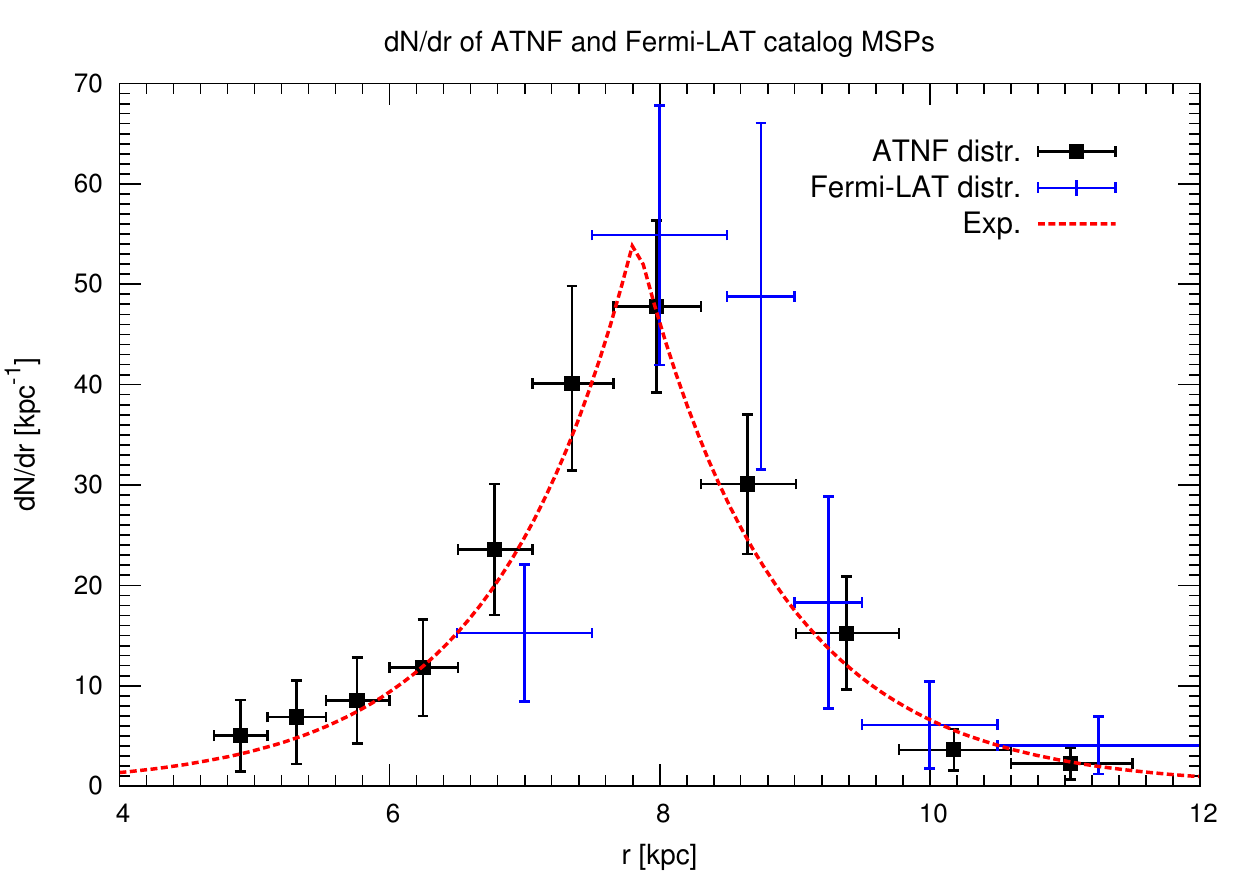}
        }%
        \subfigure[]{%
           \label{fig:distr_dNdz}
           \includegraphics[width=0.5\textwidth]{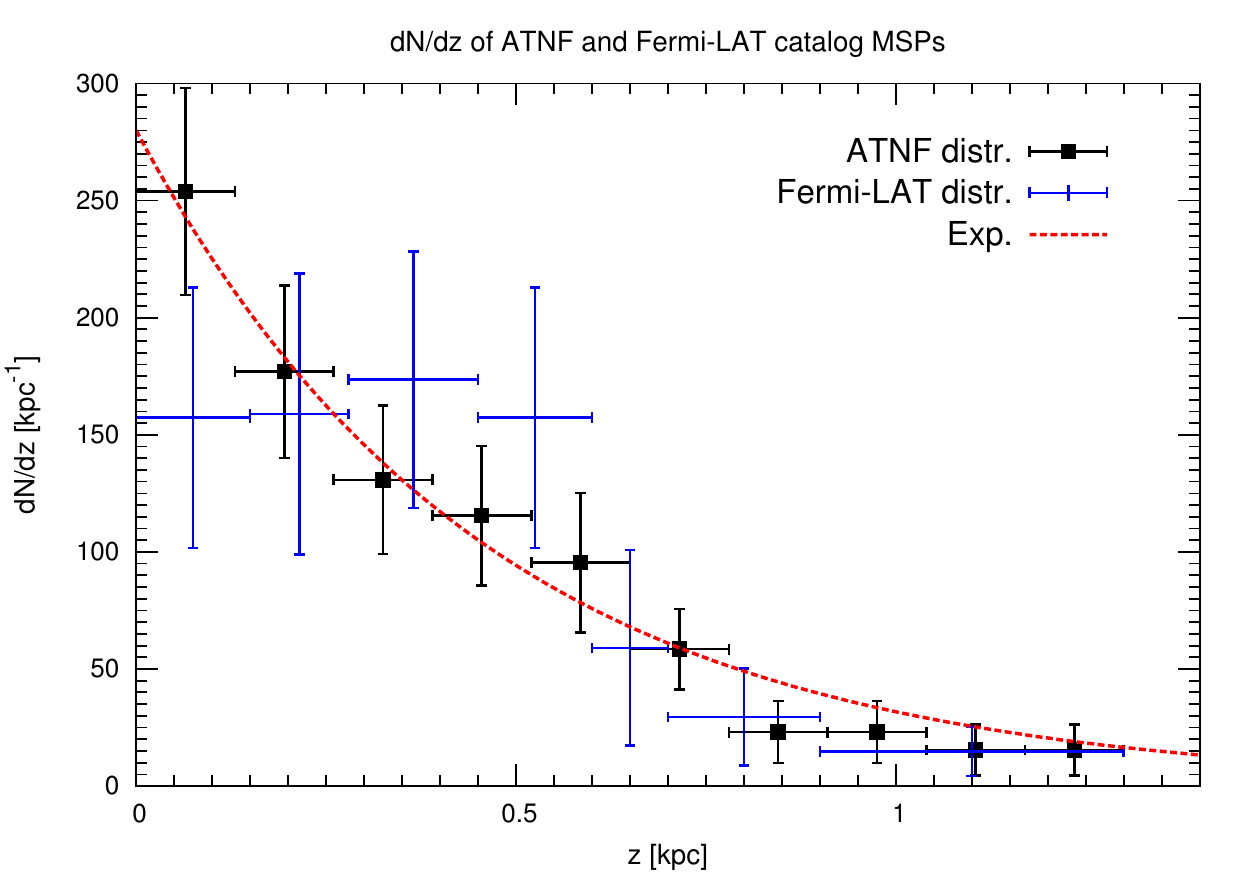}
        }
    \end{centering}
    \caption{The distribution of radial distance $r$ (left panel) and height from the galactic plane $z$ (right panel) of MSPs in the $\gamma$-ray 2FPC (Tab.~\ref{tab:sample}) is represented by the blue data points, while black points and dashed red line refer to the distribution and best fit of MSPs in the ATNF radio catalog.}
   \label{fig:distr_rz}
\end{figure*}

As for the case of the ATNF catalog sample, we derive the distribution of the quantities $r$ and $z$ for the MSPs detected by the {\it Fermi}-LAT (Tab.~\ref{tab:sample}). 
In order to directly compare the distributions of sources derived from the ATNF and {\it Fermi}-LAT catalogs we have used the following method. We have renormalised the $dN/dr$ and $dN/dz$ distributions in each bin by taking into account the different number of sources in the two catalogs. The sample of 2FPC is made of 40 sources, while the number of MSPs in the ATNF catalog with measured distance is 128. Hence, the renormalisation factor we have used is $128/40=3.2$. The $r$ and $z$ number distributions of {\it Fermi}-LAT sources are shown in Fig.~\ref{fig:distr_rz} together with the data and the best fit functions derived from the ATNF catalog sample of sources. 
The radial distribution of MSPs in the 2FPC catalog follows the radio one, despite of the poor statistics, showing again a peak at the Earth distance from the galactic center. The vertical height distribution $dN/dz$ is well compatible with radio observations far from the galactic plane, while it is observationally biased at very low distances, showing a flattening mainly due to the {\it Fermi}-LAT sensitivity suppression, as we will see in Sec.~\ref{subsec:IDGRB}.

We assume as intrinsic $dN/dz$ distribution of MSPs in the Galaxy the one deduced from the ATNF sample (see Eq.~\ref{eq:zdistrone}). We will show in Sec.~\ref{subsec:IDGRB} that the convolution of the ATNF distribution with the {\it Fermi}-LAT sensitivity gives a $dN/dz$ compatible with the one observed for the {\it Fermi}-LAT MSPs.
\newline

\section{Monte Carlo simulation of the $\gamma$-ray galactic MSP population}
\label{sec:MC}
The sample of Tab.~\ref{tab:sample} is  too poor to derive the $\gamma$-ray luminosity function directly from $\gamma$-ray data, as it has been possible, for example, for blazars, \citep{DiMauro:2013zfa,2014ApJ...780...73A,2012ApJ...751..108A}. 
On the other hand, it is not even possible to rely on some correlation between $\gamma$-ray luminosity and luminosities in other wavelengths, e.g. radio one, due to the high uncertainty on the $\gamma$-ray production mechanisms in MSPs, as it has been done for example for mis-aligned AGN \citep{RG} and for star-forming galaxies \citep{2012ApJ...755..164A}.
Therefore, since we are able to describe the space, period and magnetic field distributions of galactic MSPs, we build a MC simulation of the MSP population in order to analyse the properties of this source class in $\gamma$ ray. We can indeed use the general properties of the MSP population to construct a mock set of sources 
and to find the ensuing $\gamma$-ray diffuse emission.
This approach has been used, e.g.~, in \cite{2011MNRAS.415.1074S,2013AA...554A..62G}.

A MSP population is generated according to the $P$, $B$, $r$ and $z$ distributions discussed in Sec.~\ref{sec:distribution} and \ref{sec:gammaMSP}.
The position of each simulated source is assigned by randomly drawing pairs of $r$, $z$ from the spatial distribution (derived from Eqs.~\ref{eq:zdistrone} and \ref{eq:rhogauss}):
\begin{eqnarray}
\label{eq:rhoone}
  \displaystyle \frac{d^2N}{dr dz} \propto \exp{\left(-\frac{r^2}{2\sigma_r^2}-\frac{|z|}{z_0}\right)}
\end{eqnarray}
with $\sigma_r = 10$ kpc and $z_0 = 0.67$ kpc. We normalise it as a probability distribution function. The same holds for the normalisation of every other distribution assumed here.

\subsection{The $\gamma$-ray luminosity relation with the spin-down luminosity}
\label{sec:LgammaEdot}
\begin{figure*}
\begin{centering}
\includegraphics[width=0.65\textwidth]{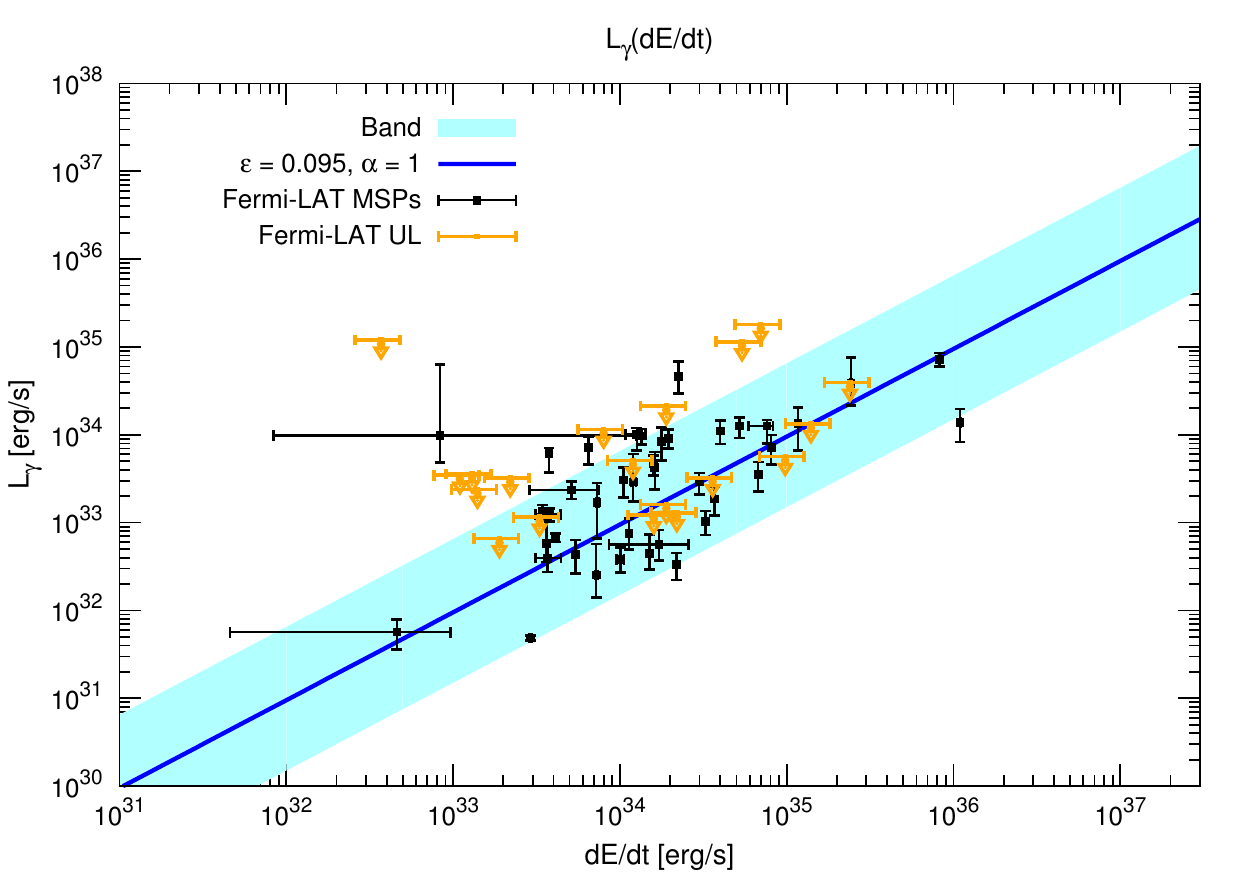}
\caption{$L_{\gamma} -  \dot{E}$ relation of the {\it Fermi}-LAT detected sources in the 2FPC (black points), together with the ULs on $\gamma$-ray luminosity (orange points) derived from the 95$\%$ C.L. ULs on the $\gamma$-ray flux for the sources listed in Tab.~\ref{tab:sampleUL}. The light blue band represents a reasonable range of uncertainty for the $L_{\gamma} (\dot{E})$ correlation and the average value is drawn (blue solid line) fixing the parameters $\alpha=1$ and $\eta=0.095$.}
\label{fig:lumedot}
\end{centering} 
\end{figure*}

As for the modelling of the $\gamma$-ray emission, we assume that the energy loss $\dot{E}$, Eq.~\ref{eq:edot}, due to the magnetic-dipole braking, is converted into $\gamma$ radiation.
We therefore extract a value for the spin period $P$ and the magnetic field $B$ from the distributions in Eqs.~\ref{eq:Pdistrlog10} and \ref{eq:bdistr} (with best fit parameters as in Tab.~\ref{tab:par}), and derive the corresponding loss energy rate $\dot{E}$ for each simulated source from Eq.~\ref{eq:edot}.
The conversion of $\dot{E}$ into $\gamma$-rays luminosity is parametrised by an empirical relationship \citep{2011MNRAS.415.1074S,2013AA...554A..62G,2010JCAP...01..005F}:
\begin{equation}
\label{eq:lgamma}
L_{\gamma}= \eta \dot{E}^{\alpha} \, ,
\end{equation}
where $L_{\gamma}$ and $\dot{E}$ are in unit of erg/s and $\eta$ is the conversion efficiency of spin-down luminosity into $\gamma$-ray luminosity, hereafter conversion efficiency.
Eq.~\ref{eq:lgamma} is an effective way to model the MSP $\gamma$-ray emission and it represents a general expression of the correlation between these two quantities often used in the literature.
Usually $\alpha$ has been empirically chosen to be 0.5 \citep{2010JCAP...01..005F,2011MNRAS.415.1074S} or 1.0 \citep{2013AA...554A..62G}, although the former value $\alpha$ might be 0.5 theoretically motivated in the framework of the outer gap models of high energy $\gamma$-ray emission.
We display in Fig.~\ref{fig:lumedot} the values of $L_{\gamma}$ and $\dot{E}$ for the 40 MSPs of the 2FPC catalog. Horizontal error bars are associated with the uncertainties on the measured period $P$ and $\dot{P}$ (see Eq.~\ref{eq:edot}), while vertical error bars are derived by propagating the uncertainties on the $\gamma$-ray parameters $\Gamma$, the normalisation of the spectrum $dN/dE$, the energy cutoff $E_{\rm{cut}}$ and the measured distance of the source.
The criticality of the relation Eq.~\ref{eq:lgamma} $L_{\gamma}(\dot{E})$ is visible by the scatter of the data points in Fig.~\ref{fig:lumedot}. This scatter prevents us to find a statistically meaningful relation $L_{\gamma}(\dot{E})$.
In order to probe Eq.~\ref{eq:lgamma} by means of further $\gamma$-ray information, we derive 95$\%$ C.L. upper limits (ULs) on the $\gamma$-ray flux of a sample of 19 sources non-detected by the {\it Fermi}-LAT. Those sources have been selected in the ATNF catalog as the ones expected to be the most powerful $\gamma$-ray emitters if standard values of $\alpha  = 1$ and $\eta = 0.1$ are assumed. We select only sources with latitudes $|b| \geq 10^{\circ}$ in order to avoid the strong contamination from the galactic foreground. For this purpose we use the {\it Fermi}-LAT Science Tools\footnote{http://fermi.gsfc.nasa.gov/ssc/data/analysis/documentation. Software version v9r32p5, Instrumental Response Functions (IRFs) P7$\_$V15}. 

\begin{table*}
\begin{centering}
\begin{tabular}  {||c|c|c|c|c|c|c||}
\hline
\hline
PSR &   $l[^\circ]$ & $b[^\circ]$  &  $d[\rm{kpc}]$ & $\dot{E}[\rm s^{-1}]$ & $F^{UL}_{\gamma}[\rm ph \, cm^{-2} \,s^{-1}]$ & $L^{UL}_{\gamma}[\rm erg \,s^{-1}]$ \\
\hline
\hline
J0218+4232  &	139.51 	  &	-17.53 &	2.64  &		$2.4\cdot 10^{+35}$	  &	$4.55\cdot 10^{-08}$  &	$3.94\cdot 10^{+34}$	\\ 
J0514-4002A & 	244.51 	  &	-35.04 &	12.6  &		$3.7\cdot 10^{+32}$	  &	$6.07\cdot 10^{-09}$	&	$1.20\cdot 10^{+35}$	\\ 
J1017-7156  &  	291.56 	  &	-12.55 &	3.00  &		$8.0\cdot 10^{+33}$	  &	$1.03\cdot 10^{-08}$	&	$1.15\cdot 10^{+34}$	\\ 
J1023+0038  &  	243.49 	  &	45.78  &	1.37  &		$9.8\cdot 10^{+34}$	  &	$2.42\cdot 10^{-08}$	&	$5.65\cdot 10^{+33}$	\\ 
J1300+1240  &  	311.31 	  &	75.41  &	0.62  &		$1.9\cdot 10^{+34}$	  &	$3.37\cdot 10^{-08}$	&	$2.12\cdot 10^{+34}$	\\ 
J1327-0755  &  	318.38 	  &	53.85  &	1.70  &		$3.6\cdot 10^{+34}$	  &	$9.02\cdot 10^{-09}$	&	$3.24\cdot 10^{+33}$	\\ 
J1342+2822B &	42.22  	  &	78.71  &	10.40 & 		$5.4\cdot 10^{+34}$	  &	$8.52\cdot 10^{-09}$	&	$1.15\cdot 10^{+35}$	\\ 
J1455-3330  &  	330.72 	  &	22.56  &	0.75  &		$1.9\cdot 10^{+33}$	  &	$9.65\cdot 10^{-09}$	&	$1.61\cdot 10^{+33}$	\\ 
J1544+4937  &  	79.17  	  &	50.17  &	2.20  &		$1.2\cdot 10^{+34}$	  &	$2.86\cdot 10^{-08}$	&	$5.12\cdot 10^{+33}$	\\ 
J1623-2631  &  	350.98 	  &	15.96  &	2.20  &		$1.9\cdot 10^{+34}$	  &	$3.53\cdot 10^{-08}$	&	$1.61\cdot 10^{+33}$	\\ 
J1709+2313  &	44.52  	  &	32.21  &	1.83  &		$1.4\cdot 10^{+33}$	  &	$5.74\cdot 10^{-09}$	&	$2.39\cdot 10^{+33}$	\\ 
J1740-5340A & 	338.16 	  &	-11.97 &	3.20  & 		$1.4\cdot 10^{+35}$  	  &	$1.05\cdot 10^{-08}$	&	$1.33\cdot 10^{+34}$	\\
J1909-3744  &  	359.73 	  &	-19.60 &	0.46  &		$2.2\cdot 10^{+34}$	  &	$4.92\cdot 10^{-09}$	&	$1.23\cdot 10^{+33}$	\\ 
J1933-6211  &  	334.43 	  &	-28.63 &	0.63  &		$3.3\cdot 10^{+33}$	  &	$2.34\cdot 10^{-08}$	&	$1.16\cdot 10^{+33}$	\\ 
J2010-1323  &  	29.45  	  &	-23.54 & 	1.29  &		$1.3\cdot 10^{+33}$	  &	$1.73\cdot 10^{-08}$	&	$3.58\cdot 10^{+33}$	\\ 
J2129+1210E &  	65.01  &	-27.31 &	10.0  &		$7.0\cdot 10^{+34}$	  &	$1.46\cdot 10^{-08}$	&	$1.82\cdot 10^{+35}$	\\ 
J2129-5721  &  	338.01 	  &	-43.57 &	0.40  &		$1.6\cdot 10^{+34}$	  &	$6.16\cdot 10^{-09}$	&	$1.23\cdot 10^{+33}$	\\ 
J2229+2643  &  	87.69  	  &	-26.28 &	1.43  &		$2.2\cdot 10^{+33}$	  &	$1.27\cdot 10^{-08}$	&	$3.23\cdot 10^{+33}$	\\ 
J2236-5527  &  	334.17 	  &	-52.72 &	2.03  &		$1.1\cdot 10^{+33}$	  &	$6.77\cdot 10^{-09}$	&	$4.67\cdot 10^{+33}$	\\ 
 \hline
 \hline
\end{tabular}
\caption{95\% C.L ULs on the $\gamma$-ray integrated photon flux and luminosity.
Column 1: Pulsar name. 
Columns 2 and 3: galacto-centric longitude and latitude. 
Column 4: distance. 
Column 5: spin-down luminosity.
Column 6: 95\% C.L. UL photon flux in the 0.1 to 100 GeV energy band.
Column 7: derived 95\% C.L. UL $\gamma$-ray luminosity.}
\label{tab:sampleUL}
\end{centering}
\end{table*} 

We choose for the analysis the same data taking period of the 2FPC catalog, namely from the starting time of the mission 2008 August 4 until 2011 August 4. 
The Mission Elapsed Time (MET) interval runs from 239557414 to 334713602.
Data are extracted from a region of interest (ROI) of radius = $10^{\circ}$ centered at the position of the source, where we select the $\gamma$ rays in the energy range 100 MeV-100 GeV.
We use the P7REP$\textunderscore$SOURCE$\textunderscore$V15 Event Selection model, we take into account the common cut of the rocking angle, selected to be less than 52$^{\circ}$, and apply
a cut on the zenith angle of 100$^{\circ}$. 

A binned maximum-likelihood analysis is performed.
Upper limits of the MSP $\gamma$-ray flux are derived by the aid of the
LATAnalysisScripts\footnote{User contributions
http://fermi.gsfc.nasa.gov/ssc/data/analysis/user/}, which make use of the
UpperLimits.py module.

The MSP source spectrum is modeled by a power-law with an exponential cutoff as in Eq.~\ref{eq:expcut}. $E_{\rm cut}$, $\Gamma$ and the flux normalization are considered as free parameters.
Besides the spectral parameters for the sources of the 2FGL catalog close to the investigated source (inside $13\,^{\circ}$ about the source position), additional free parameters are the normalizations of the diffuse backgrounds, namely the galactic diffuse emission (gll\_iem\_v05.fits) and the isotropic background (iso\_source\_v05.txt).

The 95$\%$ C.L.~ULs are computed with a profile likelihood method for the 19 sources since no evidence of detection is found (the Test
Statistic (TS) is less than 25 for all sources).
The 95$\%$ C.L.~ULs on the integrated flux in the energy range 100 MeV-100 GeV are listed in Tab.~\ref{tab:sampleUL}.
We have taken the values of the longitude $l$, the latitude $b$, the distance $d$ from the ATNF catalog and the spin-down luminosity $\dot{E}$ as derived from measurements according to Eq.~\ref{eq:edot}. Given the ULs on $F_{\gamma}$, we have then derived the ULs on the $\gamma$-ray luminosity using Eq.~\ref{eq:SgammafromLgamma}, fixing $\Gamma$ and $E_{\rm{cut}}$ to the average values derived in Sec.~\ref{sec:gammaMSP}. 

We display in Fig.~\ref{fig:lumedot} the ULs of the $\gamma$-ray luminosities.
Taking into account the MSPs of the 2FPC catalog and the derived ULs of the sample in Tab.~\ref{tab:sampleUL}, we identify a benchmark relation with $\alpha=1$ and $\eta=0.095$ and an uncertainty band defined by $\eta=\{0.015,0.65\}$ around $\alpha=1$. This assumption is based on the scatter in the $L_{\gamma} - \dot{E}$ plane when considering the 2FPC sources and the derived ULs. 
We have built the band in such a way that almost all the data were within it.
We thus stress that it does not correspond to any statistical uncertainty on the correlation but it is a reasonable way to describe it. Nevertheless, such a scatter represents a fundamental systematic uncertainty that can not be neglected by simply fixing $\alpha$ and  $\eta$ a priori to single values.
We will study the uncertainty brought by our ignorance on the real relation $L_{\gamma}(\dot{E})$ in Sec.~\ref{sec:results} and in the Appendix \ref{app:uncertainties}.

\subsection{The $\gamma$-ray diffuse emission}
\label{subsec:GRDE}
The simulated sources are characterised by randomly drawn positions in the Galaxy ($r$, $z$) as well as by $P$ and $B$ values extracted from the corresponding distributions. With the latter two quantities, it is possible to derive the energy loss rate $\dot{E}$, Eq.~\ref{eq:edot}, and then the $\gamma$-ray luminosity, Eq.~\ref{eq:lgamma}.
For each source we then compute the energetic flux as:
\begin{equation}
 S_{\gamma}=L_{\gamma}/(4\pi d^2) \, , 
\label{eq:SgammafromLgamma}
\end{equation}
 where $S_{\gamma}$ is defined in the range 0.1 - 100 GeV, according to Eq.~15 in \cite{2013ApJS..208...17A}. 
By computing the energetic flux from Eq.~\ref{eq:SgammafromLgamma}, for all simulated sources we derive the $\gamma$-ray flux in the same energy range, $F_{\gamma}$, by assuming that the single source spectral distribution $dN/dE$ is expressed by Eq.~\ref{eq:expcut}. Spectral index $\Gamma$ and energy cutoff $E_{\rm{cut}}$ are drawn from the distributions of Eqs.~\ref{eq:gauss1} and \ref{eq:gauss2} for each source.
From the definitions \citep{2013ApJS..208...17A}:
 \begin{eqnarray}
\label{eq:FSgamma}
S_{\gamma} \equiv \int_{E_1}^{E_2} \frac{dN}{dE} \, E \, dE   \;\;\; \text{and} \;\;\; F_{\gamma} \equiv \int_{E_1}^{E_2} \frac{dN}{dE} dE \, ,
\end{eqnarray}
where $E_1=0.1$ GeV, $E_2=100$ GeV, $E_0 = 1$ GeV, 
we can write $F_{\gamma}$ as a function of $S_{\gamma}$, $\Gamma$ and $E_{\rm{cut}}$ as:
\begin{equation}
\label{eq:Fgamma}
F_{\gamma}= \frac{S_{\gamma}}{E_0} \frac{\left[ \left(\frac{E_1}{E_0}\right)^{1-\Gamma} \mathcal{E}_{\Gamma}\left(\frac{E_1}{E_c}\right)- 
\left(\frac{E_2}{E_0}\right)^{1-\Gamma} \mathcal{E}_{\Gamma}\left(\frac{E_2}{E_c}\right) \right]}{\left[ \left(\frac{E_1}{E_0}\right)^{2-\Gamma} \mathcal{E}_{\Gamma-1}\left(\frac{E_1}{E_c}\right)- 
\left(\frac{E_2}{E_0}\right)^{2-\Gamma} \mathcal{E}_{\Gamma-1}\left(\frac{E_2}{E_c}\right) \right]} \, , 
\end{equation}
where $\mathcal{E}_{n}(t)$ is the exponential integral function arising from the $dN/dE$ integration.

We then classify sources in ``non-detected" and ``detected" objects. In order to discriminate if the simulated source would have been seen by the LAT or not, we compare its $S_{\gamma}$ value with the {\it Fermi}-LAT detection sensitivity curve $S_t(b)$ displayed in Fig.~17 of \cite{2013ApJS..208...17A}. The LAT sensitivity depends on the position of the source in the Galaxy; e.g., at $|b|=30^{\circ}$ it is about $3.2\cdot 10^{-12}$ erg cm$^{-2}$s$^{-1}$, while  at $|b|=5^{\circ}$ it is about $ 7 \cdot 10^{-12}$ erg cm$^{-2}$s$^{-1}$.  MSPs that have an $S_{\gamma}$ above the sensitivity curve are classified as ``detected". We simulate sources until ``detected" objects reach the number of {\it Fermi}-LAT observed MSPs above $|b| = 2^{\circ}$, i.e.~39. In general, we simulate about 1000 - 1500 sources, out of which $\sim$ 60 - 100 unresolved objects are found at $|b|\geq 10^{\circ}$.

For the set of ``non-detected", i.e.~unresolved, sources simulated by our MC procedure, we compute the total $\gamma$-ray flux in the energy range  0.1 - 100 GeV as:
\begin{eqnarray}
\label{eq:diffsim}
  \displaystyle I_{\rm{MSP}} =\frac{1}{\Delta \Omega} \sum_{|b|\geq b_{\rm{min}}} F_{\gamma} \, ,
\end{eqnarray}
where the sum is made over all the sources with $|b|\geq b_{\rm{min}}$, $\Delta \Omega$ is the solid angle corresponding to $| b |\geq b_{\rm{min}}$, with $b_{\rm{min} }= 10^{\circ}$, and $F_{\gamma}$ is the $\gamma$-ray flux for each source, Eq.~\ref{eq:Fgamma}.
$I_{\rm{MSP}}$ represents the unresolved contribution of the simulated MSP population to the IDGRB.

Moreover, by knowing the photon index $F_{\gamma}$, the spectral energy distribution of the single sources, we can get with Eq.~\ref{eq:FSgamma} the spectrum of the total $dN/dE$ by adding up all sources contributions at a given energy and we can draw the total spectrum of the unresolved population of MSPs:
\begin{eqnarray}
\label{eq:dndesim}
 \displaystyle \left( \frac{dN}{dE}(E) \right)_{\rm{MSP}}  =\frac{1}{\Delta \Omega} \sum_{|b|\geq b_{\rm{min}}} \frac{dN}{dE} (E)\,.
\end{eqnarray}

\subsection{The $\gamma$-ray anisotropy}
\label{subsec:anisotropy}
A general prediction, e.g. \cite{2011MNRAS.415.1074S}, is that a population of $\gamma$-ray sources contributes to the $\gamma$-ray anisotropy.
The {\it Fermi}-LAT Collaboration has measured the anisotropy of the IDGRB for latitude $|b| > 30^{\circ}$ in four energy bins spanning from 1 to 50 GeV, namely 1-2 GeV, 2-5 GeV, 5-10 and 10-50 GeV \citep{Ackermann:2012uf}. At multipoles $l \geq 155$ an angular power above the photon noise level is detected at $> 99 \%$ C.L.~in all the four energy bins, with approximately the same value $C_P/\langle I\rangle^2 = 9.05 \pm 0.84 \cdot 10^{-6}$ sr,  where $\langle I \rangle$ indicates the average integrated intensity in a given energy range. 
This result suggests that the anisotropy might originate from the contribution of one or more point-like source populations.

We derive the anisotropy arising from the unresolved MSPs and we compare this value with the {\it Fermi}-LAT data.
The angular power $C_p$ produced in the energy range $E \in [E'_1,E'_2]$ by the unresolved flux of $\gamma$-ray emitting MSPs is derived using the following equation \citep{Cuoco:2012yf,Ackermann:2012uf}:
\begin{eqnarray}
     \label{Cpdef}
        C_p(E'_1 \leq E \leq E'_2) =\int^{E_{\rm{cut},\rm{max}}}_{E_{\rm{cut},\rm{min}}} dE_{\rm{cut}} \int^{\Gamma_{\rm{min}}}_{\Gamma_{\rm{max}}} d\Gamma \cdot \nonumber  \\
 \cdot \int^{F'_t(S_t,\Gamma,E_{\rm{cut}},E'_1,E'_2)}_0 {F'_{\gamma}}^2 \frac{d^2N}{dF'_{\gamma} d\Gamma dE_{\rm{cut}}} dF'_{\gamma},
    \end{eqnarray} 
where $F'_{\gamma}$ is the photon flux of the source integrated in the range $E'_1 \leq E \leq E'_2$ in units of ph cm$^{-2}$s$^{-1}$. $E_{\rm{cut}},\rm{min}$ and $E_{\rm{cut}},\rm{max}$ are fixed respectively to 2.0 and 4.6 while $\Gamma_{\rm{min}}$ and $\Gamma_{\rm{max}}$ to 0.1 and 2.5.
The results do not depend on a slight modification of the limits of integration because these variables are parametrized with Gaussian distributions (see Eq.~\ref{eq:gauss1} and \ref{eq:gauss2}).
$F'_t$ is the flux sensitivity threshold which separates the {\it Fermi}-LAT detected and undetected MSPs. 
This quantity depends on the threshold energy flux $S_t$ of the sensitivity map given in the 2FPC catalog. $S_t$ is defined in units of the energy flux integrated in the range 0.1 GeV-100 GeV. We choose for $S_t$ a fixed value independent on the latitude. This assumption is justified by the fact that the anisotropy data are valid for $|b| > 30^{\circ}$, where the LAT sensitivity varies at most by $20\%$ around its average, $3.2\cdot 10^{-12}$ erg cm$^{-2}$s$^{-1}$. Therefore we fix $S_t = 3.2\cdot 10^{-12}$ erg cm$^{-2}$s$^{-1}$. This flux is integrated in the range 0.1 GeV-100 GeV but the measured $C_P$ are given in an other energy range $E'_1 \leq E \leq E'_2$. Hence, given the photon index $\Gamma$ and the energy cutoff $E_{\rm{cut}}$ in the $F'_{\gamma}$ integration of Eq.~\ref{Cpdef}, we can find the flux threshold $F'_{t}$ in the $C_P$ energy range $E'_1 \leq E \leq E'_2$ by taking into account the relation between the photon flux $F'_{\gamma}$ and the energy flux $S_{\gamma}$ (see Eq.~\ref{eq:FSgamma})
 in the range $E_1 \leq E \leq E_2$:
\begin{eqnarray}
     \label{conv}
        &F'_t&(S_t,\Gamma,E_{\rm{cut}},E'_1,E'_2) = \frac{S_t}{\int^{E_2}_{E_1} E (\frac{E}{E_0})^{-\Gamma} \exp{\left(-\frac{E}{E_{\rm{cut}}}\right)} dE}  \cdot \nonumber \\ 
\cdot &\int&^{E'_2}_{E'_1} \left(\frac{E}{E_0}\right)^{-\Gamma} \exp{\left(-\frac{E}{E_{\rm{cut}}}\right)} dE.
    \end{eqnarray} 
In Eq.~\ref{Cpdef} $d^3N/(dF'_{\gamma}d\Gamma dE_{\rm{cut}})$ is the differential distribution with respect to the flux, photon index and energy cutoff and it is usually factorised by three independent functions \citep{Cuoco:2012yf,Ackermann:2012uf}:
\begin{equation}
     \label{dNdFdGdEc}
	\frac{d^3N}{dF'_{\gamma}d\Gamma dE_{\rm{cut}}} = \frac{dN}{dF'_{\gamma}} \frac{dN}{d\Gamma} \frac{dN}{dE_{\rm{cut}}},
    \end{equation} 
where $dN/d\Gamma$ and $dN/dE_{\rm{cut}}$ are given by Gaussian distributions as in Sec.~\ref{sec:gammaMSP}. 
$dN/dF'_{\gamma}$ is usually (see e.g. \cite{Cuoco:2012yf, Ackermann:2012uf,Collaboration:2010gqa}) described by a broken power-law:
\begin{equation}
     \label{dNdF}
	\frac{dN}{dF'_{\gamma}} =
	\left\{
	\begin{array}{rl}
	A {F'}_{\gamma}^{-\beta} & F'_{\gamma} \geq F_b \\
	A {F'}_{\gamma}^{-\alpha} F_b^{\alpha-\beta} & F'_{\gamma} < F_b
	\end{array}
	\right.
    \end{equation} 
where $A$ is a normalisation factor in units of cm$^2$ s$^{-1}$ sr$^{-1}$, $F_b$ is the break flux and $\alpha$ and $\beta$ are the slopes of $dN/dF'_{\gamma}$ below and above the break respectively.
As we will see in Sec.~\ref{subsec:res_anisotropy} the broken power-law is adequate to parametrize the flux distribution of MSPs.
\newline
In order to find the values of $A$, $\beta$, $\alpha$ and $F_b$ we fit the MC MSP flux distribution with the theoretical flux distribution given by the following equation:
\begin{equation}
     \label{dNdFdt}
	{\frac{dN}{dF'_{\gamma}}}_{\rm{fit}} = \int^{E_{\rm{cut,max}}}_{E_{\rm{cut,min}}} \int^{\Gamma_{\rm {max}}}_{\Gamma_{\rm{min}}} \frac{dN}{dF'_{\gamma}d\Gamma dE_{\rm{cut}}} d \Gamma d E_{\rm{cut}}\,.
    \end{equation} 
\section{Results}
\label{sec:results}

\subsection{Contribution to the high-latitude $\gamma$-ray diffuse background}
\label{subsec:IDGRB}
By following the MC procedure highlighted in Sec.~\ref{sec:MC}, we generate a MSP population that follows the assumed distributions in period, magnetic field and distance; the $\gamma$-ray spectral properties of the simulated sources reflect the distributions of observed $\gamma$-ray MSPs (Eqs.~\ref{eq:gauss1} - \ref{eq:gauss2}).
\newline
To account for the uncertainty due to the conversion of the spin-down luminosity into the $\gamma$-ray luminosity, for each simulated source, we extract the conversion efficiency $\eta$ (Eq.~\ref{eq:lgamma}) from a $\log_{10}$ uniform distribution in the range $\log_{10}{(0.015)} - \log_{10}{(0.65)}$, with mean value 0.095, as represented in Fig.~\ref{fig:lumedot}.
Moreover, we parametrise the sensitivity of the LAT to MSPs detection using a latitude-dependent function $S_t(b)$ with a normalisation that corresponds to the best fit value of the \Fermi-LAT sensitivity curve and with a dispersion given by the 10\% and 90\% percentile sensitivity (see Fig.~17 of \cite{2013ApJS..208...17A}). In the MC procedure, we randomly extract for each source a value for the normalisation of the flux sensitivity function $S_t(b)$, bracketing this uncertainty. 
\newline
As an additional source of uncertainty, we consider the dependence of the prediction on the single MC realisation. Taking into account all the above cited distributions, i.e.~for the vertical and radial distances, the spin period, the surface magnetic field, the spectral $\gamma$-ray parameters $\Gamma$ and $E_{\rm{cut}}$, the conversion efficiency $\eta$, and the normalisation of the sensitivity function, we perform 1000 MC realisations of the MSP population in the Galaxy.
Our final goal is to compute the average differential energy distribution $dN/dE$ from unresolved MSPs and its uncertainty. 
The total differential energy spectrum of unresolved MSPs is given by the sum of the single source spectra at any energy (see Eq.~\ref{eq:dndesim}). Since, for each source with spectrum as in Eq.~\ref{eq:expcut}, we do not assume a universal value for the spectral parameters but rather we assign randomly $\Gamma$ and $E_{\rm{cut}}$ to each single simulated source from the distributions of Eqs.~\ref{eq:gauss1} and \ref{eq:gauss2}, the total differential energy spectral shape may slightly vary from one realisation to another. To guarantee a precise reconstruction of the average total spectral distribution in the whole energy range and taking into account that the spectrum rapidly varies at low energies (E $\lesssim$ 3 GeV), we derive the probability distribution function (PDF) of the integrated fluxes (see Eq.~\ref{eq:diffsim}) above three fixed energy thresholds, 0.1 GeV, 0.7 GeV and 1.5 GeV. The PDFs are consistent with a Gaussian for all the three energy thresholds, for which we derive mean and dispersion $\
sigma$ of the corresponding integrated flux distribution.
The MC realisations that have integrated fluxes above 0.1 GeV, 0.7 GeV and 1.5 GeV equal to the mean flux, mean flux $\pm$ 1$\sigma$ for all the three integrated flux distributions corresponding to the three fixed threshold energies, are identified, respectively, as our best fit, $\pm$ 1$\sigma$ configurations.
The average prediction on the total diffuse emission is computed from the best fit configuration, while the uncertainty band is delimited by the $\pm$ 1$\sigma$ realisations.
Therefore, the ``1$\sigma$" of the 1$\sigma$ uncertainty band is primarily meant to indicate the dispersion on the integrated flux distributions.

\begin{figure*}
     \begin{centering}
        \subfigure[]{%
            \label{fig:map_res}
            \includegraphics[width=0.5\textwidth]{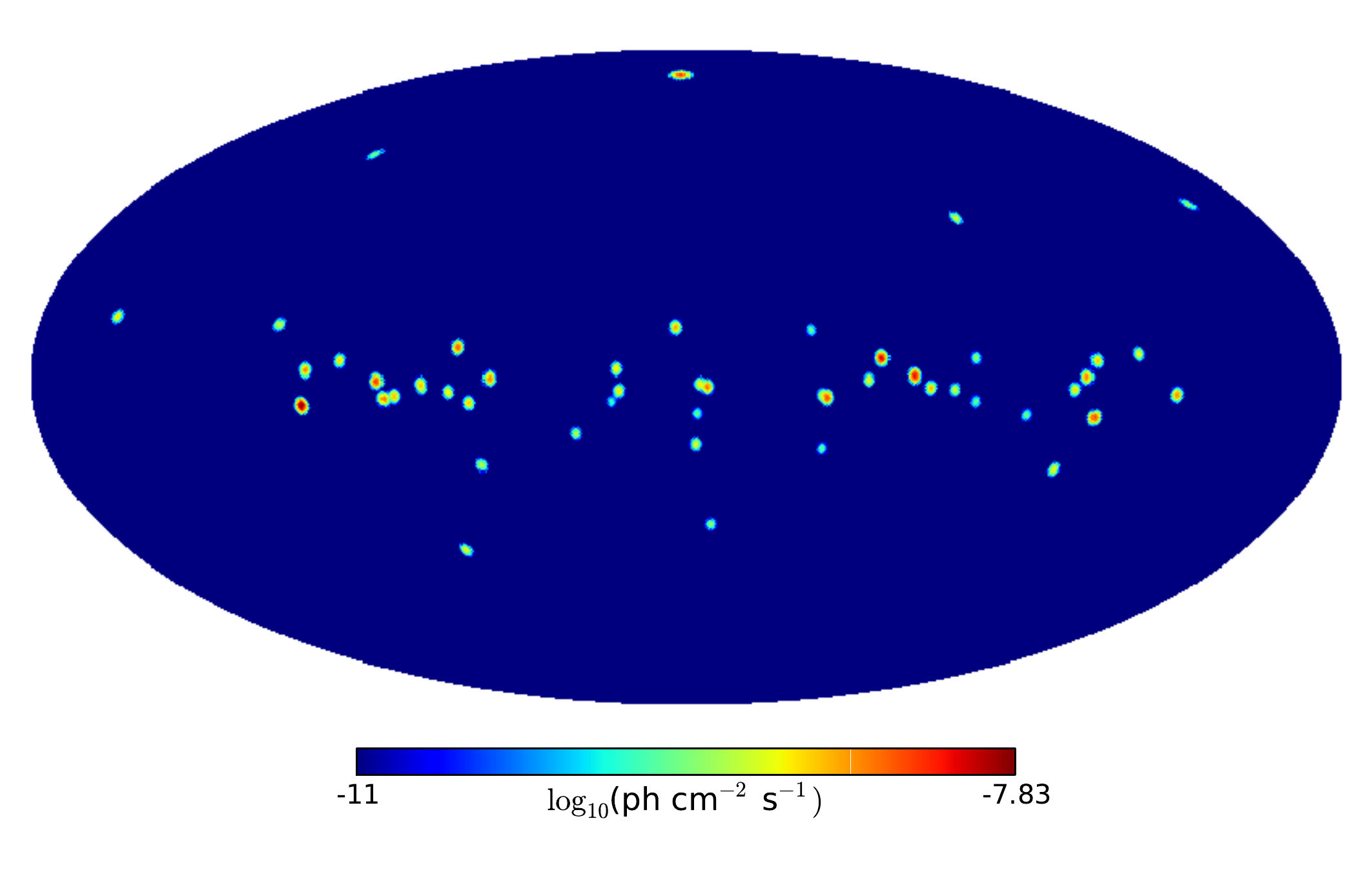}
        }%
        \subfigure[]{%
           \label{fig:map_unres}
           \includegraphics[width=0.5\textwidth]{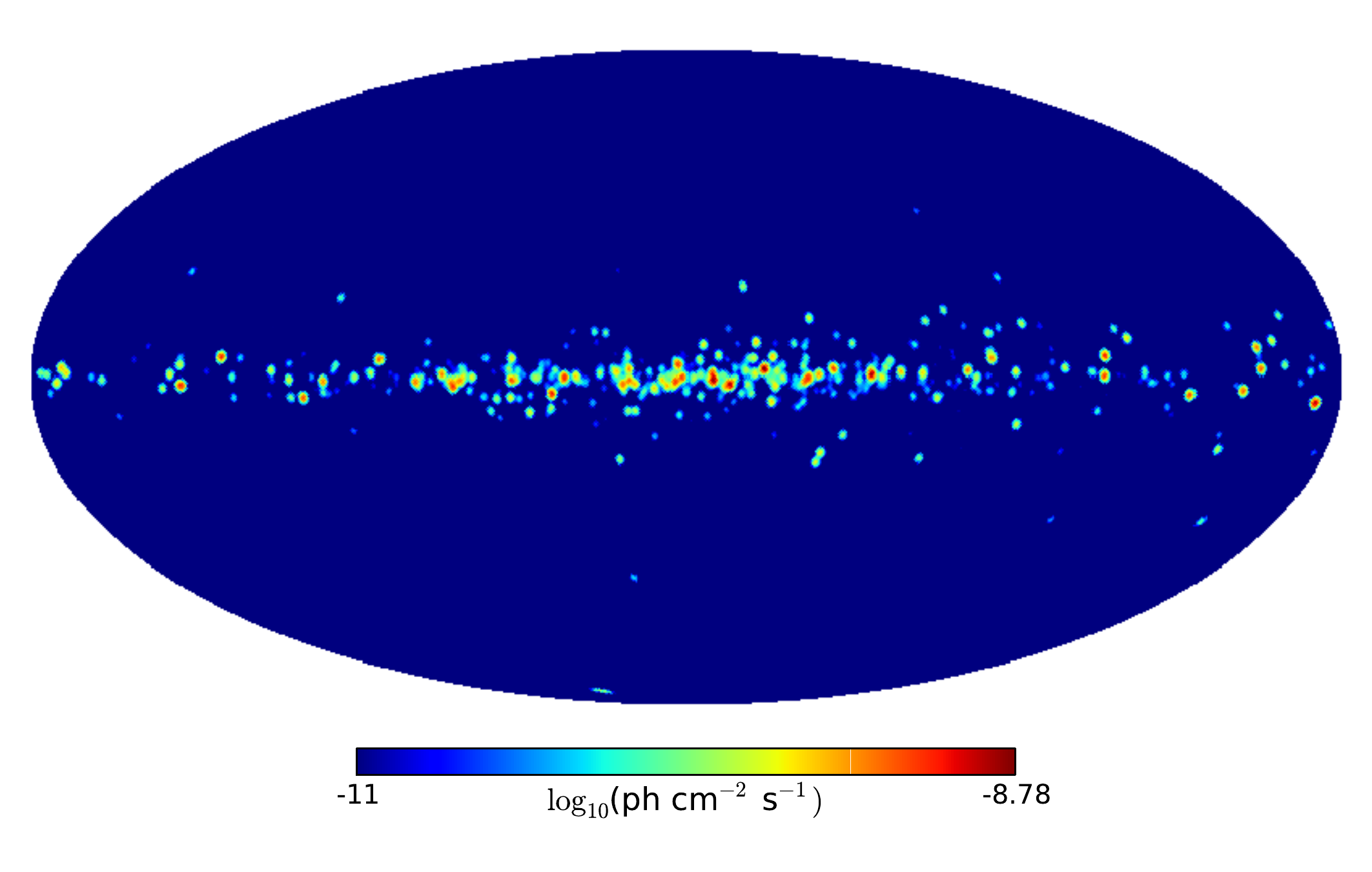}
        }
    \caption{All-sky $\gamma$-ray map of the MC simulated MSP population for the best fit realisation. The left panel shows the sources that would be detected by the LAT, while the unresolved counterpart is displayed in the right panel. The different color scale is chosen to allow a better visual effect and a smoothing of 1.5$^{\circ}$ is applied to the maps.}
   \label{fig:map}
   \end{centering}
\end{figure*}

A typical $\gamma$-ray all-sky map\footnote{The $\gamma$-ray intensity maps have been generated by using the
HEALPix software \citep{2005ApJ...622..759G}. } of our simulated population is shown in Fig.~\ref{fig:map}. We have chosen the best fit realisation which gives the best fit curve for the diffuse $\gamma$-ray emission in Fig.~\ref{fig:BMflux}. In order to highlight the properties of the population, we decide to display separately the resolved (left panel) and unresolved (right panel) components. The color scale is chosen such to allow the reader to see most of the sources, even when they have very low fluxes. The detected sources are determined by the implementation of the {\it Fermi}-LAT 2FPC sensitivity as explained above.

\begin{figure*}
\begin{centering}
\includegraphics[width=0.50\textwidth]{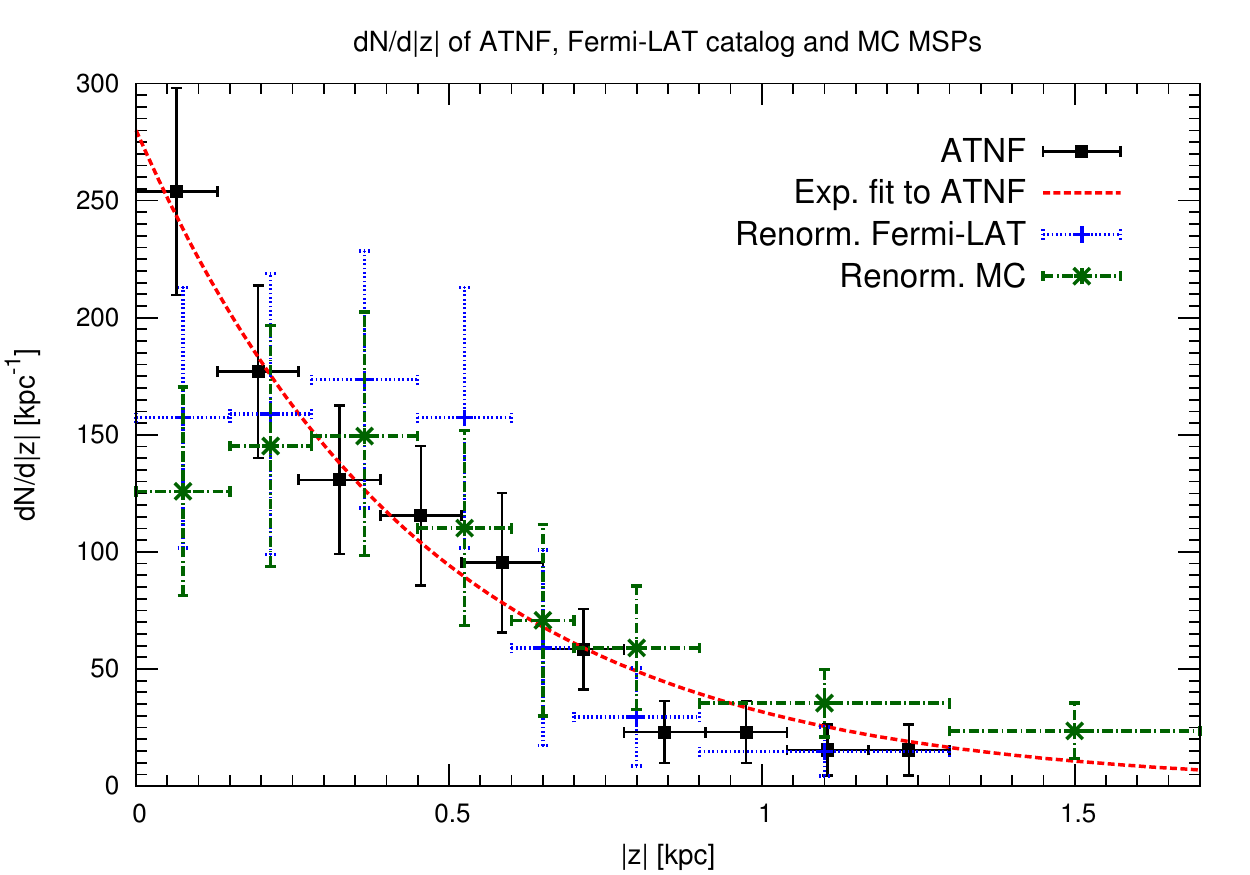}
\caption{The renormalised distribution of the height $z$ from the galactic plane of {\it Fermi}-LAT MSPs in Tab.~\ref{tab:sample} is represented by the blue dashed data points, while solid black points and dotted line refer to radio distribution and best fit. We display also the renormalised distribution for the detected sources in the MC realisation which gives the best fit of Fig.~\ref{fig:BMflux} (dot-dashed green points).}
\label{fig:dNdzfin}
\end{centering} 
\end{figure*}

\begin{figure*}
\begin{centering}
\includegraphics[width=0.65\textwidth]{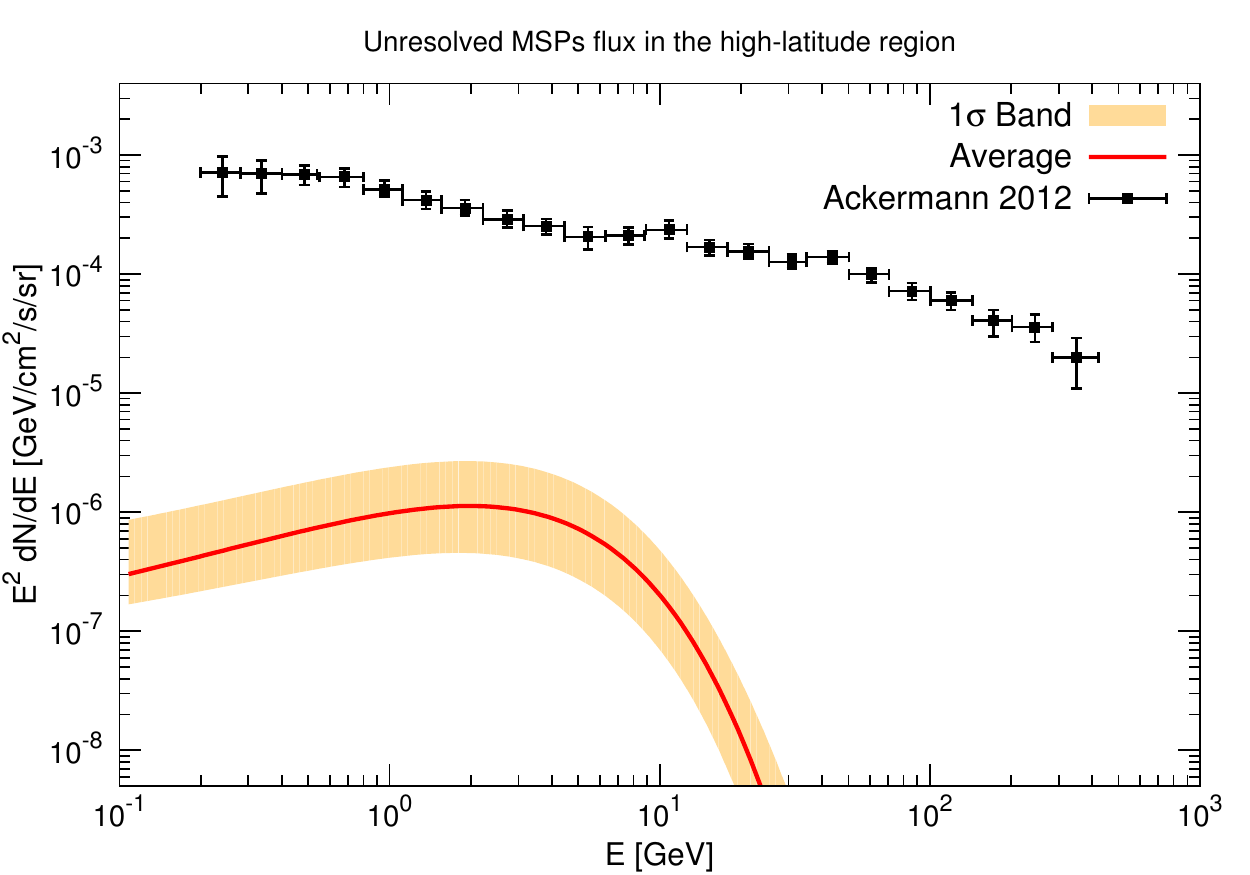}
\caption{Prediction of the diffuse $\gamma$-ray flux at $|b|\geq 10^{\circ}$ from the unresolved population of MSPs as derived from 1000 MC simulations of the MSP galactic population. The red solid line represents the mean spectral distribution (see text for further details), while the light orange band corresponds to the  1$\sigma$ uncertainty band. The black points refer to the IDGRB preliminary data taken from \cite{ackermann2012}.}
\label{fig:BMflux}
\end{centering} 
\end{figure*}

We have already shown that the intrinsic distribution of $z$, assumed to be the one derived from the ATNF catalog, and the one calculated from the {\it Fermi}-LAT MSPs are different close to the galactic plane (see Fig.~\ref{fig:distr_rz}). We can demonstrate that the reason is associated to the {\it Fermi}-LAT sensitivity flux, which is at least a factor of two larger in the galactic center with respect to high-latitude regions ($|b| \geq10^{\circ}$).
In order to verify this assumption we have derived the $dN/dz$ for 
a MC realisation of the MSP population which represents the theoretical result deduced from the intrinsic ATNF distribution convolved with the {\it Fermi}-LAT sensitivity. We have renormalised the $dN/dz$ in each bin in order to take into account the different number of sources between the MC and the ATNF catalog. In Fig.~\ref{fig:dNdzfin} the $dN/dz$ is shown for the sources in the ATNF, {\it Fermi}-LAT catalogs and for MC simulated sources. We display also the theoretical best fit for the ATNF catalog MSPs. It is clear that, considering the $dN/dz$ derived from the best fit of the ATNF catalog sources and convolving this intrinsic distribution with the {\it Fermi}-LAT sensitivity, we obtain a distribution which is compatible with {\it Fermi}-LAT data. The rescaled sets of data ``Renorm.~\Fermi-LAT" and ``Renorm.~MC", when compared, have a $\chi^2/d.~o.~f.=0.67$, indicating the good agreement between data and MC. Therefore, we do expect in the galactic center region a large number of sources, whose 
detection is prevented from the decreased instrumental sensitivity.

The contribution to the IDGRB at $|b| \geq 10^{\circ}$ from the unresolved MSPs generated by our MC method is shown in Fig.~\ref{fig:BMflux}.
The data points refer to the preliminary IDGRB data as taken from \cite{ackermann2012}.
The line corresponds to the mean prediction of our 1000 MC realisations, while the band covers the uncertainty due to the choice of the distributions for the vertical and radial distances, the spin period, the surface magnetic field, the spectral $\gamma$-ray parameters $\Gamma$ and $E_{\rm{cut}}$, $\eta$, and the normalisation of the LAT sensitivity function, and it is derived as explained above.
\newline
The MSP total differential energy spectrum in Fig.~\ref{fig:BMflux} follows a power-law with an exponential cutoff as it is peculiar of the single source spectra of which it is the sum (see Eq.~\ref{eq:dndesim}).  At the peak of the spectral emission, $\sim$ 2 GeV, the fraction of the IDGRB due to MSPs is about 0.3\% (0.1\%, 0.9\%) for the best fit (lower, upper) curve, at higer energies the spectrum exponentially decreases, giving almost zero contribution above $\sim$ 20 GeV. The MSP spectrum is always more than 2 orders of magnitude suppressed with respect to the IDGRB data.
We notice that the uncertainty is a factor of about 5 at 0.1 GeV as well as at 10 GeV. 
\newline
The integrated intensity in the range 0.1 - 100 GeV (above $b = 10^{\circ}$), Eq.~\ref{eq:diffsim}, for the mean curve in Fig.~\ref{fig:BMflux} is $5.07 \cdot10^{-9}$ ${\rm ph}\, {\rm cm}^{-2} {\rm s}^{-1} {\rm sr}^{-1}$, which corresponds to $0.05\%$ of the IDGRB integrated flux of \cite{IDGRB}. 
Upper (Lower) edge of the band accounts for $0.13\%$ ($0.02 \%$), with an integrated intensity of $1.32 \cdot10^{-8}$ ($2.43 \cdot10^{-9}$) ${\rm ph}\, {\rm cm}^{-2} {\rm s}^{-1} {\rm sr}^{-1}$.
We point out that modelling the latitude dependence of the LAT detection sensitivity instead of using a sharp threshold (as adopted usually in previous works) affects significantly the final result. We quantify this discrepancy to be about a factor of 2 in the differential flux when using a universal threshold of $10^{-8}$ ph cm$^{-2}$ s$^{-1}$.

The contribution  of unresolved MSPs to the IDGRB is here found to be smaller than what previously estimated, e.g.~in \cite{2011MNRAS.415.1074S,2013AA...554A..62G}.
In \cite{2013AA...554A..62G} the contribution at $|b| \geq 40^{\circ}$ is estimated to be about $1.5 \cdot10^{-8}$ ${\rm ph}\, {\rm cm}^{-2} {\rm s}^{-1} {\rm sr}^{-1}$, while we find $0.6 \cdot10^{-9}$ ${\rm ph}\, {\rm cm}^{-2} {\rm s}^{-1} {\rm sr}^{-1}$. The main difference is due to the values of the assumed  parameters $z_0$ and $\langle \log_{10}(B/G) \rangle$. Both parameters are indeed, in our model, significantly lower than in \cite{2013AA...554A..62G}, implying a strong depression of the final flux. We have checked explicitly that with $z_0  = 1.8$ kpc and  $\langle \log_{10}(B/G)\rangle = 8$ (as in their FG1 reference model that is directly translatable into ours), we get an integrated flux above $b = 40^{\circ}$ of $1.8 \cdot 10^{-8}$ ${\rm ph}\, {\rm cm}^{-2} {\rm s}^{-1} {\rm sr}^{-1}$. 

\subsection{Contribution of pulsars to the inner Galaxy and to the galactic center $\gamma$-ray diffuse background}
\label{subsec:Inner}
As we have already seen in Sec.~\ref{sec:distribution}, MSPs are concentrated along the galactic disk and their number decreases as far as the latitude grows. 
Therefore, although it is possible to find about $75\%$ of MSPs in the 2FPC (30 out of a total of 40 sources) at high latitudes ($|b|\geq 10^{\circ}$), we do expect a large number of sources in the inner region of the Galaxy. 
Moreover, at low latitudes, close to the disk, the population of young $\gamma$-ray pulsars is very abundant in number. Indeed, despite of the sensitivity threshold, the number of young pulsars near the galactic center is about a factor of 11 larger than at high latitudes, where only 7 out of 77 objects are found in the 2FPC.
\newline
This implies that the $\gamma$-ray emission from unresolved pulsars, both young and millisecond, in the innermost part of the Galaxy might be significant and cover an important fraction of the diffuse emission at low latitudes. In this region of the Galaxy recently an excess emission in the \Fermi-LAT $\gamma$-ray data, with respect to standard astrophysical foregrounds and backgrounds, has been claimed by different groups \citep{2013PDU.....2..118H, Gordon:2013vta, Abazajian:2014fta}.
\newline
We therefore derive the $\gamma$-ray flux from the unresolved young pulsars and MSPs in the inner part of our Galaxy, by analysing two different regions of the sky.
\newline
In order to model the distribution of young pulsars in the Galaxy and their $\gamma$-ray emission, we follow the same method explained in Secs.~\ref{sec:distribution}-\ref{sec:MC}.
The ATNF catalog contains about 2000 of sources with $P> 15$ ms \citep{2005AJ....129.1993M} that, according to the adopted convention, are classified as young pulsars.
Those objects have a space distribution similar to the one of MSPs but they are more concentrated towards the galactic center. 
We study the spatial distribution of young pulsars as we have done for MSPs in Sec.~\ref{sec:distribution}. 
We deduce that the spatial vertical height distribution $dN/dz$ of sources is well described by an exponential (Eq.~\ref{eq:zdistrone}) with $z_0 = 0.10$ kpc. Such value for $z_0$ indicates that young pulsars are more concentrated along the galactic plane with respect to MSPs ($z_0=0.67$ kpc).
The average radial distance of the $dN/dr$ is $\langle r \rangle =$ 6.5 kpc and the distribution does not show a clear peak.
It has been demonstrated in \cite{Lorimer:2006qs,FaucherGiguere:2005ny,Yusifov:2004fr} that the intrinsic distribution of young pulsars should  peak at about 3-4 kpc from the galactic center because these sources are born in the galactic spiral arms. Following the evolution of this population by starting from its birth distribution could broaden the evolved distribution and move its peak to 6-7 kpc \citep{FaucherGiguere:2005ny}.
Hence, we consider the same radial distribution derived by \cite{Yusifov:2004fr} which is compatible with the model derived in \cite{Lorimer:2006qs}. This distribution has a peak at $3.2 \pm 0.4$ kpc, a scale-length of $3.8 \pm 0.4$ kpc and a depletion of the number of sources when moving towards the galactic center:
\begin{eqnarray}
\label{eq:dNdSMC}
  \displaystyle \frac{dN}{dx} \propto x^{\alpha} \exp{\left( - \beta \left( \frac{x-x_{\rm{sun}}}{x_{\rm{sun}}} \right) \right)} \, ,
\end{eqnarray}
with $\alpha=1.64$, $\beta=4.01$, $x=r+r_1$, $x_{\rm{sun}}= r_{\rm{sun}}+r_1$ and $r_1=0.55$.
This radial distribution is compatible with the observed distribution of young pulsars of the ATNF catalog \citep{2005AJ....129.1993M} and Parkes catalog \citep{Lorimer:2006qs}.

The surface magnetic field and the rotational period are adequately fitted by a $\log_{10}$ Gaussian distribution (see Eq.~\ref{eq:Pdistrlog10} and \ref{eq:bdistr}) with, $\langle \log_{10} (B/\rm G) \rangle = 12.1$, $\sigma_{\log_{10} B}=0.6$, and $\langle \log_{10} (P/\rm s) \rangle=-0.21$, $\sigma_{\log_{10} P}=0.38$.
The radial distance $r$, the vertical height $z$, the rotational period $P$ and the surface magnetic field distributions we derive for the young pulsars in the ATNF catalog are compatible with the ones deduced in \cite{Lorimer:2006qs,FaucherGiguere:2005ny,Yusifov:2004fr,2010JCAP...01..005F}.
\newline
The $\gamma$-ray spectral properties are derived from the sample of 77 young pulsars of the 2FPC. We fit the cutoff energy $E_{\rm{cut}}$ and the photon index $\Gamma$ with a logarithmic base 10 and a linear Gaussian distribution and we find the following best fit parameters: $\langle \log_{10} (E_{\rm{cut}}/\rm{MeV}) \rangle=3.35$, $\sigma_{\log_{10} E_{\rm{cut}}}=0.26$, $\langle \Gamma \rangle=1.51$ and $\sigma_{\Gamma}=0.32$.
We show in Tab.~\ref{tab:pary} the best fit values and the uncertainties for the parameters of our young pulsars sample.

\begin{table*}
\begin{centering}
\begin{tabular}  {||c|c|c||c|c|c||c|c||}
\hline
\hline
$B$ &     $\langle \log_{10}( B /\rm G) \rangle$   &   $\sigma_{\log_{10} B}$   &   $P$   &   $\langle \log_{10} (P /\rm s) \rangle$  &  $\sigma_{\log_{10} P}$  &   $\langle \log_{10} (E_{\rm{cut}} /\rm MeV) \rangle$  &  $\sigma_{\log_{10} E_{\rm{cut}}}$   \\
\hline
       &     $12.06\pm0.06$    &    $0.55\pm0.08$                   &             &      $-0.21\pm0.03$  &  $0.38\pm0.04$ &  $3.35\pm0.09$  &  $0.26\pm0.04$                 \\ 
\hline
\hline
$r$  &     $\langle r \rangle [{\rm kpc}]$  & $r_0  [{\rm kpc}]$    &   $z$         &  $\langle z \rangle  [{\rm kpc}]$ &     $z_0  [{\rm kpc}]$    &   $\langle \Gamma \rangle$  &  $\sigma_{\Gamma}$                                \\
\hline
       &     $6.4\pm0.2$    &   $3.1\pm0.3$                   &             &      $0.00\pm0.09$    &  $0.10\pm0.05$ &  $1.51\pm0.04$  &  $0.32\pm0.05$		\\
\hline
\hline
\end{tabular}
\caption{Best fit parameters for a Log$_{10}$ Gaussian distribution 
for the magnetic field $B$ and the period $P$, for an exponential function for the distance from the galactic plane $z$ and the radial distance $r$ and for a a Log$_{10}$ Gaussian distribution for the cutoff energy $E_{\rm{cut}}$ and a 
Gaussian for the photon index $\Gamma$ for our sample of young pulsars taken from the ATNF catalog \citep{2005AJ....129.1993M} and from the 2FPC \citep{2013ApJS..208...17A}.}
\label{tab:pary}
    \end{centering}
\end{table*} 

As for the $\gamma$-ray emission, we consider again the relation $L_{\gamma}$ and $\dot{E}$  (see Eq.~\ref{eq:lgamma}), by adapting the parameters $\alpha$ and $\eta$ to the data for young pulsars. In this case, we fix $\alpha = 1$, $\langle \eta \rangle$ = 0.080 and $\eta=[0.009,0.75]$.

\begin{figure*}
     \begin{centering}
        \subfigure[]{%
            \label{fig:map_res}
            \includegraphics[width=0.5\textwidth]{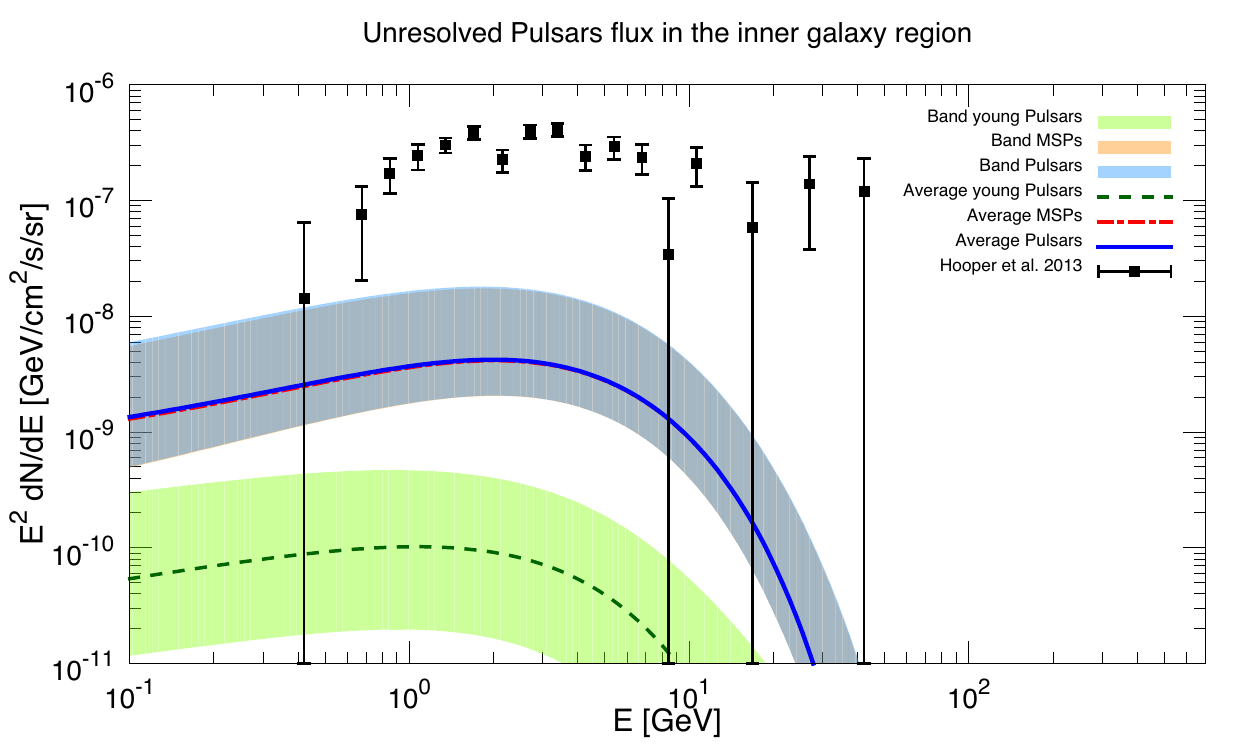}
        }%
        \subfigure[]{%
           \label{fig:map_unres}
           \includegraphics[width=0.5\textwidth]{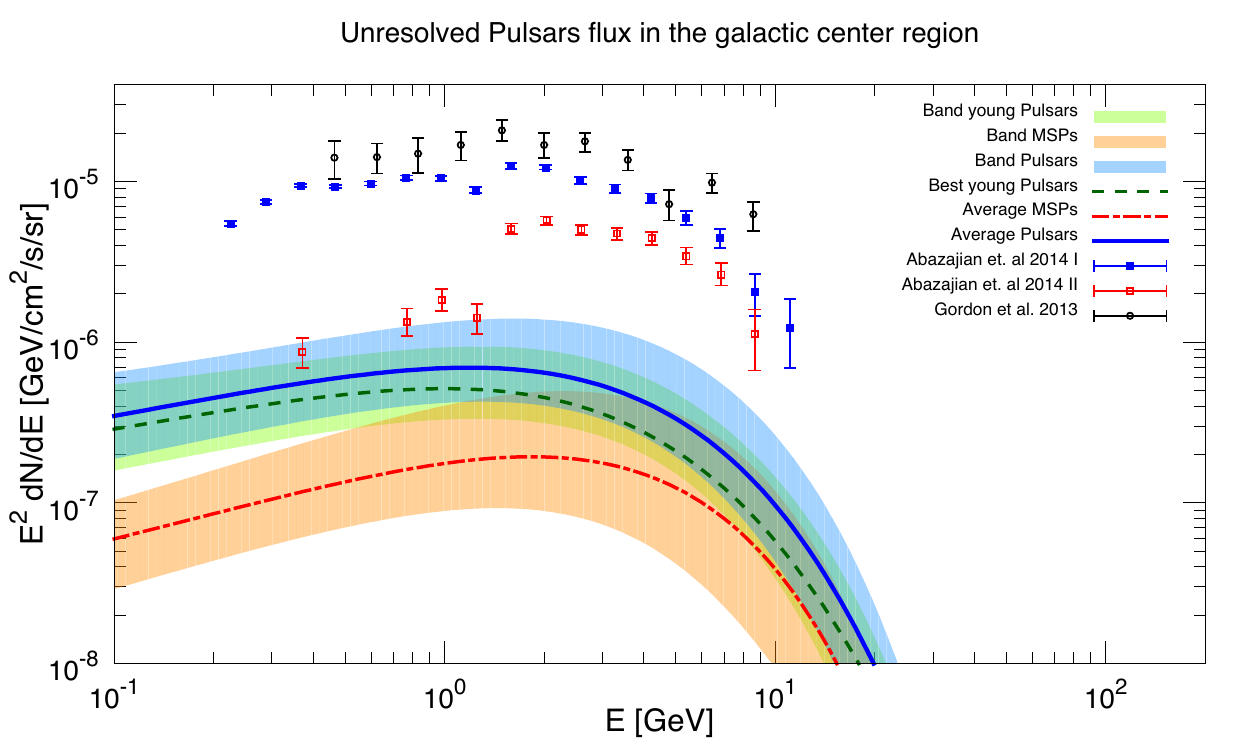}
        }
    \caption{The best fit (band) of the $\gamma$-ray flux from the unresolved young pulsars is shown with the dashed green line (green band), from the unresolved MSPs with the dot-dashed red line (orange band) and from the whole pulsars population with the blue solid line (blue band). On the left side, we display the pulsar population contribution in the inner Galaxy region ($10^{\circ} \leq |b| \leq 20^{\circ}$ and $l \in [-180^{\circ},180^{\circ}]$), while in the right panel the contribution to the galactic center diffuse emission ($|b| \leq 3.5^{\circ}$ and $|l| \leq 3.5^{\circ}$). The black points in the left panel have been taken from
\cite{2013PDU.....2..118H}.  The black dots in the right panel have been taken from \cite{Gordon:2013vta}, 
while the red and blue ones correspond to two different analysis carried by \cite{Abazajian:2014fta}.
}
   \label{fig:Illat}
   \end{centering}
\end{figure*}

By adopting the same method used for MSPs (Secs.~\ref{sec:MC}-\ref{subsec:IDGRB}), we derive the $\gamma$-ray emission from the unresolved counterpart of MSPs and young pulsars populations in different regions at intermediate and low latitudes. In order to compare with the previous literature we consider: $10^{\circ} \leq |b| \leq 20^{\circ}$ and $l \in [-180^{\circ},180^{\circ}]$, i.e.~hereafter, the inner Galaxy region, and $|b| \leq 3.5^{\circ}$ and $|l| \leq 3.5^{\circ}$, i.e.~hereafter, the galactic center region.
The result is displayed in Fig.~\ref{fig:Illat}, where we show the emission from unresolved young pulsars, MSPs and from their sum.

The left panel refers to the contribution to the inner Galaxy region, where the MSP contribution dominates the pulsars diffuse $\gamma$-ray flux because at those latitudes only few young pulsars are present. For comparison, the excess emission in the same region derived by \cite{2013PDU.....2..118H} is shown.
Instead, in the galactic center region (right panel of  Fig.~\ref{fig:Illat}) the MSPs and young pulsars populations give about the same contribution to the diffuse emission.
The overlaid data sets refer to three analysis performed in the galactic center region and finding similar results \citep{Gordon:2013vta, Abazajian:2014fta}. The two different results for the excess from \cite{Abazajian:2014fta} are obtained by differently modelling the background and the templates for point sources. The discrepancy, in particular, at low energies is emblematic of the great uncertainty in the determination of the excess spectrum.
The spectral shape of the young pulsars contribution differs from the one of the MSPs because the different best fit values of the $\Gamma$ and $E_{\rm{cut}}$ distributions. The uncertainty band of young pulsars is, in general, wider with respect to the one of MSPs because of the larger uncertainty on the $\eta$ parameter.
\newline
Considering the inner Galaxy region we find that only about 5\% of the excess in \cite{2013PDU.....2..118H} at 1-3 GeV (where the emission peaks) can be covered by the total unresolved pulsars population. 
This result, although more conservative, is in agreement with the estimate of \cite{2013PhRvD..88h3009H} and the differences may be traced back in a slightly higher value of the magnetic field and the detection threshold assumed.
In the innermost region of the Galaxy, i.e.~the galactic center region, we find that the flux from unresolved pulsars may account for about 8\% of the excess emission at the peak ($\sim$ 1 GeV) derived by \cite{Gordon:2013vta} and \cite{Abazajian:2014fta} (model I). 
Additionally, \cite{Gordon:2013vta} give an interpretation of the excess as due to an unresolved population of pulsars with the spectral energy distribution given by Eq.~\ref{eq:expcut} with $\Gamma=1.6\pm 0.2$ and $E_{\rm{cut}}= 4000 \pm 1500$ MeV. To make a comparison with this result, we derive $\Gamma$ and $E_{\rm{cut}}$ for the total pulsars emission with the function of Eq.~\ref{eq:expcut}  (fixing $E_0 = 1176$ MeV as in  \cite{Gordon:2013vta}) and we find best fit spectral index $\Gamma$ and $E_{\rm{cut}}$ equal to, respectively, $-1.53$ and $3390$ MeV, which are compatible with   \cite{Gordon:2013vta} with, however, a different normalisation factor.

Our result is not in contrast with  \cite{2014arXiv1404.2318Y}, that could explain the whole excess in terms of MSPs. Indeed, we do not consider here the MSP population from the bulge but only the one from the disk. Adding a second MSP component, that is, by definition, more concentrated towards the galactic center, would increase the contribution from MSPs at low latitudes.
We caution, however, that in those two regions the determination of the excess is very delicate and it is a matter of a huge debate because of the great impact of systematics. Future and independent analysis of {\it Fermi}-LAT data could confirm or refute this result. In particular, the normalisation, spectral shape and morphology of the excess strongly depend on the method used to analyse the data, namely the source catalog, the point source subtraction method, the modelling of the galactic diffuse foreground, and the templates for the different components of the fits.

\begin{figure*}
\begin{centering}
\includegraphics[width=0.50\textwidth]{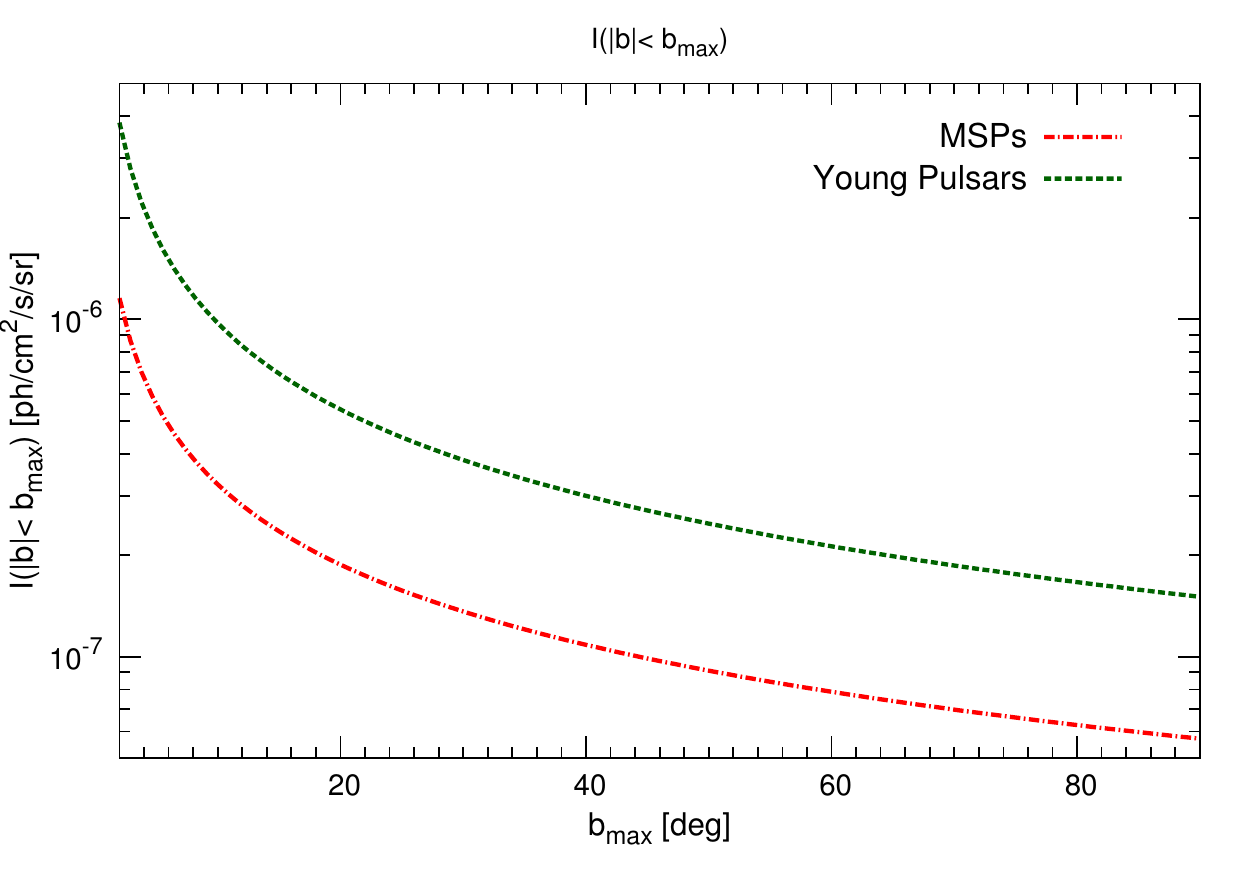}
\caption{The integrated flux $I(|b|<b_{\rm{max}})$ for latitudes smaller than $b_{\rm{max}}$, derived with Eq.~\ref{eq:diffsim} in the energy range 0.1 - 100 GeV, is shown for MSPs (blue solid line) and young pulsars (red dashed line).}
\label{fig:Ilat}
\end{centering} 
\end{figure*}

Finally, in Fig.~\ref{fig:Ilat}, we compare the integrated emission from unresolved MSPs and young pulsars as a function of the latitude, $b$.
The contribution of unresolved pulsars to the IDGRB strongly depends on the latitude, as a consequence of the spatial distribution of the sources. The integrated flux in the energy range 0.1 - 100 GeV, Eq.~\ref{eq:diffsim}, for different regions of the sky, $|b| < b_{\rm {max}}$ is displayed as a function of the maximal latitude $I(|b|<b_{\rm{max}})$, with $b_{\rm {max}}$ ranging from 2$^{\circ}$ to 90$^{\circ}$. Most of the $\gamma$-ray flux coming from the MSPs (young pulsars) is concentrated within $20^{\circ}$ ($10^{\circ}$) since the most of the emission comes from low latitudes. For high latitudes ($b_{\rm{max}}>20^{\circ}$) the integrated flux is about one order of magnitude smaller for both populations.

\subsection{Contribution of MSPs to the $\gamma$-ray anisotropy}
\label{subsec:res_anisotropy}
\begin{figure*}
\begin{centering}
\includegraphics[width=0.35\textwidth]{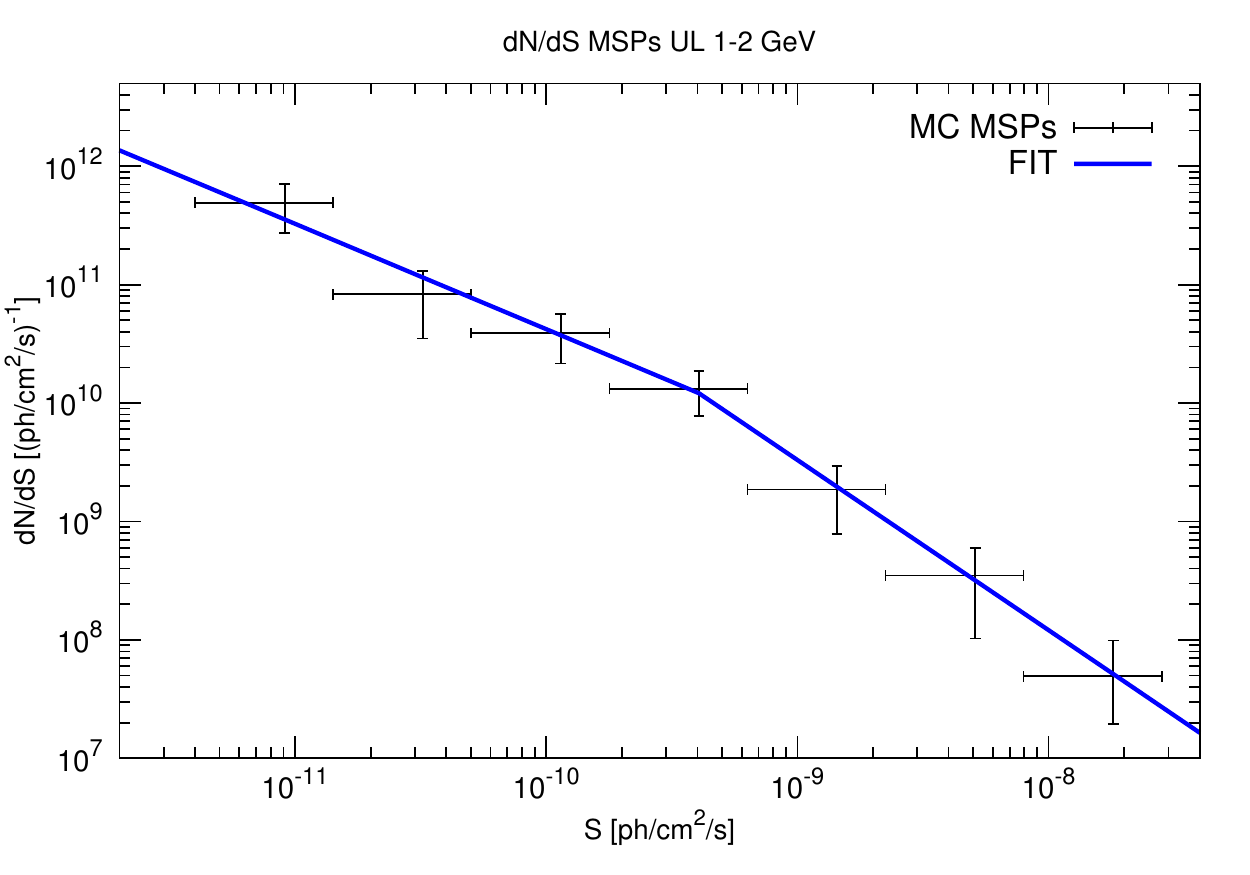}
\includegraphics[width=0.35\textwidth]{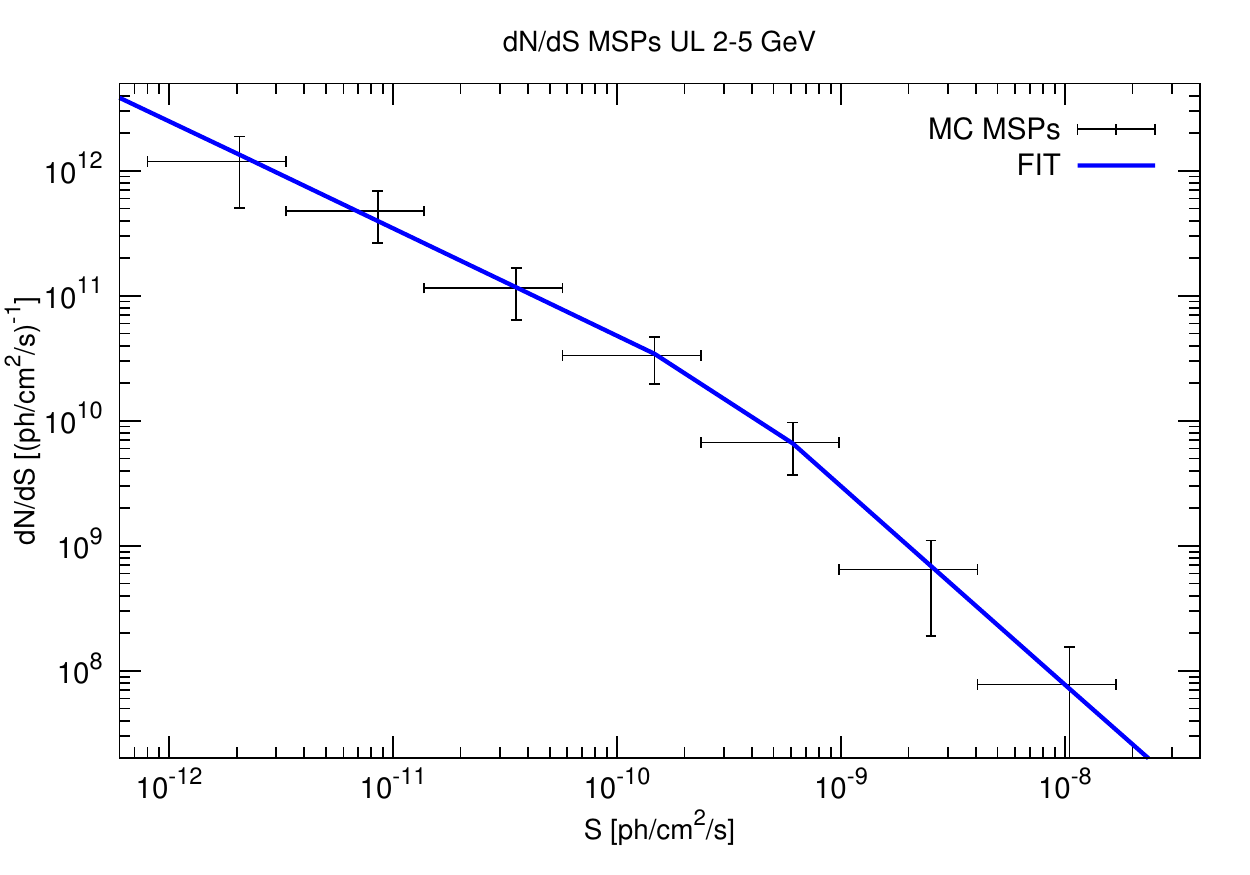}
\includegraphics[width=0.35\textwidth]{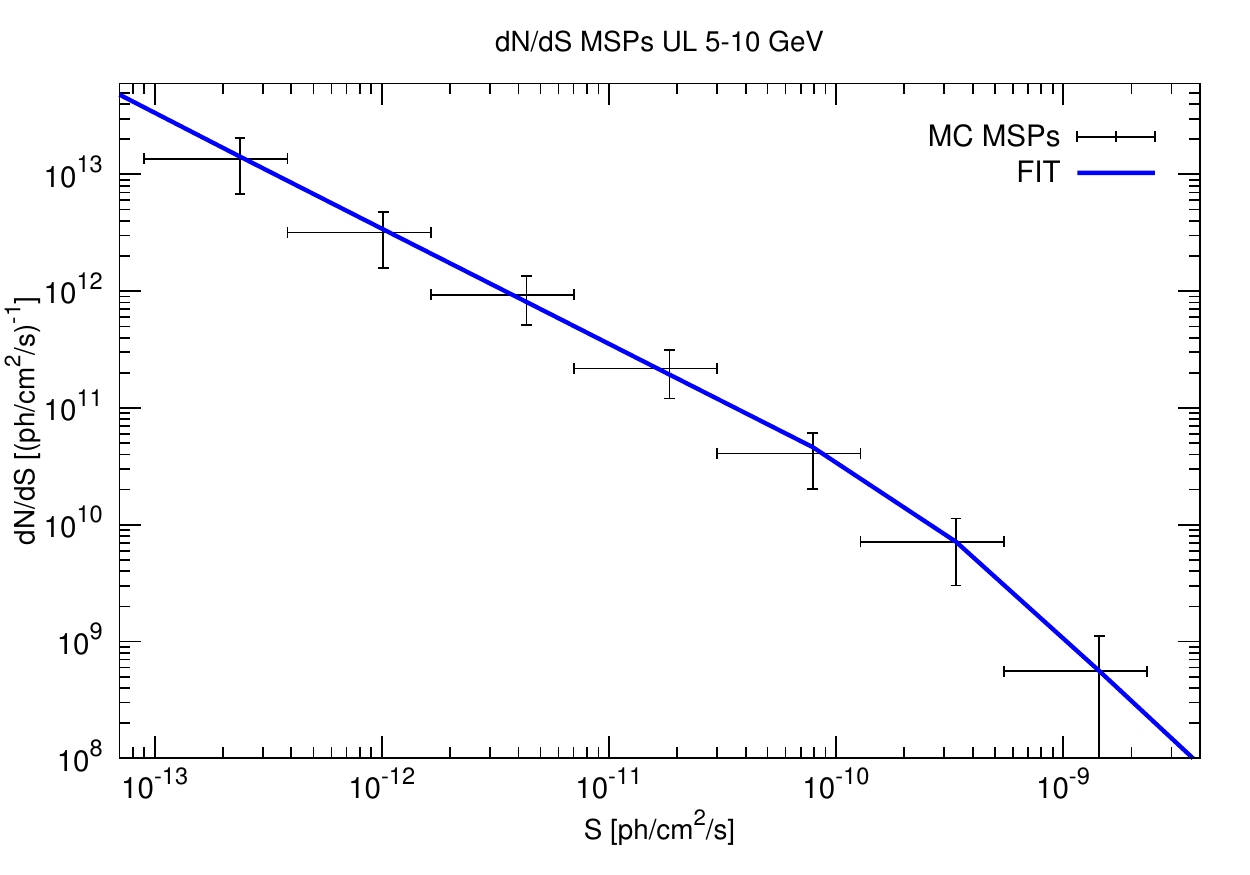}
\includegraphics[width=0.35\textwidth]{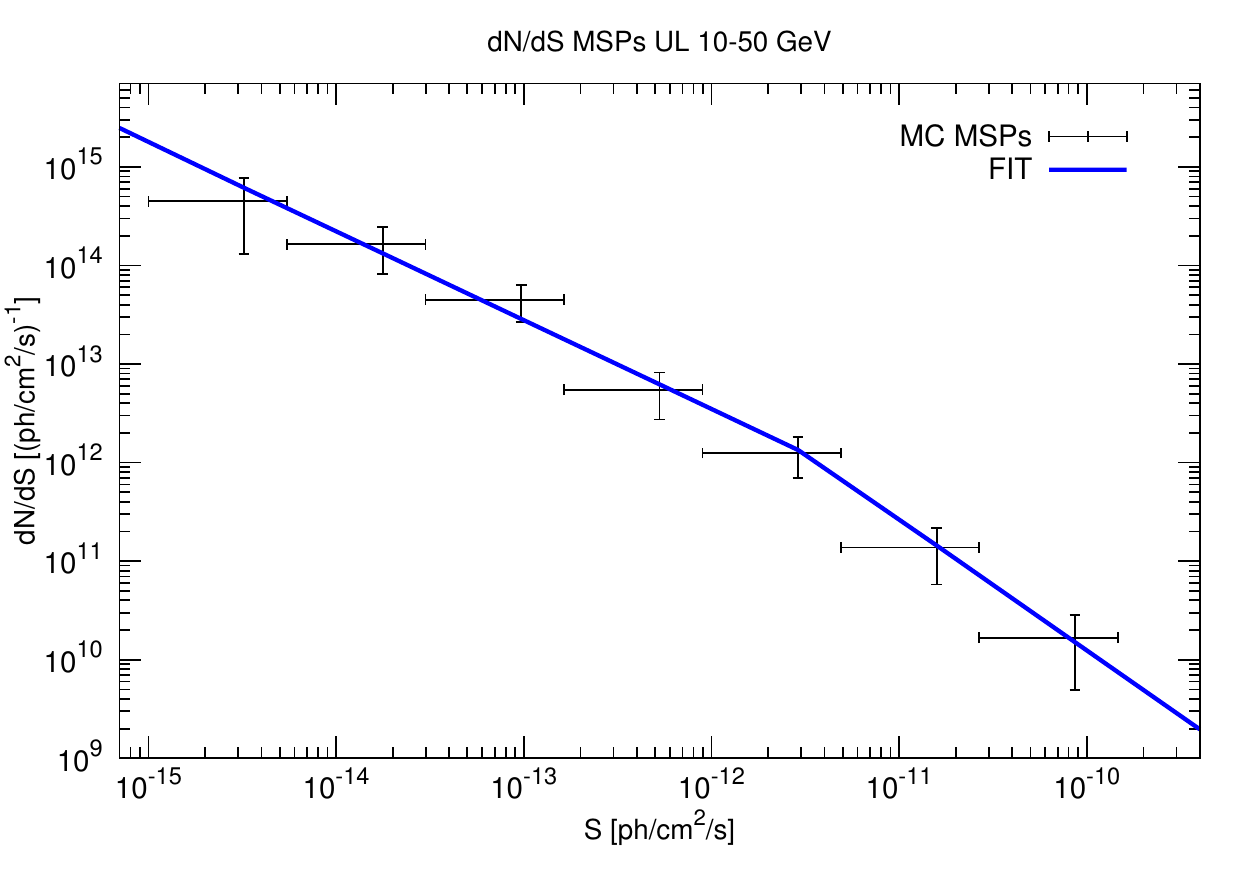}
\caption{In this figure we display, for the four energy bins quoted in the text, the flux distributions of the MSPs of the + 1$\sigma$ MC realisation (black points) and the best fit flux distribution derived using Eq.~\ref{dNdFdt} (blue solid line).}
\label{fig:dnds} 
\end{centering}
\end{figure*}

In this section we present our result on the anisotropy originating from the unresolved MSP population.
We follow the method explained in Sec.~\ref{subsec:anisotropy} and we derive the 1$\sigma$ upper limit angular power, $C_P$. 
We consider the MC realisations which give best fit, lower and upper edges of the 1$\sigma$ uncertainty band of Fig.~\ref{fig:BMflux} (Sec.~\ref{subsec:IDGRB}).
For each of these realisations we derive the differential flux distribution $({dN/dF'_{\gamma}})_{\rm{MC}}$ of the unresolved MSPs. We divide the range of flux $F'_{\gamma}$ in N bins. For each bin $({dN/dF'_{\gamma}})_{\rm{MC}}$ is given by the following equation:
\begin{eqnarray}
\label{eq:dNdSMC}
  \displaystyle \left( \frac{dN}{dF'_{\gamma}} (F'_{\gamma} \in [F^i_{\rm{min}},F^i_{\rm{max}}])  \right)_{\rm{MC}} = \frac{N^i}{\Delta {F'}_{\gamma}^{i}}\, ,
\end{eqnarray}
where $dN/{dF'_{\gamma}}_{\rm{MC}}$ is calculated in the i-th bin defined as $F'_{\gamma} \in [F^i_{\rm{min}},F^i_{\rm{max}}]$, $N^i$ is the number of unresolved sources in the bin and $\Delta F_{\gamma}^{i} = F^i_{\rm{max}}-F^i_{\rm{min}}$ is the width of the bin.
$({dN/dF'_{\gamma}})_{\rm{MC}}$ represents the differential flux distribution of the unresolved part of the considered MC realisation. In Fig.~\ref{fig:dnds} we show the result for the realisation corresponding to the upper limit of the band in Fig.~\ref{fig:BMflux}. The uncertainties are Poissonian errors, namely $\sqrt{N^i}$, associated to the number of sources in each bin $N^i$.
The realisation which gives the lower part of the 1$\sigma$ uncertainty band of the diffuse $\gamma$-ray emission has a too small number of unresolved sources for computing the MC MSP flux distribution and therefore we are not able to derive the best fit and 1$\sigma$ uncertainty values of $C_P$.
Nevertheless, we do compute the 1$\sigma$ upper limit of $C_P$ considering the realisation which gives the upper bound of the flux band.
We calculate the differential distribution $d^3N/(dF'_{\gamma}d\Gamma dE_{\rm{cut}})$, defined in Eq.~\ref{dNdFdGdEc}, performing a fit of the theoretical flux distribution given in Eq.~\ref{dNdFdt} with the same quantity derived from the MC realisation with Eq.~\ref{eq:dNdSMC}.
In Fig.~\ref{fig:dnds} we overlap, for the four energy bins, the flux distribution of the MC realisation and the best fit for the theoretical differential distribution.

\begin{figure*}
\begin{centering}
\includegraphics[width=0.55\textwidth]{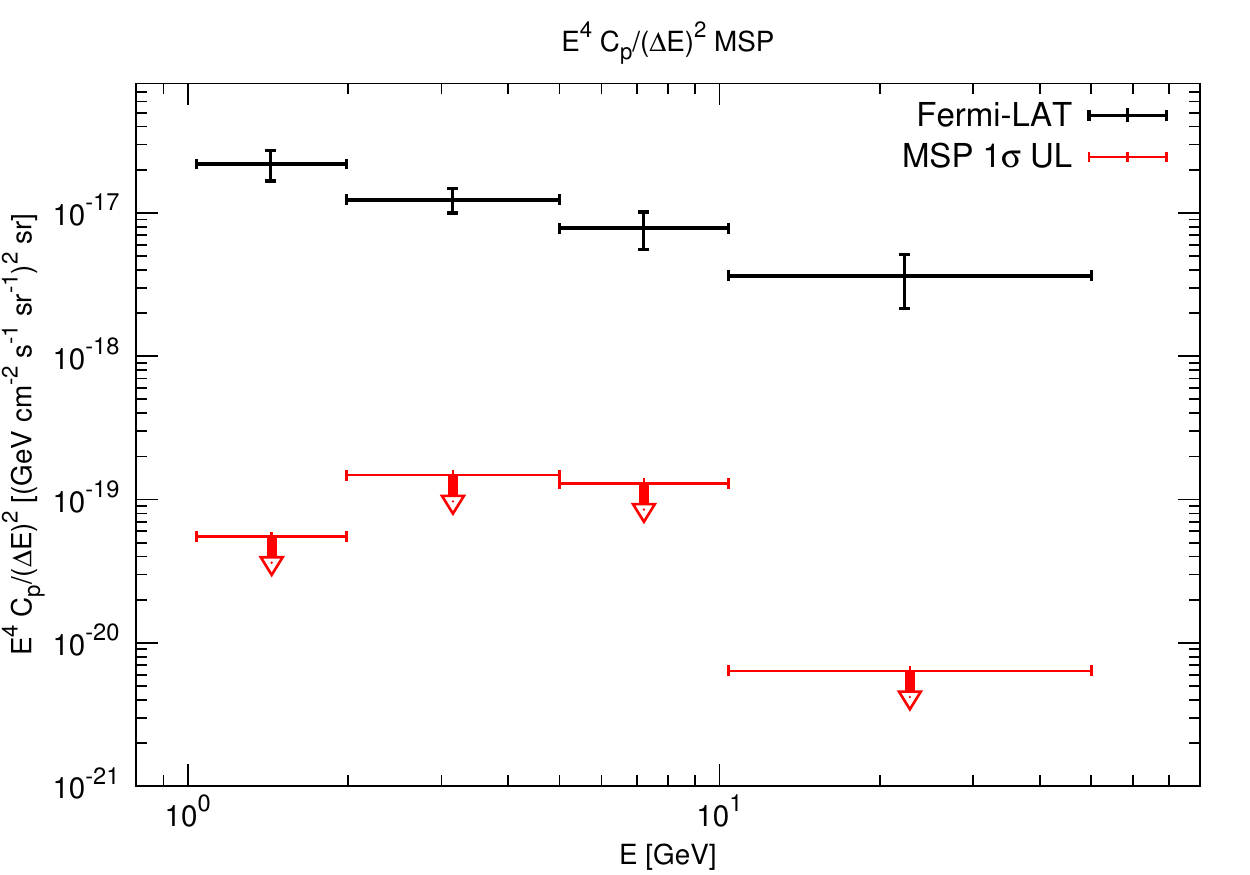}
\caption{The 1$\sigma$ upper limit angular power $C_P$ from the unresolved MSPs (red points) is shown together with {\it Fermi}-LAT data \citep{Ackermann:2012uf} (black points).}
\label{fig:cp}
\end{centering} 
\end{figure*}

Taking into account the best fit distributions $(dN/{dF'_{\gamma}})_{\rm{fit}}$ derived as above, we calculate the 1$\sigma$ upper limit of the angular power $C_P$ using Eq.~\ref{Cpdef}. 
We show the result in Fig.~\ref{fig:cp} together with {\it Fermi}-LAT results for the DATA:CLEANED event class \citep{Ackermann:2012uf}. For the sake of completeness, we quote the 1$\sigma$ upper limit of the angular power $C_P$ in Tab.~\ref{tab:Cp} for the four energy bins.
\begin{table*}
\begin{centering}
\begin{tabular}  {||c|c|c|c|c||}
\hline
\hline
 Energy range [GeV] &     1.04 -- 1.99  & 1.99 -- 5.0  & 5.0 -- 10.4 & 10.4 -- 50.0    \\
\hline
   $C_P$[(cm$^{-2}$ s$^{-1}$ sr$^{-1}$)$^{2}$ sr]    & $1.16 \cdot 10^{-20}$  & $1.35 \cdot 10^{-20}$  &   $1.40 \cdot 10^{-21}$ &    $3.40 \cdot 10^{-23}$        \\ 
\hline
\hline
\end{tabular}
\caption{ The 1$\sigma$ upper limit of the angular power $C_P$ in the four energy bins derived from the simulation of the MSP population.}
\label{tab:Cp}
\end{centering}
\end{table*} 

The anisotropy from unresolved MSPs is at least a factor of 60 smaller than {\it Fermi}-LAT data and therefore MSPs turn out to represent a population which gives a negligible contribution also to the \Fermi-LAT anisotropy.  This conclusion is compatible with the expectations of the $\gamma$-ray anisotropy produced by radio-loud AGN and blazars \citep{Cuoco:2012yf,dimauroale2104}, according to which blazars would dominate the angular power, so that any other population should give a negligible anisotropy contribution.

Moreover, we stress that our result represents the first and most up-to-date estimate of anisotropy from MSPs based on $\gamma$-ray data. \cite{2011MNRAS.415.1074S} has been indeed based on the first 9 months {\it Fermi}-LAT observations of eight MSPs. In particular, at latitudes above $30^{\circ}$, we find that $C_P/\langle I \rangle^2$ is about 0.2 sr in the first energy bin, while \cite{2011MNRAS.415.1074S} found $C_P/\langle I \rangle^2 \sim 0.03 - 0.04$ at $|b|\geq 40^{\circ}$.
The discrepancy of $C_P/\langle I \rangle^2$ is due to different  assumptions about the modeling of the $\gamma$-ray MSP population,  the $\gamma$-ray detection sensitivity (assumed there as a fixed threshold) and the region of the sky considered (above $b=40^{\circ}$). Such differences imply different values of both the dimensionful $C_P$ and of the mean average intensity $\langle I \rangle$.  
We finally notice that the energy dependence of the MSPs' $C_P$ in Fig.~\ref{fig:cp} is significantly different from the power-law-like behaviour of the {\it Fermi}-LAT data. This strengthen  the fact that MSPs can not be the dominant contributor to the IDGRB anisotropy.

\section{Conclusions}
\label{sec:conclusions}
Motivated by the possibility that MSPs might be an important contributor to the intensity and anisotropy of the IDGRB we have derived the properties of the MSP population in order to build a model consistent with radio and $\gamma$-ray observations.
Spatial and luminosity distributions have been calibrated on radio observations of 132 sources from the ATNF pulsars catalog. This represents, to our knowledge, the first systematic analysis of such a large MSP sample. The best fit distributions of the surface magnetic field $B$, the spin period $P$, the distance height from the galactic plane $z$ and the radial distance $r$ have been derived from radio observations.

Spectral and luminosity properties of the $\gamma$-ray MSP population have been derived from the most up-to-date $\gamma$-ray {\it Fermi}-LAT pulsars catalog, the 2FPC. In particular, we have used the information from the 95\% C.L. ULs on the $\gamma$-ray flux of about 20 MSPs in order to constrain the critical relation $L_{\gamma} - \dot{E}$ and derive an empirical uncertainty band around the scattered data points.

With the aid of the derived distributions, we have been able to simulate the $\gamma$-ray MSP population. To do so, we have relied on an MC simulation of the MSP distribution in the Galaxy. We have run $\sim 1000$ MC simulations to take into account the dispersion on the single MC realisation as well as other main theoretical uncertainties, such as the uncertainty on the coefficient of proportionality of the  $L_{\gamma} - \dot{E}$ relation, i.e. $\eta$.
By averaging over 1000 MC realisations, we have computed the contribution to the IDGRB at high latitudes due to the unresolved counterpart of the simulated MSP population, together with the 1$\sigma$ uncertainty band. We find that this source class might account for $0.13\%$-$0.02\%$ of the measured integrated flux of the IDGRB from 100 MeV. At the peak located at about 2 GeV the MSPs contribute at most for a 0.9$\%$ to the spectra of the IDGRB. We notice that we shrink the uncertainty on this contribution to be $\mathcal{O}(10)$.

The observed {\it Fermi}-LAT and ATNF MSP height distributions differ significantly in the inner part 
 of the Galaxy ($z\leq$ 0.3 kpc), being the sources detected in the radio band more numerous than the 
$\gamma$-ray counterparts.  It is demonstrated in this paper that the discrepancy is fully explained 
by the role of the LAT sensitivity at low latitudes. 

Given the distribution of MSPs, which predicts  that they are more abundant  in the innermost part of the Galaxy, we may expect a non-negligible contribution of the unresolved population to the $\gamma$-ray emission at low latitudes.
Moreover, close to the galactic disk the number of young pulsars overcomes the one of MSPs. We have
therefore  considered both young and millisecond pulsar populations in order to assess the total $\gamma$-ray emission at low latitudes.
The young pulsars population  has been modeled by following the method used for MSPs: we have derived the distributions for radial and vertical distances, spin period and surface magnetic field from the data of the ATNF catalog, while we have relied on the 2FPC for the computation of best fit $\gamma$-ray spectral parameters.
We  have then derived the $\gamma$-ray emission from the unresolved young pulsars and MSPs in the inner part of our Galaxy, by analysing the inner Galaxy and the galactic center regions, where recently an excess in the \Fermi-LAT data has been claimed by different groups \citep{2013PDU.....2..118H, Gordon:2013vta, Abazajian:2014fta}.
Going from the inner Galaxy to the galactic center, the contribution of young pulsar increases with respect
to the MSP one.  
We find that the peak of the $\gamma$-ray emission (at 2 GeV) from the unresolved MSP (young pulsar) 
is $4.1\cdot 10^{-9}$ ${\rm GeV}\, {\rm cm}^{-2} {\rm s}^{-1} {\rm sr}^{-1}$ ($9.1\cdot 10^{-11}$ ${\rm GeV}\, {\rm cm}^{-2} {\rm s}^{-1} {\rm sr}^{-1}$) for the inner Galaxy, and $1.9\cdot 10^{-7}$ ${\rm GeV}\, {\rm cm}^{-2} {\rm s}^{-1} {\rm sr}^{-1}$ ($4.5\cdot 10^{-7}$ ${\rm GeV}\, {\rm cm}^{-2} {\rm s}^{-1} {\rm sr}^{-1}$) for the galactic center.  

Moreover, we have calculated the $\gamma$-ray anisotropy arising from this source class, by presenting the 1$\sigma$ upper limit on the angular power, $C_P$. The result is that MSPs can not be a sizeable contributor to the $\gamma$-ray anisotropy measured by the LAT at $|b|>30^{\circ}$, which should be therefore dominated by other source classes, e.g. radio-loud AGN \cite{dimauroale2104}. 

We have finally demonstrated that although the population model is still affected by several uncertainties, those systematics have a little impact on the final prediction of the $\gamma$-ray flux from unresolved MSPs.

\appendix
\section{Discussion of systematic uncertainties}
\label{app:uncertainties}
As stressed in the text, the final results depend on how the MSP population has been built up, i.e. on the choices of the parameters characterizing the distributions assumed to model the source population.
We dedicate this Appendix to the discussion of the systematics uncertainties affecting the results, showing how the diffuse emission at high latitudes changes when other distributions are assumed.

\begin{figure*}
\centering
\includegraphics[width=0.45\textwidth]{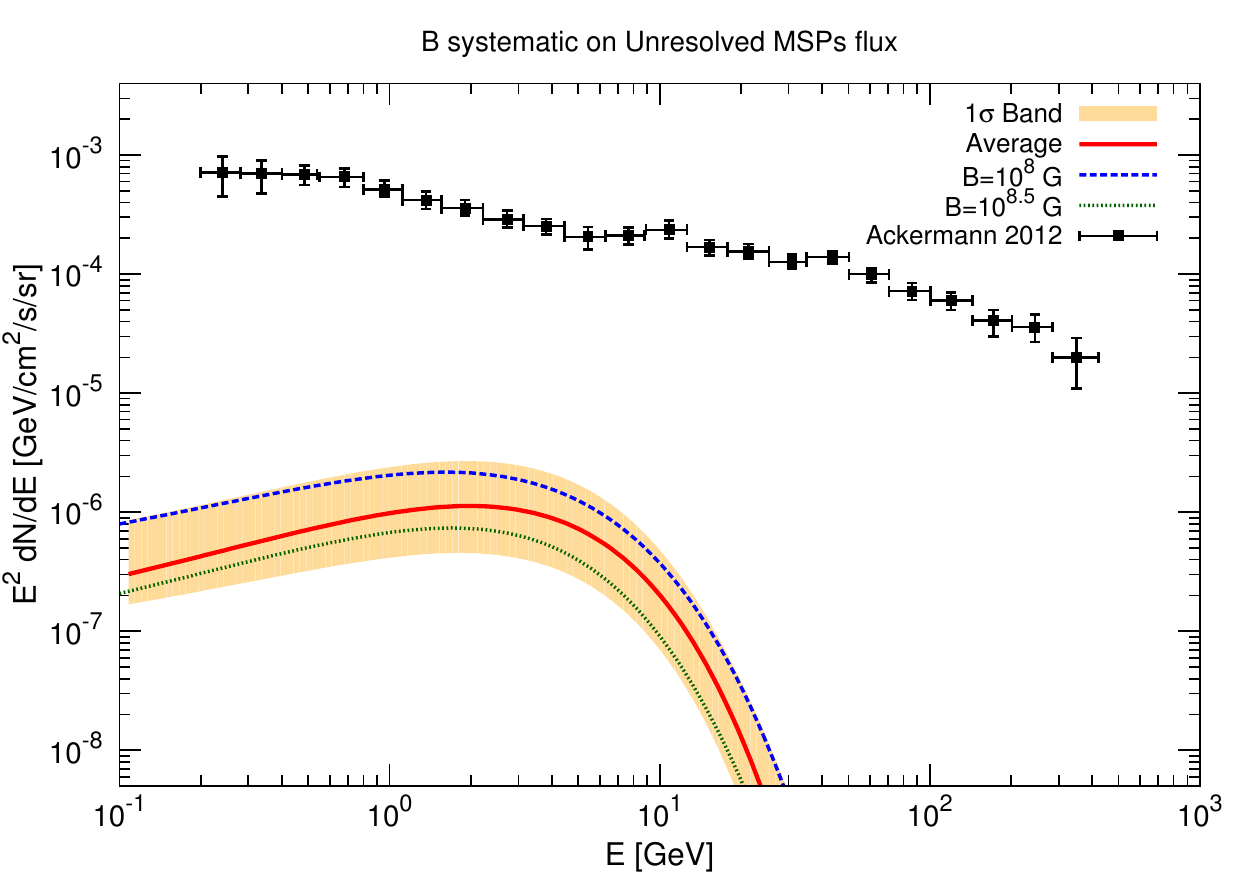}
\includegraphics[width=0.45\textwidth]{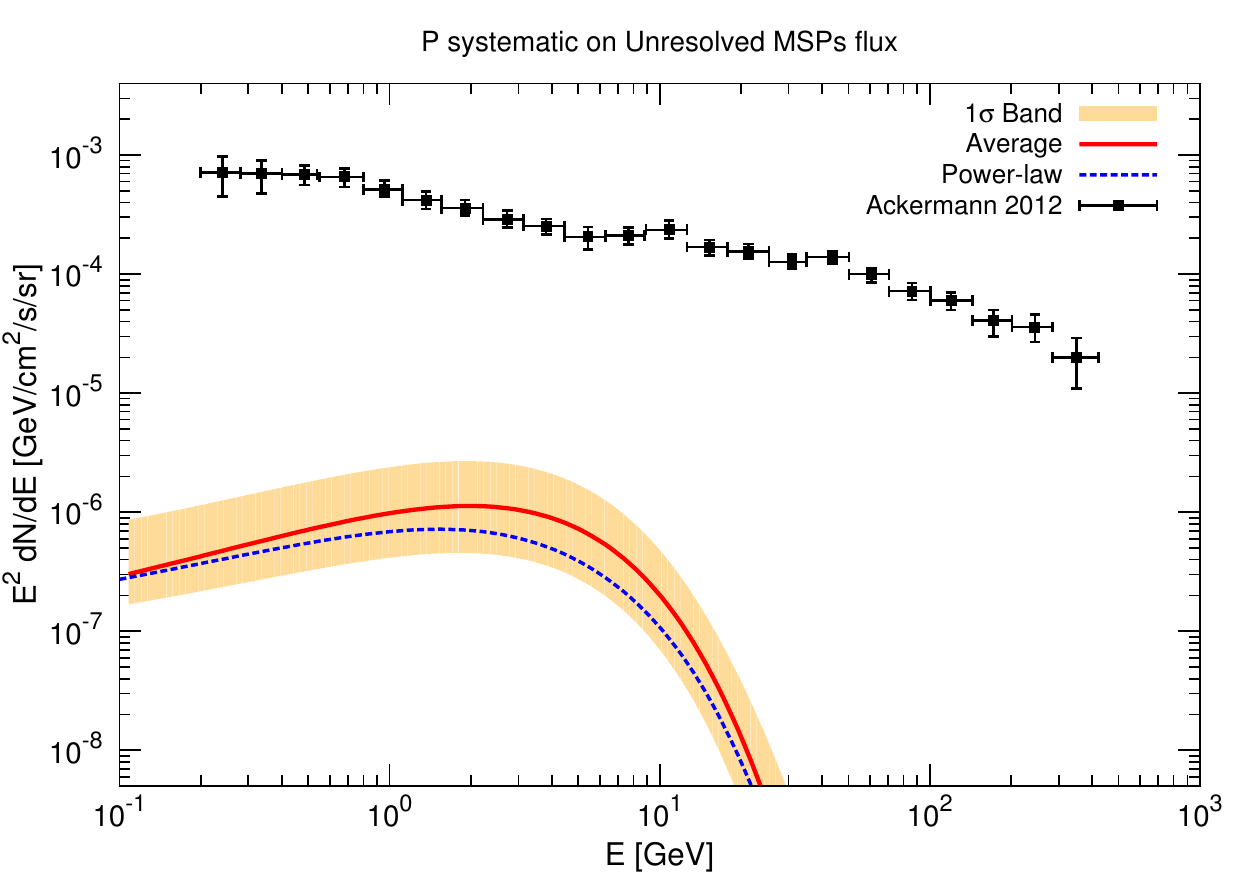}
\includegraphics[width=0.45\textwidth]{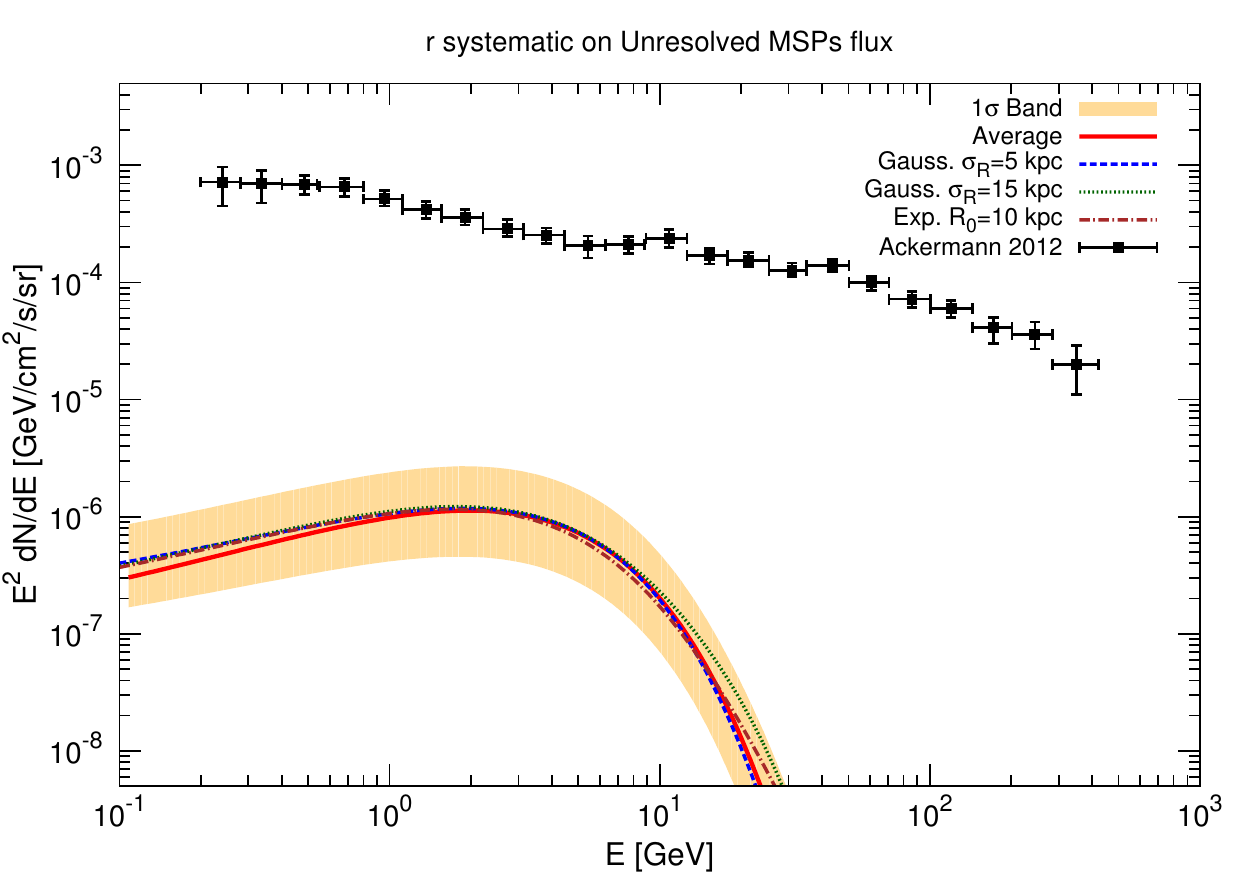}
\includegraphics[width=0.45\textwidth]{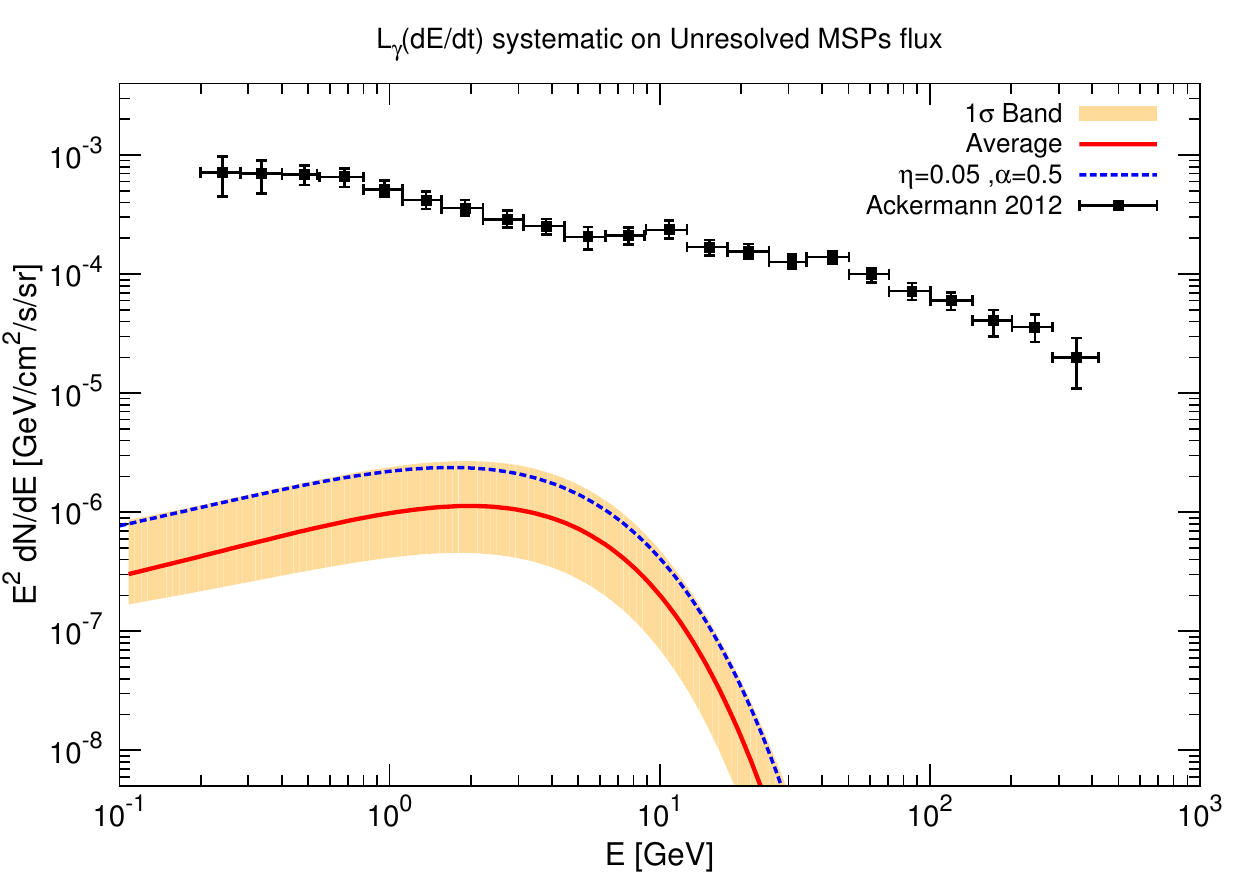}
\caption{The systematic uncertainty on the MSP IDGRB flux  is quantified,  from the top-left to the bottom-right panels, by varying the assumption on the distributions for the magnetic field value, the period $P$ distribution, the galactic distance of the galactic plane $r$ distributions and on the correlation between $L_{\gamma}$ and $\dot{E}$. In each figure the best fit curve and the 1$\sigma$ band of Fig.~\ref{fig:BMflux} are shown together with the cases explained in the text.}
\label{fig:systematics} 
\end{figure*}

Firstly, we discuss how the final spectrum is altered when assuming different choices for magnetic field $B$, period $P$ and spatial radial $r$ distributions. 
The results are displayed by Fig.~\ref{fig:systematics}. For each varied parameter, we derive the corresponding contribution to the diffuse emission as the average over 1000 MC realisations.
The top-left panel shows the impact of assuming different values of $\langle \log_{10}( B/G) \rangle$ for the magnetic field distribution of Eq.~\ref{eq:bdistr}. The cases with $\langle \log_{10}( B/G) \rangle = 8$ and $\langle \log_{10}( B/G) \rangle = 8.5$ are displayed. The integrated flux in the 0.1 - 100 GeV band is respectively about twice and a factor of 0.7 of the best fit realisation. The corresponding spectra are displayed in the panel together with the 1$\sigma$ uncertainty band.
In the top-right panel, the reader can see how the choice of a power-law distribution for the rotation period $P$ in the form  $N(P)\propto P^{-\alpha}$ (with $\alpha = 2$), as assumed by \cite{2011MNRAS.415.1074S,2013AA...554A..62G,2010JCAP...01..005F}, only marginally affects the results.
We also notice that the final results are independent from the spatial (radial) distribution adopted; the bottom-left panel shows that using a Gaussian distribution, Eq.~\ref{eq:rdistrtwo}, with $\sigma_r = 5, 10, 15$ kpc or an exponential distribution, Eq.~\ref{eq:rdistrone}, with $r_0 = 10$ kpc (from \cite{2013AA...554A..62G}), does not change the result and we are still close to the best fit spectrum.
Therefore, varying the surface magnetic field, rotation period and space distribution does not alter our conclusions and in all cases the ensuing systematic uncertainty is inside the 1$\sigma$ uncertainty band.

In Sec.~\ref{sec:LgammaEdot}, we have discussed the criticality and uncertainty on the efficiency conversion of spin-down luminosity $\dot{E}$ into $\gamma$-ray luminosity.
Several papers \citep{2011MNRAS.415.1074S,2013AA...554A..62G,2010JCAP...01..005F,2013ApJS..208...17A} quote as possible $L_{\gamma} - \dot{E}$ relation:
\begin{equation}
L_{\gamma} \propto \sqrt{\dot{E}}.
\end{equation}
We found that assuming this relation, does not impact on our final result as it is evident from the bottom-right panel of Fig.~\ref{fig:systematics}.

\begin{acknowledgements}
We acknowledge S.~Ando, G.~Bertone, and  P.~D.~Serpico for a careful reading of the manuscript and useful comments. 
F.C. acknowledges support from the European Research Council through
the ERC starting grant WIMPs Kairos, P.I. G. Bertone.
This work is supported by the research grant {\sl TAsP (Theoretical Astroparticle Physics)}
funded by the Istituto Nazionale di Fisica Nucleare (INFN), by the  {\sl Strategic Research Grant: Origin and Detection of Galactic and Extragalactic Cosmic Rays} funded by Torino University and Compagnia di San Paolo. At LAPTh this activity was supported by the Labex grant ENIGMASS.
\end{acknowledgements}

\vspace{2cm}

\bibliography{paper_tmp}

\begin{thebibliography}{67}
\expandafter\ifx\csname natexlab\endcsname\relax\def\natexlab#1{#1}\fi

\bibitem[{{Abazajian} {et~al.}(2010){Abazajian}, {Agrawal}, {Chacko}, \&
  {Kilic}}]{2010JCAP...11..041A}
{Abazajian}, K.~N., {Agrawal}, P., {Chacko}, Z., \& {Kilic}, C. 2010, \jcap,
  11, 41

\bibitem[{{Abazajian} {et~al.}(2011){Abazajian}, {Blanchet}, \&
  {Harding}}]{2011PhRvD..84j3007A}
{Abazajian}, K.~N., {Blanchet}, S., \& {Harding}, J.~P. 2011, \prd, 84, 103007

\bibitem[{Abazajian {et~al.}(2014)Abazajian, Canac, Horiuchi, \&
  Kaplinghat}]{Abazajian:2014fta}
Abazajian, K.~N., Canac, N., Horiuchi, S., \& Kaplinghat, M. 2014

\bibitem[{{Abdo} {et~al.}(2010{\natexlab{a}}){Abdo}, {Ackermann}, {Ajello},
  {Antolini}, {Baldini}, {Ballet}, {Barbiellini}, {Bastieri}, {Baughman},
  {Bechtol}, {Bellazzini}, {Berenji}, {Blandford}, {Bloom}, {Bonamente},
  {Borgland}, {Bouvier}, {Bregeon}, {Brez}, {Brigida}, {Bruel}, {Burnett},
  {Buson}, {Caliandro}, {Cameron}, {Caraveo}, {Carrigan}, {Casandjian},
  {Cavazzuti}, {Cecchi}, {{\c C}elik}, {Charles}, {Chekhtman}, {Cheung},
  {Chiang}, {Ciprini}, {Claus}, {Cohen-Tanugi}, {Conrad}, {Costamante},
  {Cutini}, {Dermer}, {de Angelis}, {de Palma}, {Silva}, {Drell}, {Dubois},
  {Dumora}, {Farnier}, {Favuzzi}, {Fegan}, {Focke}, {Fukazawa}, {Funk},
  {Fusco}, {Gargano}, {Gasparrini}, {Gehrels}, {Germani}, {Giglietto},
  {Giommi}, {Giordano}, {Glanzman}, {Godfrey}, {Grenier}, {Grove}, {Guiriec},
  {Hadasch}, {Hayashida}, {Hays}, {Healey}, {Horan}, {Hughes}, {Itoh},
  {J{\'o}hannesson}, {Johnson}, {Johnson}, {Johnson}, {Kamae}, {Katagiri},
  {Kataoka}, {Kawai}, {Kn{\"o}dlseder}, {Kuss}, {Lande}, {Latronico}, {Lee},
  {Lemoine-Goumard}, {Llena Garde}, {Longo}, {Loparco}, {Lott}, {Lovellette},
  {Lubrano}, {Madejski}, {Makeev}, {Mazziotta}, {McConville}, {McEnery},
  {Meurer}, {Michelson}, {Mitthumsiri}, {Mizuno}, {Monte}, {Monzani},
  {Morselli}, {Moskalenko}, {Murgia}, {Nolan}, {Norris}, {Nuss}, {Ohsugi},
  {Omodei}, {Orlando}, {Ormes}, {Ozaki}, {Paneque}, {Panetta}, {Parent},
  {Pelassa}, {Pepe}, {Pesce-Rollins}, {Piron}, {Porter}, {Rain{\`o}}, {Rando},
  {Razzano}, {Reimer}, {Reimer}, {Ritz}, {Rochester}, {Rodriguez}, {Romani},
  {Roth}, {Sadrozinski}, {Sander}, {Saz Parkinson}, {Scargle}, {Sgr{\`o}},
  {Shaw}, {Smith}, {Spandre}, {Spinelli}, {Starck}, {Strickman}, {Strong},
  {Suson}, {Tajima}, {Takahashi}, {Takahashi}, {Tanaka}, {Thayer}, {Thayer},
  {Thompson}, {Tibaldo}, {Torres}, {Tosti}, {Tramacere}, {Uchiyama}, {Usher},
  {Vasileiou}, {Vilchez}, {Vitale}, {Waite}, {Wang}, {Winer}, {Wood}, {Yang},
  {Ylinen}, {Ziegler}, \& {Fermi LAT Collaboration}}]{Collaboration:2010gqa}
{Abdo}, A.~A., {Ackermann}, M., {Ajello}, M., {et~al.} 2010{\natexlab{a}},
  \apj, 720, 435

\bibitem[{{Abdo} {et~al.}(2010{\natexlab{b}}){Abdo}, {Ackermann}, {Ajello},
  {Atwood}, {Baldini}, {Ballet}, {Barbiellini}, {Bastieri}, {Baughman},
  {Bechtol}, {Bellazzini}, {Berenji}, {Blandford}, {Bloom}, {Bonamente},
  {Borgland}, {Bregeon}, {Brez}, {Brigida}, {Bruel}, {Burnett}, {Buson},
  {Caliandro}, {Cameron}, {Caraveo}, {Casandjian}, {Cavazzuti}, {Cecchi}, {{\c
  C}elik}, {Charles}, {Chekhtman}, {Cheung}, {Chiang}, {Ciprini}, {Claus},
  {Cohen-Tanugi}, {Cominsky}, {Conrad}, {Cutini}, {Dermer}, {de Angelis}, {de
  Palma}, {Digel}, {di Bernardo}, {do Couto e Silva}, {Drell}, {Drlica-Wagner},
  {Dubois}, {Dumora}, {Farnier}, {Favuzzi}, {Fegan}, {Focke}, {Fortin},
  {Frailis}, {Fukazawa}, {Funk}, {Fusco}, {Gaggero}, {Gargano}, {Gasparrini},
  {Gehrels}, {Germani}, {Giebels}, {Giglietto}, {Giommi}, {Giordano},
  {Glanzman}, {Godfrey}, {Grenier}, {Grondin}, {Grove}, {Guillemot}, {Guiriec},
  {Gustafsson}, {Hanabata}, {Harding}, {Hayashida}, {Hughes}, {Itoh},
  {Jackson}, {J{\'o}hannesson}, {Johnson}, {Johnson}, {Johnson}, {Johnson},
  {Kamae}, {Katagiri}, {Kataoka}, {Kawai}, {Kerr}, {Kn{\"o}dlseder}, {Kocian},
  {Kuehn}, {Kuss}, {Lande}, {Latronico}, {Lemoine-Goumard}, {Longo}, {Loparco},
  {Lott}, {Lovellette}, {Lubrano}, {Madejski}, {Makeev}, {Mazziotta},
  {McConville}, {McEnery}, {Meurer}, {Michelson}, {Mitthumsiri}, {Mizuno},
  {Moiseev}, {Monte}, {Monzani}, {Morselli}, {Moskalenko}, {Murgia}, {Nolan},
  {Norris}, {Nuss}, {Ohsugi}, {Omodei}, {Orlando}, {Ormes}, {Paneque},
  {Panetta}, {Parent}, {Pelassa}, {Pepe}, {Pesce-Rollins}, {Piron}, {Porter},
  {Rain{\`o}}, {Rando}, {Razzano}, {Reimer}, {Reimer}, {Reposeur}, {Ritz},
  {Rochester}, {Rodriguez}, {Roth}, {Ryde}, {Sadrozinski}, {Sanchez}, {Sander},
  {Parkinson}, {Scargle}, {Sellerholm}, {Sgr{\`o}}, {Shaw}, {Siskind}, {Smith},
  {Smith}, {Spandre}, {Spinelli}, {Starck}, {Strickman}, {Strong}, {Suson},
  {Tajima}, {Takahashi}, {Takahashi}, {Tanaka}, {Thayer}, {Thayer}, {Thompson},
  {Tibaldo}, {Torres}, {Tosti}, {Tramacere}, {Uchiyama}, {Usher}, {Vasileiou},
  {Vilchez}, {Vitale}, {Waite}, {Wang}, {Winer}, {Wood}, {Ylinen}, {Ziegler},
  \& {Fermi LAT Collaboration}}]{IDGRB}
{Abdo}, A.~A., {Ackermann}, M., {Ajello}, M., {et~al.} 2010{\natexlab{b}},
  Physical Review Letters, 104, 101101

\bibitem[{{Abdo} {et~al.}(2009){Abdo}, {Ackermann}, {Atwood}, {Axelsson},
  {Baldini}, {Ballet}, {Barbiellini}, {Bastieri}, {Battelino}, {Baughman},
  {Bechtol}, {Bellazzini}, {Berenji}, {Bloom}, {Bonamente}, {Borgland},
  {Bregeon}, {Brez}, {Brigida}, {Bruel}, {Burnett}, {Caliandro}, {Cameron},
  {Caraveo}, {Casandjian}, {Cecchi}, {Charles}, {Chekhtman}, {Cheung},
  {Chiang}, {Ciprini}, {Claus}, {Cognard}, {Cohen-Tanugi}, {Cominsky},
  {Conrad}, {Cutini}, {Dermer}, {de Angelis}, {de Palma}, {Digel}, {Dormody},
  {Silva}, {Drell}, {Dubois}, {Dumora}, {Farnier}, {Favuzzi}, {Focke},
  {Frailis}, {Fukazawa}, {Funk}, {Fusco}, {Gargano}, {Gasparrini}, {Gehrels},
  {Germani}, {Giebels}, {Giglietto}, {Giordano}, {Glanzman}, {Godfrey},
  {Grenier}, {Grondin}, {Grove}, {Guillemot}, {Guiriec}, {Hanabata}, {Harding},
  {Hayashida}, {Hays}, {Hughes}, {J{\'o}hannesson}, {Johnson}, {Johnson},
  {Johnson}, {Johnson}, {Kamae}, {Katagiri}, {Kataoka}, {Kawai}, {Kerr},
  {Kn{\"o}dlseder}, {Kocian}, {Komin}, {Kuehn}, {Kuss}, {Lande}, {Latronico},
  {Lee}, {Lemoine-Goumard}, {Longo}, {Loparco}, {Lott}, {Lovellette},
  {Lubrano}, {Madejski}, {Makeev}, {Marelli}, {Mazziotta}, {McConville},
  {McEnery}, {Meurer}, {Michelson}, {Mitthumsiri}, {Mizuno}, {Moiseev},
  {Monte}, {Monzani}, {Morselli}, {Moskalenko}, {Murgia}, {Nolan}, {Nuss},
  {Ohsugi}, {Omodei}, {Orlando}, {Ormes}, {Pancrazi}, {Paneque}, {Panetta},
  {Parent}, {Pepe}, {Pesce-Rollins}, {Piron}, {Porter}, {Rain{\`o}}, {Rando},
  {Razzano}, {Reimer}, {Reimer}, {Reposeur}, {Ritz}, {Rochester}, {Rodriguez},
  {Romani}, {Ryde}, {Sadrozinski}, {Sanchez}, {Sander}, {Parkinson},
  {Sgr{\`o}}, {Siskind}, {Smith}, {Smith}, {Spandre}, {Spinelli}, {Starck},
  {Strickman}, {Suson}, {Tajima}, {Takahashi}, {Tanaka}, {Thayer}, {Thayer},
  {Theureau}, {Thompson}, {Tibaldo}, {Torres}, {Tosti}, {Tramacere},
  {Uchiyama}, {Usher}, {Van Etten}, {Vilchez}, {Vitale}, {Waite}, {Watters},
  {Webb}, {Wood}, {Ylinen}, \& {Ziegler}}]{2009ApJ...699.1171A}
{Abdo}, A.~A., {Ackermann}, M., {Atwood}, W.~B., {et~al.} 2009, \apj, 699, 1171

\bibitem[{{Abdo} {et~al.}(2013){Abdo}, {Ajello}, {Allafort}, {Baldini},
  {Ballet}, {Barbiellini}, {Baring}, {Bastieri}, {Belfiore}, {Bellazzini}, \&
  et~al.}]{2013ApJS..208...17A}
{Abdo}, A.~A., {Ajello}, M., {Allafort}, A., {et~al.} 2013, apjs, 208, 17

\bibitem[{{Ackermann}(2012)}]{ackermann2012}
{Ackermann}, M. 2012, {4th Fermi Symposium,
  http://galprop.stanford.edu/resources.php}

\bibitem[{{Ackermann} {et~al.}(2012){Ackermann}, {Ajello}, {Allafort},
  {Baldini}, {Ballet}, {Bastieri}, {Bechtol}, {Bellazzini}, {Berenji}, {Bloom},
  {Bonamente}, {Borgland}, {Bouvier}, {Bregeon}, {Brigida}, {Bruel}, {Buehler},
  {Buson}, {Caliandro}, {Cameron}, {Caraveo}, {Casandjian}, {Cecchi},
  {Charles}, {Chekhtman}, {Cheung}, {Chiang}, {Cillis}, {Ciprini}, {Claus},
  {Cohen-Tanugi}, {Conrad}, {Cutini}, {de Palma}, {Dermer}, {Digel}, {Silva},
  {Drell}, {Drlica-Wagner}, {Favuzzi}, {Fegan}, {Fortin}, {Fukazawa}, {Funk},
  {Fusco}, {Gargano}, {Gasparrini}, {Germani}, {Giglietto}, {Giordano},
  {Glanzman}, {Godfrey}, {Grenier}, {Guiriec}, {Gustafsson}, {Hadasch},
  {Hayashida}, {Hays}, {Hughes}, {J{\'o}hannesson}, {Johnson}, {Kamae},
  {Katagiri}, {Kataoka}, {Kn{\"o}dlseder}, {Kuss}, {Lande}, {Longo}, {Loparco},
  {Lott}, {Lovellette}, {Lubrano}, {Madejski}, {Martin}, {Mazziotta},
  {McEnery}, {Michelson}, {Mizuno}, {Monte}, {Monzani}, {Morselli},
  {Moskalenko}, {Murgia}, {Nishino}, {Norris}, {Nuss}, {Ohno}, {Ohsugi},
  {Okumura}, {Omodei}, {Orlando}, {Ozaki}, {Parent}, {Persic}, {Pesce-Rollins},
  {Petrosian}, {Pierbattista}, {Piron}, {Pivato}, {Porter}, {Rain{\`o}},
  {Rando}, {Razzano}, {Reimer}, {Reimer}, {Ritz}, {Roth}, {Sbarra}, {Sgr{\`o}},
  {Siskind}, {Spandre}, {Spinelli}, {Stawarz}, {Strong}, {Takahashi}, {Tanaka},
  {Thayer}, {Tibaldo}, {Tinivella}, {Torres}, {Tosti}, {Troja}, {Uchiyama},
  {Vandenbroucke}, {Vianello}, {Vitale}, {Waite}, {Wood}, \&
  {Yang}}]{2012ApJ...755..164A}
{Ackermann}, M., {Ajello}, M., {Allafort}, A., {et~al.} 2012, \apj, 755, 164

\bibitem[{Ackermann {et~al.}(2012)}]{Ackermann:2012uf}
Ackermann, M. {et~al.} 2012, Phys.Rev., D85, 083007

\bibitem[{Aharonian {et~al.}(2006)}]{Aharonian:2006pe}
Aharonian, F. {et~al.} 2006, Astron.Astrophys., 457, 899

\bibitem[{{Ajello} {et~al.}(2014){Ajello}, {Romani}, {Gasparrini}, {Shaw},
  {Bolmer}, {Cotter}, {Finke}, {Greiner}, {Healey}, {King}, {Max-Moerbeck},
  {Michelson}, {Potter}, {Rau}, {Readhead}, {Richards}, \&
  {Schady}}]{2014ApJ...780...73A}
{Ajello}, M., {Romani}, R.~W., {Gasparrini}, D., {et~al.} 2014, \apj, 780, 73

\bibitem[{{Ajello} {et~al.}(2012){Ajello}, {Shaw}, {Romani}, {Dermer},
  {Costamante}, {King}, {Max-Moerbeck}, {Readhead}, {Reimer}, {Richards}, \&
  {Stevenson}}]{2012ApJ...751..108A}
{Ajello}, M., {Shaw}, M.~S., {Romani}, R.~W., {et~al.} 2012, \apj, 751, 108

\bibitem[{{Alpar} {et~al.}(1982){Alpar}, {Cheng}, {Ruderman}, \&
  {Shaham}}]{1982Natur.300..728A}
{Alpar}, M.~A., {Cheng}, A.~F., {Ruderman}, M.~A., \& {Shaham}, J. 1982, \nat,
  300, 728

\bibitem[{{Backer} {et~al.}(1982){Backer}, {Kulkarni}, {Heiles}, {Davis}, \&
  {Goss}}]{Backer82}
{Backer}, D.~C., {Kulkarni}, S.~R., {Heiles}, C., {Davis}, M.~M., \& {Goss},
  W.~M. 1982, \nat, 300, 615

\bibitem[{{Berezinsky} {et~al.}(2011){Berezinsky}, {Gazizov}, {Kachelrie{\ss}},
  \& {Ostapchenko}}]{2011PhLB..695...13B}
{Berezinsky}, V., {Gazizov}, A., {Kachelrie{\ss}}, M., \& {Ostapchenko}, S.
  2011, Physics Letters B, 695, 13

\bibitem[{{Bergstr{\"o}m} {et~al.}(1998){Bergstr{\"o}m}, {Ullio}, \&
  {Buckley}}]{1998APh.....9..137B}
{Bergstr{\"o}m}, L., {Ullio}, P., \& {Buckley}, J.~H. 1998, Astropart. Phys.,
  9, 137

\bibitem[{{Blasi} {et~al.}(2007){Blasi}, {Gabici}, \&
  {Brunetti}}]{2007IJMPA..22..681B}
{Blasi}, P., {Gabici}, S., \& {Brunetti}, G. 2007, International Journal of
  Modern Physics A, 22, 681

\bibitem[{Bringmann {et~al.}(2014)Bringmann, Calore, Di~Mauro, \&
  Donato}]{Calore:2013yia}
Bringmann, T., Calore, F., Di~Mauro, M., \& Donato, F. 2014, Phys.Rev., D89,
  023012

\bibitem[{{Brunthaler} {et~al.}(2011){Brunthaler}, {Reid}, {Menten}, {Zheng},
  {Bartkiewicz}, {Choi}, {Dame}, {Hachisuka}, {Immer}, {Moellenbrock},
  {Moscadelli}, {Rygl}, {Sanna}, {Sato}, {Wu}, {Xu}, \&
  {Zhang}}]{2011AN....332..461B}
{Brunthaler}, A., {Reid}, M.~J., {Menten}, K.~M., {et~al.} 2011, Astronomische
  Nachrichten, 332, 461

\bibitem[{{Burgay} {et~al.}(2013){Burgay}, {Keith}, {Lorimer}, {Hassall},
  {Lyne}, {Camilo}, {D'Amico}, {Hobbs}, {Kramer}, {Manchester}, {McLaughlin},
  {Possenti}, {Stairs}, \& {Stappers}}]{2013MNRAS.429..579B}
{Burgay}, M., {Keith}, M.~J., {Lorimer}, D.~R., {et~al.} 2013, \mnras, 429, 579

\bibitem[{Calore {et~al.}(2014)Calore, De~Romeri, Di~Mauro, Donato, Herpich,
  {et~al.}}]{Calore:2014hna}
Calore, F., De~Romeri, V., Di~Mauro, M., {et~al.} 2014,
  Mon.Not.Roy.Astron.Soc., 442, 1151

\bibitem[{{Calore} {et~al.}(2012){Calore}, {de Romeri}, \&
  {Donato}}]{2012PhRvD..85b3004C}
{Calore}, F., {de Romeri}, V., \& {Donato}, F. 2012, \prd, 85, 023004

\bibitem[{Cholis {et~al.}(2014)Cholis, Hooper, \& McDermott}]{Cholis:2013ena}
Cholis, I., Hooper, D., \& McDermott, S.~D. 2014, JCAP, 1402, 014

\bibitem[{{Cordes} \& {Chernoff}(1997)}]{1997ApJ...482..971C}
{Cordes}, J.~M. \& {Chernoff}, D.~F. 1997, apj, 482, 971

\bibitem[{Cuoco {et~al.}(2012)Cuoco, Komatsu, \& Siegal-Gaskins}]{Cuoco:2012yf}
Cuoco, A., Komatsu, E., \& Siegal-Gaskins, J. 2012, Phys.Rev., D86, 063004

\bibitem[{Di~Mauro {et~al.}(2014{\natexlab{a}})Di~Mauro, Calore, Donato,
  Ajello, \& Latronico}]{RG}
Di~Mauro, M., Calore, F., Donato, F., Ajello, M., \& Latronico, L.
  2014{\natexlab{a}}, Astrophys.J., 780, 161

\bibitem[{Di~Mauro {et~al.}(2014{\natexlab{b}})Di~Mauro, {Donato}, {Cuoco}, \&
  {Siegal-Gaskins}}]{dimauroale2104}
Di~Mauro, M., {Donato}, F., {Cuoco}, A., \& {Siegal-Gaskins}, J.
  2014{\natexlab{b}}, In praparation

\bibitem[{Di~Mauro {et~al.}(2014{\natexlab{c}})Di~Mauro, Donato, Lamanna,
  Sanchez, \& Serpico}]{DiMauro:2013zfa}
Di~Mauro, M., Donato, F., Lamanna, G., Sanchez, D., \& Serpico, P.
  2014{\natexlab{c}}, Astrophys.J., 786, 129

\bibitem[{Faucher-Giguere \& Kaspi(2006)}]{FaucherGiguere:2005ny}
Faucher-Giguere, C.-A. \& Kaspi, V.~M. 2006, Astrophys.J., 643, 332

\bibitem[{{Faucher-Gigu{\`e}re} \& {Loeb}(2010)}]{2010JCAP...01..005F}
{Faucher-Gigu{\`e}re}, C.-A. \& {Loeb}, A. 2010, jcap, 1, 5

\bibitem[{{Fichtel} {et~al.}(1975){Fichtel}, {Hartman}, {Kniffen}, {Thompson},
  {Ogelman}, {Ozel}, {Tumer}, \& {Bignami}}]{1975ApJ...198..163F}
{Fichtel}, C.~E., {Hartman}, R.~C., {Kniffen}, D.~A., {et~al.} 1975, \apj, 198,
  163

\bibitem[{{Fornasa} {et~al.}(2013){Fornasa}, {Zavala}, {S{\'a}nchez-Conde},
  {Siegal-Gaskins}, {Delahaye}, {Prada}, {Vogelsberger}, {Zandanel}, \&
  {Frenk}}]{2013MNRAS.429.1529F}
{Fornasa}, M., {Zavala}, J., {S{\'a}nchez-Conde}, M.~A., {et~al.} 2013, \mnras,
  429, 1529

\bibitem[{{Gillessen} {et~al.}(2009){Gillessen}, {Eisenhauer}, {Trippe},
  {Alexander}, {Genzel}, {Martins}, \& {Ott}}]{2009ApJ...692.1075G}
{Gillessen}, S., {Eisenhauer}, F., {Trippe}, S., {et~al.} 2009, \apj, 692, 1075

\bibitem[{Gordon \& Macias(2013)}]{Gordon:2013vta}
Gordon, C. \& Macias, O. 2013, Phys.Rev., D88, 083521

\bibitem[{{G{\'o}rski} {et~al.}(2005){G{\'o}rski}, {Hivon}, {Banday},
  {Wandelt}, {Hansen}, {Reinecke}, \& {Bartelmann}}]{2005ApJ...622..759G}
{G{\'o}rski}, K.~M., {Hivon}, E., {Banday}, A.~J., {et~al.} 2005, \apj, 622,
  759

\bibitem[{{Gr{\'e}goire} \& {Kn{\"o}dlseder}(2013)}]{2013AA...554A..62G}
{Gr{\'e}goire}, T. \& {Kn{\"o}dlseder}, J. 2013, aap, 554, A62

\bibitem[{{Hooper} {et~al.}(2013){Hooper}, {Cholis}, {Linden},
  {Siegal-Gaskins}, \& {Slatyer}}]{2013PhRvD..88h3009H}
{Hooper}, D., {Cholis}, I., {Linden}, T., {Siegal-Gaskins}, J.~M., \&
  {Slatyer}, T.~R. 2013, \prd, 88, 083009

\bibitem[{{Hooper} \& {Slatyer}(2013)}]{2013PDU.....2..118H}
{Hooper}, D. \& {Slatyer}, T.~R. 2013, Physics of the Dark Universe, 2, 118

\bibitem[{{Inoue}(2011)}]{2011ApJ...733...66I}
{Inoue}, Y. 2011, \apj, 733, 66

\bibitem[{{Johnson} {et~al.}(2014){Johnson}, {Venter}, {Harding}, {Guillemot},
  {Smith}, {Kramer}, {Celik}, {den Hartog}, {Ferrara}, {Hou}, {Lande}, \&
  {Ray}}]{2014arXiv1404.2264J}
{Johnson}, T.~J., {Venter}, C., {Harding}, A.~K., {et~al.} 2014, ArXiv e-prints

\bibitem[{Kalashev {et~al.}(2009)Kalashev, Semikoz, \& Sigl}]{Kalashev:2007sn}
Kalashev, O.~E., Semikoz, D.~V., \& Sigl, G. 2009, Phys.Rev., D79, 063005

\bibitem[{Keith {et~al.}(2011)Keith, Johnston, Bailes, Bates, Bhat,
  {et~al.}}]{Keith:2011km}
Keith, M., Johnston, S., Bailes, M., {et~al.} 2011

\bibitem[{{Kerr} \& {Fermi-LAT Collaboration}(2013)}]{2013IAUS..291..307K}
{Kerr}, M. \& {Fermi-LAT Collaboration}. 2013, in IAU Symposium, Vol. 291, IAU
  Symposium, ed. J.~{van Leeuwen}, 307--312

\bibitem[{{Kraushaar} {et~al.}(1972){Kraushaar}, {Clark}, {Garmire}, {Borken},
  {Higbie}, {Leong}, \& {Thorsos}}]{1972ApJ...177..341K}
{Kraushaar}, W.~L., {Clark}, G.~W., {Garmire}, G.~P., {et~al.} 1972, \apj, 177,
  341

\bibitem[{{Lange} {et~al.}(2001){Lange}, {Camilo}, {Wex}, {Kramer}, {Backer},
  {Lyne}, \& {Doroshenko}}]{2001MNRAS.326..274L}
{Lange}, C., {Camilo}, F., {Wex}, N., {et~al.} 2001, \mnras, 326, 274

\bibitem[{Levin {et~al.}(2013)Levin, Bailes, Barsdell, Bates, Bhat,
  {et~al.}}]{Levin:2013usa}
Levin, L., Bailes, M., Barsdell, B., {et~al.} 2013

\bibitem[{Lorimer {et~al.}(2006)Lorimer, Faulkner, Lyne, Manchester, Kramer,
  {et~al.}}]{Lorimer:2006qs}
Lorimer, D., Faulkner, A., Lyne, A., {et~al.} 2006, Mon.Not.Roy.Astron.Soc.,
  372, 777

\bibitem[{Lorimer(2012)}]{Lorimer:2012hy}
Lorimer, D.~R. 2012

\bibitem[{{Lorimer} \& {Kramer}(2004)}]{2004hpa..book.....L}
{Lorimer}, D.~R. \& {Kramer}, M. 2004, {Handbook of Pulsar Astronomy}, ed.
  R.~{Ellis}, J.~{Huchra}, S.~{Kahn}, G.~{Rieke}, \& P.~B. {Stetson}

\bibitem[{{Lyne}(2000)}]{2000RSPTA.358..831L}
{Lyne}, A.~G. 2000, in Royal Society of London Philosophical Transactions
  Series A, Vol. 358, Astronomy, physics and chemistry of H$^{+}$$_{3}$,
  831--840

\bibitem[{{Lyutikov} {et~al.}(2012){Lyutikov}, {Otte}, \&
  {McCann}}]{Lyutikov12}
{Lyutikov}, M., {Otte}, N., \& {McCann}, A. 2012, \apj, 754, 33

\bibitem[{{Magic Collaboration}(2011)}]{Aleksic11}
{Magic Collaboration}. 2011, \apj, 742, 43

\bibitem[{{Magic Collaboration}(2012)}]{Aleksic12}
{Magic Collaboration}. 2012, \aap, 540, A69

\bibitem[{{Manchester} {et~al.}(2005){Manchester}, {Hobbs}, {Teoh}, \&
  {Hobbs}}]{2005AJ....129.1993M}
{Manchester}, R.~N., {Hobbs}, G.~B., {Teoh}, A., \& {Hobbs}, M. 2005, \aj, 129,
  1993

\bibitem[{{McMillan} \& {Binney}(2010)}]{2010MNRAS.402..934M}
{McMillan}, P.~J. \& {Binney}, J.~J. 2010, \mnras, 402, 934

\bibitem[{Ng {et~al.}(2014)Ng, Takata, Leung, Cheng, \&
  Philippopoulos}]{Ng:2014qja}
Ng, C.~Y., Takata, J., Leung, G., Cheng, K., \& Philippopoulos, P. 2014

\bibitem[{{Ng} {et~al.}(2014){Ng}, {Takata}, {Leung}, {Cheng}, \&
  {Philippopoulos}}]{2014arXiv1405.2148N}
{Ng}, C.-Y., {Takata}, J., {Leung}, G.~C.~K., {Cheng}, K.~S., \&
  {Philippopoulos}, P. 2014, ArXiv e-prints

\bibitem[{{Siegal-Gaskins} {et~al.}(2011){Siegal-Gaskins}, {Reesman},
  {Pavlidou}, {Profumo}, \& {Walker}}]{2011MNRAS.415.1074S}
{Siegal-Gaskins}, J.~M., {Reesman}, R., {Pavlidou}, V., {Profumo}, S., \&
  {Walker}, T.~P. 2011, \mnras, 415, 1074

\bibitem[{{Sreekumar} {et~al.}(1998){Sreekumar}, {Bertsch}, {Dingus},
  {Esposito}, {Fichtel}, {Hartman}, {Hunter}, {Kanbach}, {Kniffen}, {Lin},
  {Mayer-Hasselwander}, {Michelson}, {von Montigny}, {Muecke}, {Mukherjee},
  {Nolan}, {Pohl}, {Reimer}, {Schneid}, {Stacy}, {Stecker}, {Thompson}, \&
  {Willis}}]{1998ApJ...494..523S}
{Sreekumar}, P., {Bertsch}, D.~L., {Dingus}, B.~L., {et~al.} 1998, \apj, 494,
  523

\bibitem[{{Story} {et~al.}(2007){Story}, {Gonthier}, \&
  {Harding}}]{2007ApJ...671..713S}
{Story}, S.~A., {Gonthier}, P.~L., \& {Harding}, A.~K. 2007, apj, 671, 713

\bibitem[{Tamborra {et~al.}(2014)Tamborra, Ando, \& Murase}]{Tamborra:2014xia}
Tamborra, I., Ando, S., \& Murase, K. 2014

\bibitem[{{Taylor} \& {Cordes}(1993)}]{1993ApJ...411..674T}
{Taylor}, J.~H. \& {Cordes}, J.~M. 1993, \apj, 411, 674

\bibitem[{{Usov}(1983)}]{Usov83}
{Usov}, V.~V. 1983, \nat, 305, 409

\bibitem[{{VERITAS Collaboration}(2011)}]{Aliu11}
{VERITAS Collaboration}. 2011, Science, 334, 69

\bibitem[{{Yuan} \& {Zhang}(2014)}]{2014arXiv1404.2318Y}
{Yuan}, Q. \& {Zhang}, B. 2014, ArXiv e-prints

\bibitem[{Yusifov \& Kucuk(2004)}]{Yusifov:2004fr}
Yusifov, I. \& Kucuk, I. 2004, Astron.Astrophys., 422, 545

\end{thebibliography}

\end{document}